\documentclass[11pt]{article}
\usepackage[utf8]{inputenc}
\usepackage{amsmath,amssymb,mathtools}    
\allowdisplaybreaks[4]  
\usepackage{geometry}[top=1cm,bottom=1cm,left=1cm,right=1cm]
\usepackage{color}
\usepackage{graphicx}
\usepackage{float}
\usepackage{subfigure}
\usepackage{cite}                
\usepackage{hyperref}           
\usepackage{multirow,makecell}   
\usepackage{textcomp}
\usepackage{wasysym}
\usepackage{bm}

\graphicspath{{newfigures/}}
\title{Thermoelectric Transport in Holographic Quantum Matter under Shear Strain}
\author{
~~Teng Ji$^{a,b}$\footnote{jiteng@itp.ac.cn}~,
~~Li Li$^{a,b,c,d}$\footnote{liliphy@itp.ac.cn}~,
~~Hao-Tian Sun $^{a,b}$\footnote{sunhaotian@itp.ac.cn}~,
\\
\small $^a$CAS Key Laboratory of Theoretical Physics, Institute of Theoretical Physics, \\
\small Chinese Academy of Sciences, Beijing 100190, China\\
\small $^b$School of Physical Sciences, University of Chinese Academy of Sciences, \\
\small No.19A Yuquan Road, Beijing 100049, China\\
\small $^c$School of Fundamental Physics and Mathematical Sciences, Hangzhou Institute for Advanced Study, \\
\small UCAS, Hangzhou 310024, China\\
\small $^d$Peng Huanwu Collaborative Center for Research and Education, Beihang University, \\
\small Beijing 100191, China.
}
\date{}
\begin{document}

\maketitle

\begin{abstract}
We study the thermoelectric transport under shear strain in two spatial dimensional quantum matter using the holographic duality. General analytic formulae for the DC thermoelectric conductivities subjected to finite shear strain are obtained in terms of the black hole horizon data. Off-diagonal terms in the conductivity matrix also appear at zero magnetic field, resembling an emergent electronic nematicity, which cannot nevertheless be identified with the presence of an anomalous Hall effect. For an explicit model study, we numerically construct a family of strained black holes and obtain the corresponding nonlinear stress-strain curves. We then compute all electric, thermoelectric, and thermal conductivities and discuss the effects of strain. While the shear elastic deformation does not affect the temperature dependence of thermoelectric and thermal conductivities quantitatively, it can strongly change the behavior of the electric conductivity. For both shear hardening and softening cases, we find a clear metal-insulator transition driven by the shear deformation. Moreover, the violation of the previously conjectured thermal conductivity bound is observed for large shear deformation.
\end{abstract}

\newpage

\tableofcontents

\newpage

\section{Introduction}
The transport coefficients characterize the intrinsic properties of a system and are fundamental for understanding the transport phenomena that are ubiquitous in nature. 
For example, the electric or thermal conductivities are associated with the conduction of a system due to the electric or heat current. Other transport coefficients, such as the shear and bulk, viscosities are related to the resistance of a system under the shear and tensile stress. While these coefficients could be quantitatively obtained in experiments, it is usually challenging to obtain them from first-principles calculations. Thanks to the development of the holographic duality, the transport coefficients of strongly coupled systems can be easily computed by analyzing small perturbations about their dual black holes. The most famous example is a universal ratio between the shear viscosity and the entropy density, known as the Kovtun-Son-Starinets (KSS) bound~\cite{Kovtun:2004de}. Interesting applications of holography can be found, \emph{e.g.} in some recent reviews~\cite{Hartnoll:2016apf,Baggioli:2022pyb,Cai:2015cya,Zaanen:2021llz,Liu:2020rrn,Landsteiner:2019kxb,Chen:2022goa}.

The study of the transport behaviors from holography has attracted much attention in connection to the ``anomalous" experimental observations in quantum matter (\emph{e.g.} strange metals~\cite{Kim:2010zq,Amoretti:2015gna,Kiritsis:2016cpm,Ge:2016sel,Blauvelt:2017koq,Cremonini:2018kla}). Although most works have concentrated on cases in the absence of mechanical deformation, stress/strain effects could be the most obvious way to statically or dynamically tune transport. From a microscopic perspective, stress/strain directly induces lattice deformation and modulates phonon propagation, thus could significantly affect material properties (electrical, optical,  mechanical, and so on). Strain engineering is now recognized as a promising approach to  generating novel effects in graphene and other two-dimensional materials~\cite{Amorim:2015bga}. It has also been widely used as a common tool in amorphous systems~\cite{Nicolas:2018}. Moreover, there is a growing interest related to the nonlinear mechanical characterization of high-temperature superconductors (see, \emph{e.g.}~\cite{Abrecht:2003,Malinowski:2020,Kostylev:2019ezg,Hameed:2005}). In holography, the nonlinear mechanical response was first introduced in~\cite{Baggioli:2020qdg}, where the nonlinear stress-strain curves, as well as an estimate of the elasticity bounds in the elastic regime were discussed. More recently, based on the effective field theory and holographic duality, a potential relationship among the nonlinear elasticity, yielding and entropy of granular matter was suggested and then verified by computer simulations of granular models~\cite{Pan:2021cux}. The thermodynamic and mechanical properties of a holographic bottom-up model for the supersolid were considered in~\cite{Baggioli:2022aft}. Interestingly, shear flows in some holographic models were investigated in~\cite{Baggioli:2021tzr} where the viscosity-entropy density ratio shows some universality even far away from equilibrium.\,\footnote{See also~\cite{Baggioli:2019mck} for the oscillatory shear tests in a holographic model with finite elastic response in the large amplitude oscillatory shear regime.}

For a strongly coupled system at finite temperature, $T$, and charge density, $\rho$, its response to the application of an electric field $E_i$ and a temperature gradient $\partial_i T$ is characterized by the thermoelectric transport coefficients in linear response:
\begin{align}\label{eq:Ohm}
\left(\begin{array}{c}J^i\\Q^i\end{array}\right)=
\left(\begin{array}{cc}\sigma^{ij}&\alpha^{ij} T\\
\bar\alpha^{ij} T&\bar \kappa^{ij} T
\end{array}\right)
\left(\begin{array}{c}
E_j\\
-\partial_i T/T
\end{array}\right)\,,
\end{align}
where $J^i$ is the electric current and $Q^i$ is the heat current. Here $i, j$ refer to space indices and sum over repeated indices is understood. In the above expression the matrix $\sigma$ is the electrical conductivity, $\alpha$ and $\bar{\alpha}$ are the thermoelectric conductivities and $\bar{\kappa}$ is the thermal conductivity. Moreover, when the system respects time-reversal symmetry, the Onsager reciprocal relation yields $\alpha=\bar{\alpha}$. To have a finite DC conductivity that is crucial for describing real materials, a mechanism for momentum dissipation is necessary. It is remarkable that an analytic expression for the DC conductivity can be obtained in terms of black hole horizon data~\cite{Donos:2015gia,Banks:2015wha}. The simplest and most convenient way to realize the momentum dissipation in holography is the so-called linear axion models~\cite{Andrade:2013gsa} (more details and generalizations can be found in a recent review~\cite{Baggioli:2021xuv}).

As an initial study of holographic transport subject to nonlinear mechanical strain, we focus on the transition between metallic and insulating behavior in the present work. As a fascinating and enduring area of study in condensed matter physics, the understanding of metal-insulator transition (MIT) is still controversial and incomplete (see~\emph{e.g.}~\cite{Imada:1998zz,MIT:2011} for reviews). A minimal holographic setup in two spatial dimensions was given in~\cite{Baggioli:2016oqk,Gouteraux:2016wxj} where the DC electric conductivity was shown a transition from metallic to insulating behavior driven by disorder. Moreover, the magnetotransport was then discussed and the temperature dependence of resistivity was found to be well scaled with a single parameter~\cite{An:2020tkn}, which agrees qualitatively with the experimental observation in the literature. We shall extend previous studies to the full thermoelectric conductivities~\eqref{eq:Ohm} and turn on nonlinear shear strain. In particular, we shall show how the effect of mechanical deformation on the thermoelectric transport for a strongly coupled system that allows a holographic dual description. We will find that there are non-vanishing off-diagonal components of conductivity tensors even in the absence of magnetic field. We will show that this phenomenon bears no relation with the anomalous Hall effect, but is a manifestation of a strong electronic nematicity due to the strain-induced anisotropy. Moreover, the electric conductivity will be shown to be sensitive to the shear deformation. For both shear hardening
and softening cases, we will observe a clear transition between an insulating phase and a metallic phase by increasing the strength of mechanical deformation. This transition will be shown to be closely associated with the electronic nematicity and pressure anisotropy. Interestingly, we will find that the previously conjectured thermal conductivity bound~\cite{Grozdanov:2015djs} can be violated by increasing the shear deformation.

The organization of this paper is as follows. In Section \ref{model}, we introduce the holographic setup. We derive the general formulae for the DC electric, thermoelectric, and thermal conductivities at finite elastic mechanical deformation in Section~\ref{sec:cond}. In particular, we classify the physical origin of the off-diagonal components of the thermoelectric transport. In Section \ref{softeningandhardening}, we illustrate two kinds of behaviors under shear deformation, \emph{i.e.} the shear hardening and softening. In section~\ref{transport}, we study in detail the effect of shear deformation on the DC transport. We conclude with further discussions in Section~\ref{conclusion}. The holographic renormalisation procedure is performed in detail in Appendix~\ref{appendix}. The electric transport under shear strain by dialing the charge density is summarized in Appendix~\ref{app:charge}.

\section{Holographic Model}\label{model}
We consider the following holographic model in four dimensions.
\begin{align}\label{eq:S}
\mathcal{S}_0=\int d^{4} x \sqrt{-g}\left[R-2 \Lambda-2 W(X, Z)-\frac{Y(X_0)}{4}F_{\mu\nu}F^{\mu\nu}- V(X_0)\right]\,,
\end{align}
where $\Lambda$ is the negative cosmological constant and $F_{\mu\nu}$ the field strength for the $U(1)$ gauge field encoding the charge degrees of freedom. There are two sets of massless scalar fields $\phi^I$ and $\varphi^I (I=1,2)$ known as holographic axions (see~\cite{Baggioli:2021xuv} for a recent review). The former  provides a mechanism for momentum dissipation, while the latter for the elastic mechanical deformation. $Y$ and $V$ are functions of the translation breaking sector $X_0\equiv\frac{1}{2}g^{\mu\nu}\partial_\mu \phi^I \partial_\nu\phi^I$. The precise form of mechanical deformation is determined by the potential term $W(X, Z)$ where $\mathcal{X}_{I J} \equiv \partial_{\mu} \varphi_I \partial^{\mu} \varphi_J$, $X \equiv \frac{1}{2} \operatorname{Tr}\left(\mathcal{X}_{I J}\right)$ and $Z \equiv \operatorname{det}\left(\mathcal{X}_{I J}\right)$.\,\footnote{The idea of splitting explicit and spontaneous symmetry breaking using two sets of scalars was discussed, \emph{e.g.} in~\cite{Donos:2019hpp,Amoretti:2019kuf}.}

The general equations of motion are given by
\begin{equation} \label{eom1}
\begin{split}
&\nabla_{\mu}\left(W_{X} \nabla^{\mu} \varphi^1+2 W_{Z}\nabla^{\mu}\varphi^1 \mathcal{X}^{22}-2 W_{Z} \nabla^{\mu} \varphi^{2}\mathcal{X}^{12}\right)=0\,, \\
&\nabla_{\mu}\left(W_{X} \nabla^{\mu} \varphi^2+2 W_{Z}\nabla^{\mu}\varphi^2 \mathcal{X}^{11}-2 W_{Z} \nabla^{\mu} \varphi^{1}\mathcal{X}^{21}\right)=0\,, \\
&\nabla_\mu[Y(X_0) F^{\mu\nu}]=0\,,  \\
&\nabla_{\mu}\left[\left(\frac{Y^{\prime}(X_0)}{4} F_{\mu \nu} F^{\mu \nu}+V^{\prime}(X_0)\right) \nabla^{\mu} \phi^{I}\right]=0\,,  \\
&R_{\mu \nu}-\frac{1}{2}R-g_{\mu \nu}+\Lambda g_{\mu \nu}=\mathcal{T}_{\mu\nu}\,, 
\end{split}
\end{equation}
with $\mathcal{T}_{\mu\nu}$ the energy-momentum tensor
\begin{align}
&T_{\mu\nu}=-\frac{1}{2} g_{\mu \nu}\left[\frac{Y(X_0)}{4} F_{\mu \nu} F^{\mu \nu}+V(X_0)+2 W(X,Z)\right] \notag \\
&+\frac{1}{2}\left[\frac{Y^{\prime}(X_0)}{4} F_{\mu \nu} F^{\mu \nu}+ V^{\prime}(X_0)\right] \sum_{I=1}^{2}\partial_{\mu} \phi^{I} \partial_{\nu} \phi^{I}+\frac{Y(X_0)}{2} F_{\mu \rho} F_{\nu}^{\rho} \notag+W_X \sum_{I=1}^{2}\partial_{\mu} \varphi^{I} \partial_{\nu} \varphi^{I}\\
&+2 W_Z (\mathcal{X}^{22}\partial_{\mu} \varphi^{1} \partial_{\nu} \varphi^{1}+\mathcal{X}^{11}\partial_{\mu} \varphi^{2} \partial_{\nu} \varphi^{2}-\mathcal{X}^{12}\partial_{\mu} \varphi^{2} \partial_{\nu} \varphi^{1}-\mathcal{X}^{21}\partial_{\mu} \varphi^{1} \partial_{\nu} \varphi^{2})\,, \label{eom2}    
\end{align}
and
\begin{align}
&W_Z=\frac{\partial W(X,Z)}{\partial Z},\quad W_X=\frac{\partial W(X,Z)}{\partial X}\,.
\end{align}

To study the system at finite mechanical deformation and charge density, we take the following bulk geometry.
\begin{equation}\label{metric}
\begin{aligned}
d s^{2}&=\frac{1}{u^{2}}\left(-f(u) e^{-\chi(u)} d t^{2}+\frac{d u^{2}}{f(u)}+\gamma_{i j}(u) d x^{i} d x^{j}\right),\\
A_\mu dx^\mu&=A_t(u)dt,\\
\phi^I&=k\, {\delta^I}_j x^j,\qquad \varphi^I=O_{j}^{I} \,x^{j},
\end{aligned}
\end{equation}
where $\gamma_{ij}$ and $O^I_j$ are parameterized by
\begin{align}\label{deformation}
\gamma_{ij}=\left(\begin{array}{cc}
\cosh h(u) & \sinh h(u) \\
\sinh h(u) & \cosh h(u)
\end{array}\right),\quad
O^I_j=\beta \left(\begin{array}{cc}
\cosh \Omega/2 & \sinh \Omega/2 \\
\sinh \Omega/2 & \cosh \Omega/2
\end{array}\right)\,.
\end{align}
Here the constant $k$ captures an averaged description of disorder strength coming from a homogeneously distributed set of impurities~\cite{Baggioli:2016oqk}. The parameter $\beta$ corresponds to the bulk deformation and the parameter $\Omega$ to the shear deformation $\varepsilon$ via\,\footnote{From an effective field theory point of view, the general mechanical deformation can be parametrized by the matrix~\cite{Alberte:2018doe}
\begin{align}
O^I_j=\beta\left(\begin{array}{cc}
\sqrt{1+\varepsilon^{2} / 4} & \varepsilon / 2 \\
\varepsilon / 2 & \sqrt{1+\varepsilon^{2} / 4}
\end{array}\right)\,,\nonumber
\end{align}
with the two parameters $\beta$ and $\varepsilon$ directly related to a background bulk strain and a background shear strain. $\beta$ controls the change of volume in the system, while $\varepsilon$ implements a pure shear.
}
\begin{equation}\label{deformation}
\varepsilon=2\sinh (\Omega/2)\,.
\end{equation}
Unsheared configurations correspond to $\Omega=0$. The holographic coordinate $u$ spans from the AdS boundary $u=0$ to the black hole horizon $u=u_h$ where $f(u_h)$ vanishes. The temperature and entropy density of the background geometry~\eqref{metric} is given by
\begin{equation}\label{Tem}
T=-\frac{f'(u_h)}{4\pi}e^{-\chi(u_h)/2},\quad s=4\pi/u_h^2 \,.
\end{equation}

Substituting the ansatz~\eqref{metric} into~\eqref{eom1}, we obtain the following equations of motion.
\begin{align}
&\chi'-\frac{1}{2} u h'^2=0\,,\label{eomchi}\\
 &\left[Y(\bar X_0) e^{\chi/2}A_t'\right]'=0\,,\label{eomAt} \\
  &4 \beta ^2u \sinh (\Omega -h) W_{ X}(\bar X,\bar Z)+k^2 u \sinh (h) \left(u^4
   e^{\chi} A_t'^2 Y'(\bar X_0)-2V'(\bar X_0)\right)+2 u f' h'\notag\\
   &-\frac{1}{2}f\left(-4 u h''+u^2 h'^3+8 h'\right)=0\,,\\
   &-4 W\left(\bar X,\bar Z\right)+4 \beta^2 u^2 \cosh (h-\Omega ) W_{X}\left(\bar X,\bar Z\right)+8 \beta ^4 u^4 W_{Z}\left(\bar X,\bar Z\right)+u^4 e^{\chi } A_t'^2Y\left(\bar X_0\right)\notag\\
   &-k^2 u^6 e^{\chi}A_t'^2 \cosh (h) Y'\left(\bar X_0\right)-2u^2 f''+\frac{3}{2} u^3 f'h'^2+8 u f'-2V\left(\bar X_0\right)-4 \Lambda\notag\\
&-\frac{1}{4} f\left(u^4 h'^4+8 u^2 h'^2-8 u^3 h'h''+48\right)+2 k^2 u^2 \cosh (h) V'\left(\bar X_0\right)=0\,,\\
&4 W(\bar X,\bar Z)+u^4 e^{\chi} A_t'^2 Y(\bar X_0)-4 u f'+f \left(u^2 h'^2+12\right)+2 V\left(k^2u^2 \cosh (h)\right)+4 \Lambda=0\,,\label{constrainteom}
\end{align}
where we have introduced
\begin{align}
    \bar X_0=k^2 u^2 \cosh (h),\ \ \bar X=\beta^2 u^2
\cosh (\Omega-h),\ \ \bar Z=\beta^4 u^4\,.
\end{align}
We point out that the last equation of motion~\eqref{constrainteom} is a first order one and can be derived from other equations.

In the absence of mechanical deformation, the above equations of motion can be solved analytically. For the case with finite shear strain/stress, one has to solve the coupled system~\eqref{eomchi}-\eqref{constrainteom} numerically. When solving the above equations, we impose the regularity condition at the horizon $u=u_h$ where $f(u_h)=A_t(u_h)=0$ and both $h(u_h)$ and $\chi(u_h)$ are finite. At the UV boundary $u=0$, we have
\begin{equation}\label{adsUV}
\begin{split}
f(u)=1-\frac{k^2}{2}u^2+f^{(3)}u^3+\dots, \quad A_t(u)=\mu-\rho\, u+\dots\,,\\
\quad \chi(u)=0+\dots, \quad h(u)=h^{(3)} u^3+\dots\,,
\end{split}
\end{equation}
where dots represent higher powers of $u$. Note that the time scale of the time coordinate is fixed by requiring $\chi(u=0)=0$ such that the temperature at the UV boundary is equal to the standard Hawking temperature~\eqref{Tem}. We identify $\mu$ and $\rho$ to be the chemical potential and the charge density of the boundary system, respectively.\,\footnote{Without loss of generality, we have required $Y(X_0)=1$ at the AdS boundary $u=0$.}  From the Maxwell equation~\eqref{eomAt}, there is a conserved charge in the radial direction
\begin{equation}\label{rhoeq}
-Y(\bar X_0) e^{\chi/2}A_t'=\rho\,,
\end{equation}
which can be evaluated at any value of $u$ including $u=u_h$.

The constants $f^{(3)}$ and $h^{(3)}$ correspond to the energy density and shear stress, respectively.~\footnote{The asymptotic expansion of $h(u)$ near the boundary $u=0$ reads
\begin{align}
h(u)=h^{(0)}+\cdots+\frac{1}{6}h^{(3)} u^3+\cdots\,,\nonumber
\end{align}
where $h^{(0)}$ is identified with the source of the stress tensor $\hat T^{xy}$ and $h^{(3)}$ is related to the expectation value of $\hat T^{xy}$. We turn off the sources of spacetime deformation, \emph{i.e.} $h^{(0)}=0$, so the only possible source for the stress tensor is from the mechanical strain deformation encoded in the scalars $\varphi^I$.}
\begin{align}\label{ESig}
\mathcal{E}\equiv \left\langle \hat T_{tt}\right\rangle=-2f^{(3)},\quad \Sigma\equiv \left\langle \hat T_{xy}\right\rangle=3 h^{(3)}\,.
\end{align}
where $\left\langle \hat T_{\mu\nu}\right\rangle$ is the stress tensor of the dual boundary system.
We point out that the profile $\varphi^I$  of~\eqref{metric} is the vacuum expectation value rather than the source of the dual theory. Therefore, the spatial anisotropy of the boundary system appears spontaneously.~\footnote{For the explicit translation symmetry breaking encoded by profile $\phi^I$ of~\eqref{metric}, the resulted geometry is still isotropic. The anisotropy of the geometry is only due to the shear deformation that is given by the vacuum expectation value without any source. In this sense, we mean the rotation symmetry is broken spontaneously.}
Please consult Appendix~\ref{appendix}  for more details of the holographic renormalization.

\section{Computation of Transport Coefficients}\label{sec:cond}
Holography provides an elegant prescription for calculating transport coefficients of strongly coupled systems. In particular, there are analytic formulas for DC conductivities in terms of black hole horizon data~\cite{Donos:2015gia,Banks:2015wha}.
To study the transport properties of the dual system we apply this method to our theory at finite shear deformation. In particular, we will classify the physical origin of the off-diagonal components of the thermoelectric transport induced by shear strain.

\subsection{Conductivities from black hole horizons}
We turn on small perturbations around the background~\eqref{metric}~\footnote{From the Ward identity~\eqref{app:Ward} in Appendix~\ref{appendix}, one can check that the momentum does dissipate at the level of fluctuations due to the contribution of $\phi^I$.}
\begin{equation}\label{pertur}
\begin{split}
\delta A_i&=\left[-E_i+\zeta_i A_t\left(u\right)\right]t+\delta a_i\left(u\right)\,,\\
\delta g_{iu}&= \frac{h_{iu}(u)}{u^2},\ \ \delta g_{ti}=h_{ti}(u)-\frac{f(u)}{u^2} \zeta_i e^{-\chi(u)}  t\,,\\
\delta \phi_{i}&=\delta \phi_{i}(u)\,,\ \ 
\delta \varphi_{i}=\delta \varphi_{i}(u)\,,
\end{split}
\end{equation}
where $i=x,y$.  $E_i$ is the electric field and $\zeta_i$ is identified as the temperature gradient of the boundary theory, \emph{i.e.} $\zeta_i=-\partial_i T/T$. The next key ingredient is to introduce the following bulk currents $J_i$ and $Q_i$.
\begin{equation}\label{current}
\begin{split}
J^{i}&=\sqrt{-g}Y(\bar X_0)F^{ui}\,,\\
Q^{i}&=2\sqrt{-g}\nabla^u\xi^i-A_t(u)\sqrt{-g}Y(\bar X_0)F^{ui}\,,
\end{split}
\end{equation}
where $\xi^\mu$ is chosen to be the Killing vector field $\partial_t$. Evaluating them at the UV boundary, one can find that they are, respectively, the electric current and heat current of the dual field theory. Moreover, one can verify that both currents are radially conserved, \emph{i.e.} $\partial_u J^i=\partial_u Q^{i}=0$ by substituting into the equations of motion~\eqref{eom1}. Therefore, they can be calculated anywhere in the bulk. The strategy is to evaluate them at the event horizon $u=u_h$, where $\delta g_{ti}, \delta \phi_{i}, \delta \varphi_{i}$ are finite and $\delta g_{iu}$ can be expressed by the constraint of regularity. More precisely, one has the following relation for perturbations near $u_h$.
\begin{equation}\label{horizonrelation}
\begin{split}
\delta a_i'(u)&=\frac{E_i}{f(u)}e^\frac{\chi(u)}{2}+\dots\,,\\
h_{iu}(u)&=-\frac{u^2}{f(u)} e^{\frac{\chi(u)}{2}}h_{ti}(u)+\dots\,.
\end{split}
\end{equation}
%
Inserting~\eqref{horizonrelation} and evaluating at the horizon $u=u_h$, both currents in~\eqref{current} are thus given by their values at the black hole horizon, which in turn are fixed by $E_i$ and $\zeta_i$.\footnote{One also needs to express $h_{iu}(u)$ in terms of $E_i, \zeta_i$ as well as background fields by using the Einstein's equation.}

Finally, we are able to obtain all DC conductivities via~\eqref{eq:Ohm}. More explicitly, with the notation
  \begin{equation} 
 \begin{split}
    D_x&=\frac{C-f' h'\tanh h}{C^2-(2f' h')^2},\quad D_y=\frac{C-f' h'\coth h}{C^2-(2f' h')^2}\,,\\
    C&=\frac{4 \Lambda+4 W+2V}{u_h^2}-\left( 3 f'\chi '+8f'/u_h-2 f''+ u_h^2 e^{\chi} Y A_t'^2\right)\,,
 \end{split}
 \end{equation}
the general DC conductivity formulae are given as follows.
\begin{itemize}
  \item The electric conductivity tensor $\sigma$ with
 \begin{equation}\label{elecDC}
 \begin{split}
 \sigma_{xx}&=\sigma_{yy}=\left(Y+\frac{8\pi\rho^2}{s}D_x\right)\cosh h\,,\\
  \sigma_{xy}&=\sigma_{yx}=-\left(Y+\frac{8\pi\rho^2}{s}D_y\right)\sinh h\,.
 \end{split}
 \end{equation}
  \item The thermoelectric conductivity tensor $\alpha$ with
 \begin{equation}\label{thermelec}
 \begin{split}
 \alpha_{xx}&=\alpha_{yy}={8\pi \rho}D_x\cosh h\,,\\
  \alpha_{xy}&=\alpha_{yx}=-{8\pi \rho}D_y\sinh h\,,\\
  \bar{\alpha}&=\alpha\,.
 \end{split}
 \end{equation}

  \item The thermal conductivity tensor $\bar{\kappa}$ with
 \begin{equation}\label{thermalDC}
 \begin{split}
\bar{\kappa}_{xx}&=\bar{\kappa}_{yy}={8\pi s T}D_x\cosh h\,,\\
\bar{\kappa}_{xy}&=\bar{\kappa}_{yx}=-{8\pi s T}D_y\sinh h\,.
 \end{split}
 \end{equation}
\end{itemize} 
All functions in the above expressions will be understood to be evaluated at the horizon $u=u_h$. We have also used~\eqref{Tem} and~\eqref{rhoeq} for $T, s$ and $\rho$.

It is manifest that, in the linear response regime, the conductivity tensors $(\sigma, \alpha, \bar{\kappa})$ can be fully determined by the horizon data of the background solutions. Note that there is a simple relation between $\bar{\kappa}$ and $\alpha$ that is given by
\begin{equation}\label{ratio}
\bar{\kappa}=\frac{Ts}{\rho} \alpha\,,
\end{equation}
irrespective of the model details. By turning off the mechanical deformation, we obtain the following unstrained results:
\begin{equation}\label{nostrain}
\begin{split}
\sigma_0&\equiv\sigma_{xx}(\beta=1,\varepsilon=0)=Y+\frac{4\pi \rho^{2}}{s}\frac{1}{2 W_X+4u^2 W_Z+k^{2}\left(V^{\prime}-\frac{\rho^{2} Y^{\prime} u^{4}}{2 Y^{2}}\right)}\,,\\
\alpha_0&\equiv\alpha_{xx}(\beta=1,\varepsilon=0)=\frac{4\pi \rho}{2 W_X+4u^2 W_Z+k^{2}\left(V^{\prime}-\frac{\rho^{2} Y^{\prime} u^{4}}{2 Y^{2}}\right)}\,,\\
\bar{\kappa}_0&\equiv\bar{\kappa}_{xx}(\beta=1,\varepsilon=0)=\frac{4\pi s T}{2 W_X+4u^2 W_Z+k^{2}\left(V^{\prime}-\frac{\rho^{2} Y^{\prime} u^{4}}{2 Y^{2}}\right)}\,,\\
\end{split}
\end{equation}
with all off-diagonal components vanishing. As a consistency check, the DC conductivity precisely recovers the result in~\cite{Baggioli:2016oqk,An:2020tkn,Baggioli:2016pia} when $W=0$. 

\subsection{Origin of the off-diagonal components}
We obtain the finite off-diagonal components of conductivities subject to finite shear strain in the linear response.
Thus, a natural question is the physical origin of such off-diagonal components.

Let's focus on the electric conductivity for which we have a symmetric part $\sigma_{xy}=\sigma_{yx}$.  The symmetric part of the off-diagonal conductivity tensor was identified as anomalous Hall conductivity in~\cite{Grandi:2021bsp}. Nevertheless, we should point out that Hall conductivity necessarily requires breaking of time-reversal symmetry and is antisymmetric ($\sigma_{xy}=-\sigma_{yx}$) as a consequence of Onsager reciprocity relations\footnote{Note, however, that nonlinear Hall effect can occur in systems with time-reversal symmetry~\cite{Carmine:2014}.}. Note that $\alpha=\bar{\alpha}$~\eqref{thermelec} in our holographic model, thus respecting the time-reversal symmetry. Another way to understand the symmetric part of the off-diagonal conductivity tensor is as follows~\cite{Wu:2017}. Given that the system becomes anisotropic due to the shear strain~\eqref{deformation}. In the coordinate system $\bm{e}_a, \bm{e}_b$ defined by the principal axes (eigendirections) of the strain tensor~\eqref{straintensor}, the electric conductivity tensor becomes diagonal
\begin{equation}\label{princpsig}
\hat{\sigma}=\begin{pmatrix}
\sigma_a      &  0  \\
  0    &  \sigma_b
\end{pmatrix}=
\begin{pmatrix}
\bar{\sigma}+\triangle\sigma      &  0  \\
  0    &  \bar{\sigma}-\triangle\sigma
\end{pmatrix}\,,
\end{equation}
where the anisotropy is encoded in $\triangle\sigma\equiv(\sigma_a-\sigma_b)/2$.
Now we rotate the axes by an angle $\theta$ and measure the conductivity in this new coordinate. Then the conductivity matrix~\eqref{princpsig} changes accordingly and becomes
\begin{equation}\label{sigtheta}
\sigma(\theta)=\begin{pmatrix}
\cos\theta      & - \sin\theta  \\
  \sin\theta    &  \cos\theta
\end{pmatrix}
\hat{\sigma}
\begin{pmatrix}
\cos\theta      &  \sin\theta  \\
  -\sin\theta    &  \cos\theta
\end{pmatrix}=
\begin{pmatrix}
\bar{\sigma}+\triangle\sigma\cos(2\theta)      & \triangle\sigma\sin(2\theta)    \\
 \triangle\sigma\sin(2\theta)     &  \bar{\sigma}-\triangle\sigma\cos(2\theta) 
\end{pmatrix}\,.
\end{equation}
It is now clear that there must occur a spontaneous transverse voltage once the current is not aligned with one of the principal axes. We show a polar-coordinate plot of $\sigma(\theta)_{xy}$ in Fig.~\ref{fig:sigxyangle}, where the ‘cloverleaf’ shape characteristic of $d$-orbital symmetry is manifest.
\begin{figure}[H]
    \centering
    \includegraphics[width=0.60\linewidth]{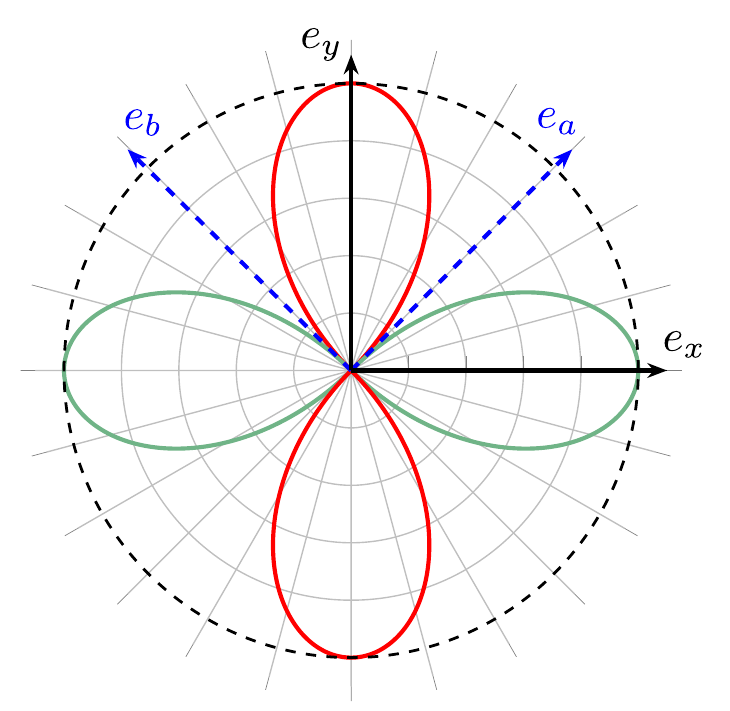}
    \caption{Illustration of the off-diagonal component of $\sigma(\theta)$ in polar coordinates. The radial distance measures its magnitude, with the positive values in red and negative in green. The two solid dark arrows denote the unit vectors $\bm{e}_x$ and $\bm{e}_y$ along the $x$ and $y$ directions, while the dashed blue arrows correspond to the unit vectors $\bm{e}_a, \bm{e}_b$ defined by the principal axes of the stress tensor~\eqref{straintensor}.}\label{fig:sigxyangle}
\end{figure}

Note that all conductivities~\eqref{elecDC}-\eqref{thermalDC} are $2\times 2$ real symmetric matrices. Therefore, they share the same eigendirections $\bm{e}_a=\frac{1}{\sqrt{2}}(\bm{e}_x+\bm{e}_y)$ and $\bm{e}_b=\frac{1}{\sqrt{2}}(-\bm{e}_x+\bm{e}_y)$, where $\bm{e}_x$ and $\bm{e}_y$ are, respectively, the unit vectors along the $x$ and $y$ directions of our original coordinate system~\eqref{metric}. Meanwhile, the strain tensor~\eqref{straintensor} is diagonalized with the pressure $P_a=\mathcal{E}/2+\Sigma$ along $e_a$ and $P_b=\mathcal{E}/2-\Sigma$ along $e_b$ where $\mathcal{E}$ and $\Sigma$ are, respectively, the energy density and shear strain given in~\eqref{ESig}. It is now manifest that $e_a$ and $e_b$ define the principal stress axes.

In our computation, we have forced the currents to flow along $\bm{e}_x$ and $\bm{e}_y$, see~\eqref{pertur}. Therefore, the rotation angle from $\bm{e}_a$ to $\bm{e}_x$ (and, equivalently, $\bm{e}_b$ to $\bm{e}_y$) is $\theta=-\pi/4$. Indeed, by taking $\theta=-\pi/4$, we can obtain our holographic results from~\eqref{sigtheta}.
\begin{equation}\label{offsigma}
\sigma=\begin{pmatrix}
\bar{\sigma}      &  -\triangle\sigma \\
  -\triangle\sigma   &  \bar{\sigma}
\end{pmatrix}=
\begin{pmatrix}
\sigma_{xx}      &  \sigma_{xy} \\
 \sigma_{xy}   &  \sigma_{xx}
\end{pmatrix}\,,
\end{equation}
with $\sigma_{xx}$ and $\sigma_{xy}$ given in~\eqref{elecDC}. It is clear that $\sigma_{xy}=\sigma_{yx}$. Similar discussion applies to $\alpha$ and $\bar{\kappa}$. Motivated by the above observation, a good order parameter quantifying the anisotropy of the system is given by the (relative) magnitude:
\begin{equation}\label{eqN}
N\equiv\frac{\triangle\sigma}{\bar{\sigma}}=-\frac{\sigma_{xy}}{\sigma_{xx}}\,,
\end{equation}
that resembles a strong electronic nematicity in the system.

Before ending this section, we emphasize that $\sigma_{xy}$ in our present theory occurs in zero magnetic field, and is not connected to Hall effect, skew spin scattering, Berry phase, loop currents, and so on~\cite{Wu:2017}.

\section{Shear Softening and Hardening}\label{softeningandhardening}

As a key representation of the material deformation response, stress-strain curves reveal the relationship between the material structure and its mechanical properties. The stress-strain curve $\Sigma(\varepsilon)$ could be easily measured by experiments while it is usually difficult to compute from the microscopic ingredients, especially in strongly coupled systems. Nevertheless, nonlinear elastic response can be obtained using holographic methods, yielding a rich phenomenology of nonlinear elasticity behaviours, see \emph{e.g.}~\cite{Pan:2021cux,Baggioli:2022aft,Baggioli:2021tzr,Baggioli:2019mck}.
Given the full stress-strain curves, one can define the nonlinear elastic shear modulus $G\equiv d\Sigma/d\varepsilon$. For sufficiently small strain, one recovers the standard linear elasticity relation $\Sigma=G_0\, \varepsilon$ where $G_0$ is the linear shear modulus that can be computed by linear response theory.  In the nonlinear regime, shear hardening corresponds to the case where the shear modulus $G$ is positively correlated with shear stress, while the shear softening is evidenced by a rapid decrease of the shear modulus.

\begin{figure}[H]
    \centering
    \includegraphics[width=0.48\linewidth]{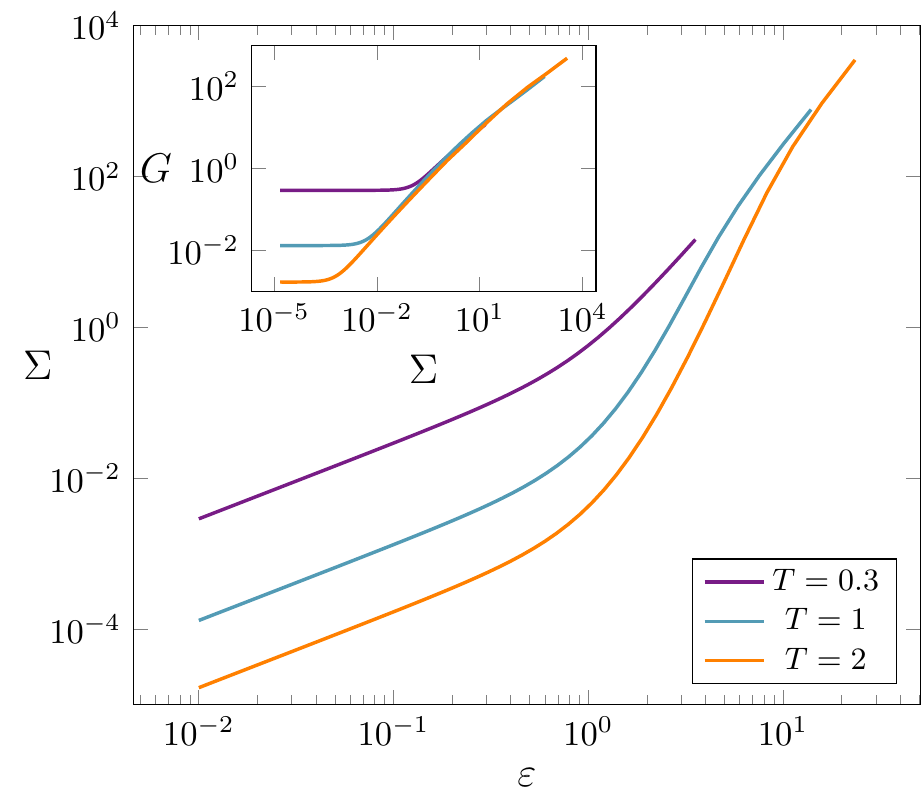}
    \includegraphics[width=0.48\linewidth]{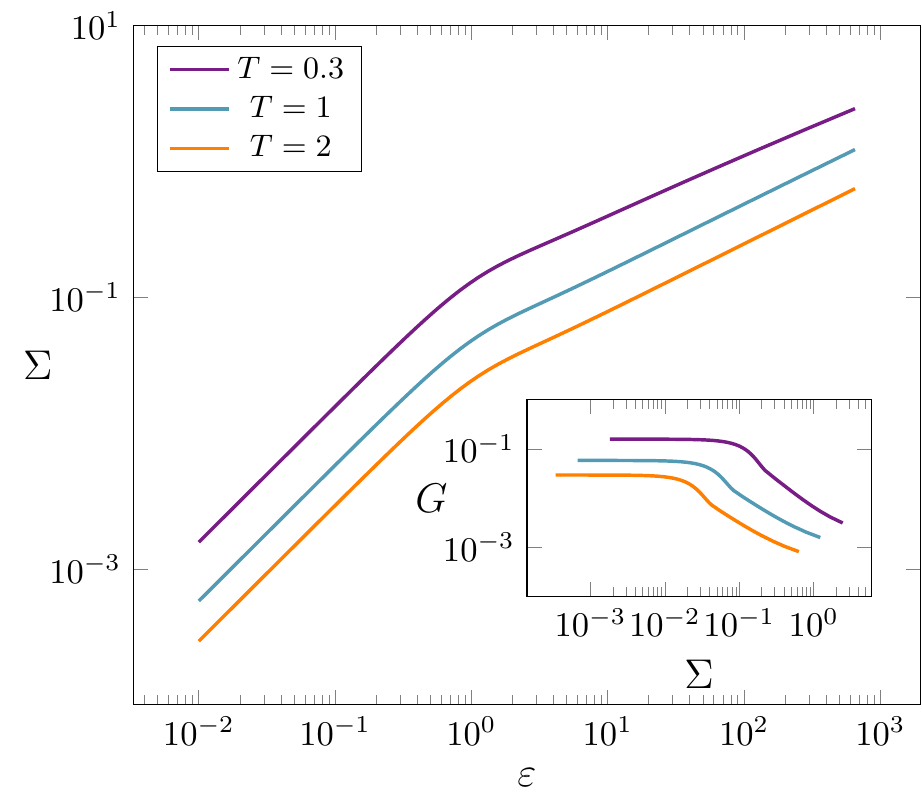}
    \caption{Illustration of shear hardening and shear softening. \textbf{Left:}  Stress-strain curves for $W(X,Z)=X^3$. \textbf{Right:} Stress-strain curves for $W(X,Z)=X^{\frac{1}{4}}Z^{\frac{7}{8}}$. \textbf{Inset:} The shear modulus $G$ in function of the shear stress. Different colors correspond to different temperatures. Other parameters are fixed to be $\beta=k=\rho=1$. }\label{fig:HardeningSoftening}
\end{figure}

We are interested in the pure shear case, so we choose $\beta=1$ \emph{i.e.} no bulk deformation.\footnote{This fixes the scaling symmetry of our system and all physical quantities below are understood to be in units of $\beta$.} In the following discussion, we shall focus on the following representative model with
\begin{align}
Y(X_0)=1-\frac{1}{6} X_0,\ \ \ V(X_0)=X_0\,.
\end{align}
A transition from metallic to insulating phases driven by charge density, disorder and magnetic field was found in the above model~\cite{An:2020tkn}.
The behavior of nonlinear elastic response is encoded in the potential $W(X,Z)$. 
To be specific, we take two benchmark models with monomial potentials $W(X,Z)=X^3$ and $W(X,Z)=X^{\frac{1}{4}}Z^{\frac{7}{8}}$. The stress-strain curves are shown in Fig.~\ref{fig:HardeningSoftening}. In both cases, at small strains $\varepsilon\ll 1$, the response is linear and the slope is given by the unstrained shear modulus $G_0$. Moving away from the linear regime, the stress $\Sigma$ increases monotonically with the strain $\varepsilon$. For the former potential, the shear modulus $G$ increases with the increase of $\varepsilon$, thus describing a shear hardening. In contrast, in the latter case, $G$ decreases with the increase of $\varepsilon$, yielding shear softening. In a sense, it means that the softening material is less sensitive to the shear strain.

\section{DC Conductivity under Shear Strain}\label{transport}
After the discussion on a class of models, we study in detail the DC conductivity under nonlinear shear strain. We will show the behavior of electric $(\sigma)$, thermoelectric $(\alpha, \bar{\alpha})$, and thermal $(\bar{\kappa})$ conductivities subject to mechanical response with both shear hardening and shear softening. 
Since we are mostly interested in the effect on the transport with applied shear strain, we mainly focus on the shear hardening case as the system is more sensitive to the shear strain, see Fig.~\ref{fig:HardeningSoftening}. We will show that similar features still appear in the shear softening case. 

\begin{figure}[H]
    \centering
    \includegraphics[width=0.32\linewidth]{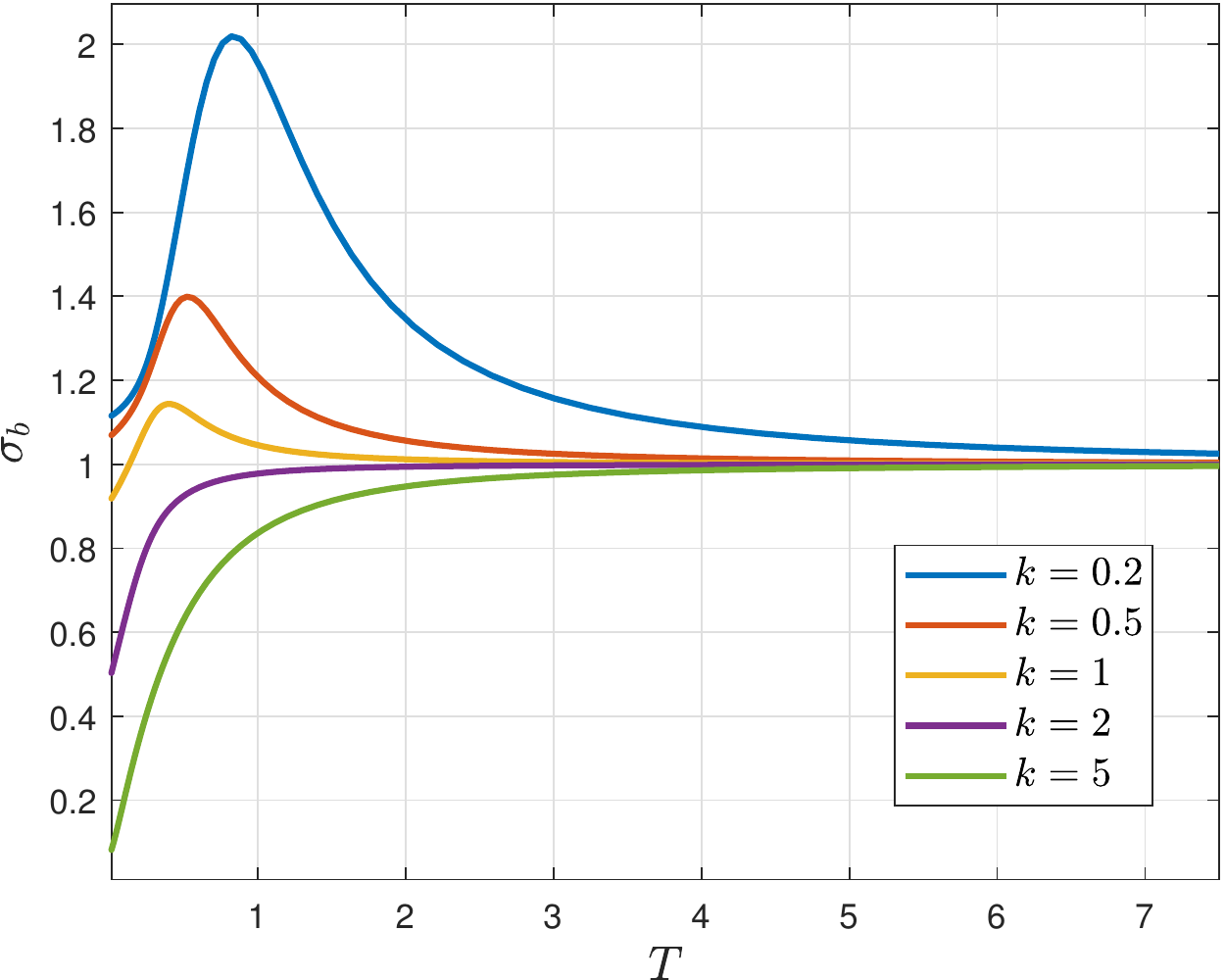}
     \includegraphics[width=0.32\linewidth]{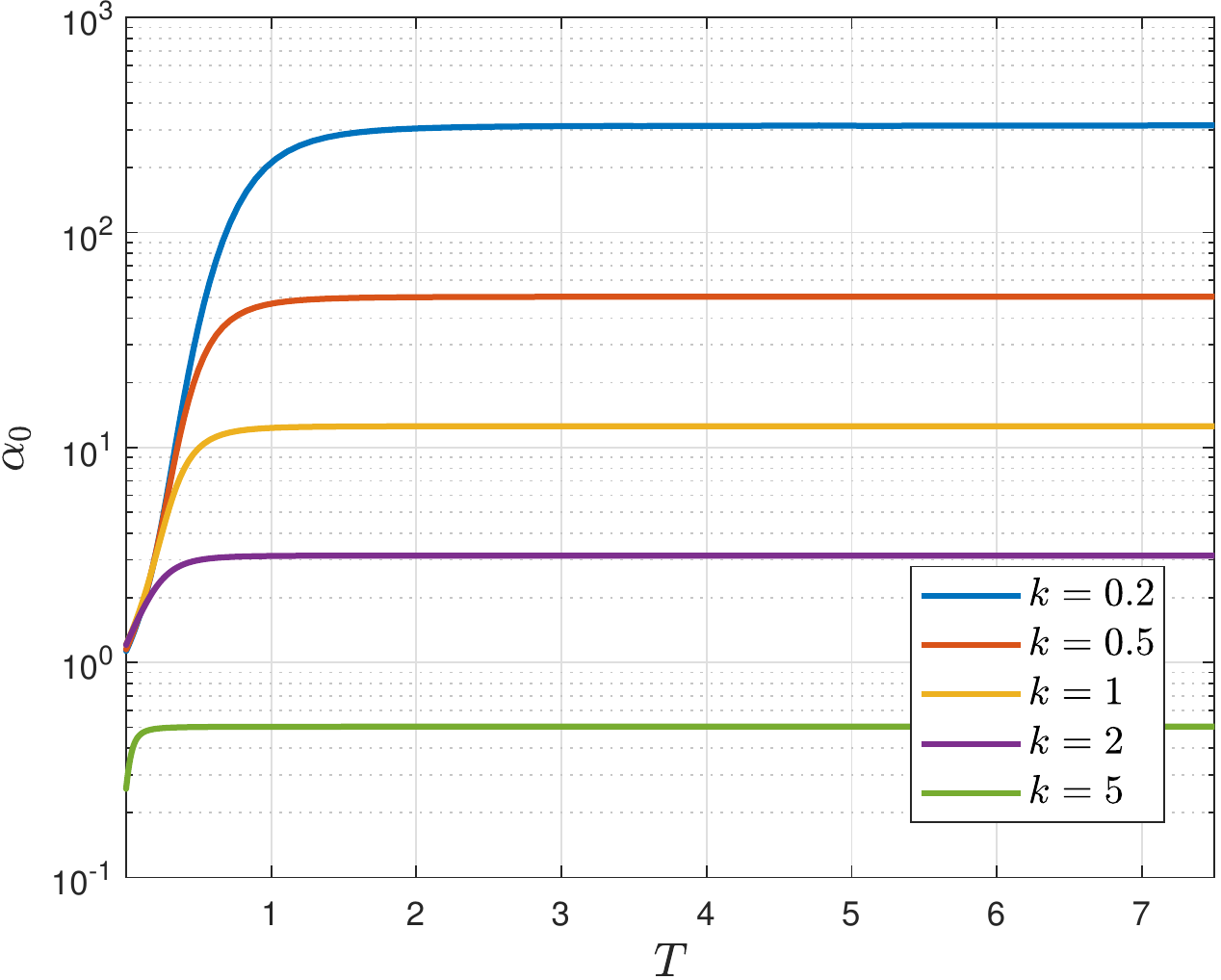}
      \includegraphics[width=0.32\linewidth]{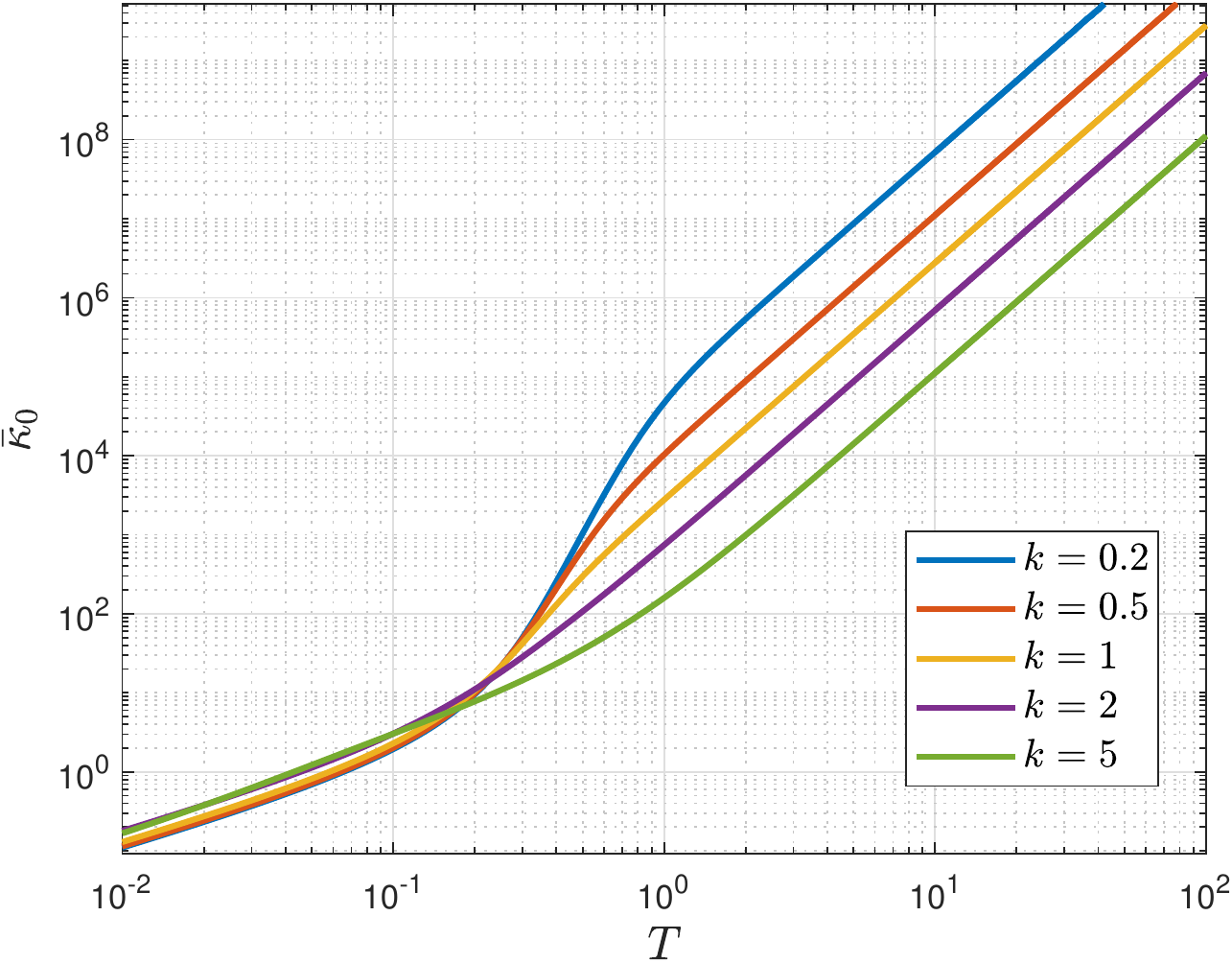}
    \caption{Transport properties in function of $T$ for various disorder strength $k$ in the absence of shear strain. \textbf{Left:} Electric conductivity $\sigma_0$. \textbf{Center:} Thermoelectric conductivity $\alpha_0$. \textbf{Right:} Thermal conductivity $\bar{\kappa}_0$. We consider the shear hardening model $W=X^3$.  Other parameters are $\rho=\beta=1$ and $\varepsilon=0$.
}\label{fig:nostrain}
\end{figure}

Let us consider the transport behavior in the absence of strain/stress before proceeding. To be specific, we take the potential $W=X^3$ and tune the disorder strength $k$ with $\rho=1$, see~\eqref{nostrain}. As can be seen from the left panel of Fig.~\ref{fig:nostrain}, the temperature dependence of the electric conductivity $\sigma_0$ has a rich behavior. 
At strong disorder strength $k$, $\sigma_0$ drops monotonically as $T$ is decreased, characterizing an insulating behavior.\,\footnote{We adopt a more realistic and phenomenological definition for an electric insulator that is basically $d\sigma/dT>0$ and for a metallic behavior by $d\sigma/dT<0$. A similar definition applies to both the thermoelectric and the thermal conductivities.} In contrast, at weak disorder, $\sigma_0$ shows a non-monotonic pattern. As the temperature is decreased, $\sigma_0$ first increases, exhibiting a metallic behavior, and then decreases. Nevertheless, there is a clear transition between the insulating and the metallic behaviors driven by the disorder above a particular temperature.

For both the thermoelectric coefficient $\alpha_0$ (center panel) and the thermal conductivity $\bar{\kappa}_0$ (right panel), one finds that they decrease monotonically by decreasing $T$, irrespective of the disorder strength. More precisely, we find the low temperature behaviors to be
\begin{equation}
 \alpha_0=A_0(k,\rho)+\mathcal{O}(T) ,\quad \bar{\kappa}_0=K_0(k,\rho)\, T+\mathcal{O}(T)\,,
\end{equation}
with $A_0$ and $K_0$ two constants that depend on $k$ and $\rho$. Therefore, the system at zero strain is a good thermal insulator. Moreover, by increasing the disorder strength $k$, the amplitude of both $\bar{\kappa}_0$ and $\alpha$ is suppressed, and both become less sensitive in the change of temperature. 

In the following numerical calculations, we focus on the effect of shear deformation on the transport presented in Fig.~\ref{fig:nostrain}.

\subsection{Electric conductivity}
\begin{figure}[H]
    \centering
    \includegraphics[width=0.49\linewidth]{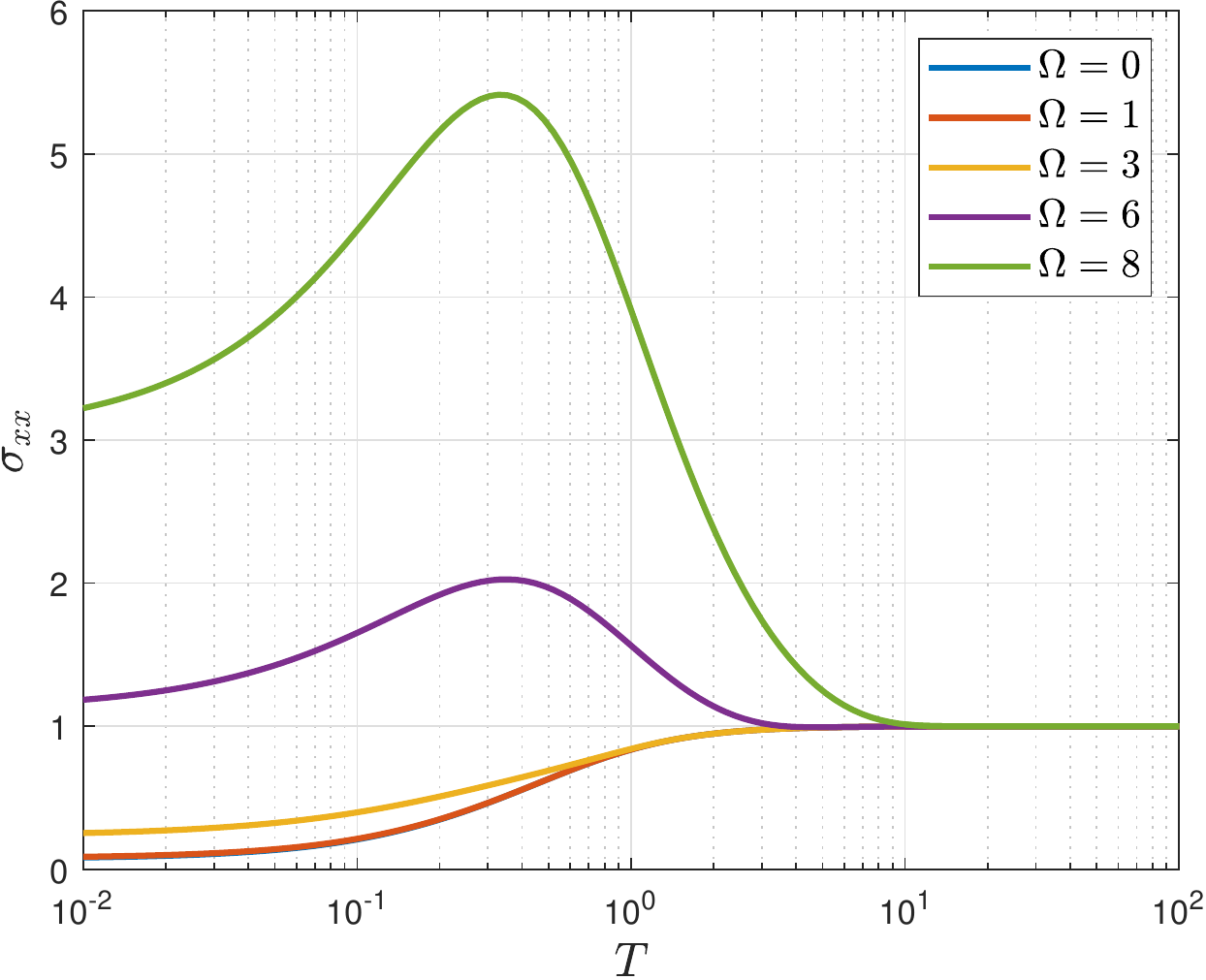}
    \includegraphics[width=0.49\linewidth]{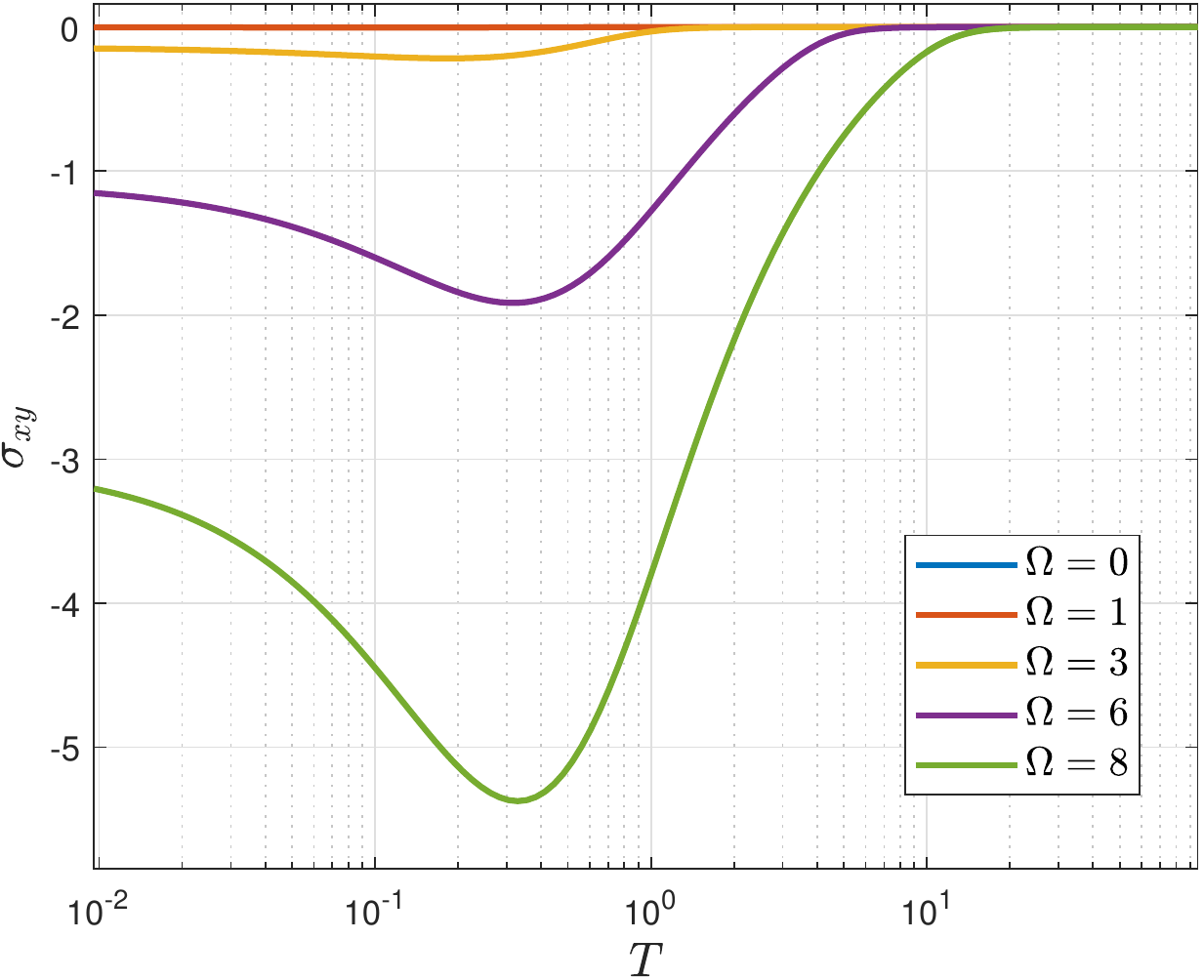}
    \caption{Temperature dependence of $\sigma_{xx}$ (left) and $\sigma_{xy}$ (right) under different shear strain $\varepsilon=2\sinh (\Omega/2)$ for $W=X^3$ and $k=5$. Other parameters are fixed to be $\rho=\beta=1$.
}\label{fig:x3sigmalk}
\end{figure}

We begin with the case at strong disorder for which the system at zero strain is a good insulator. The temperature dependence of the electric conductivity in the shear hardening case by dialing the shear deformation is presented in Fig.~\ref{fig:x3sigmalk}. 

It is clear that the off-diagonal component $\sigma_{xy}$ develops after turning on the shear deformation. Its amplitude increases with the deformation parameter $\varepsilon=2\sinh (\Omega/2)$~\eqref{deformation}. Meanwhile, with the shear strain increased, the temperature dependence of $\sigma_{xx}$ changes quantitatively. For small strain, $\sigma_{xx}(T)$ is similar to the one without deformation (the left panel of Fig.~\ref{fig:nostrain}). For sufficiently large shear deformation, there is a critical temperature $T_0$ above which $\sigma_{xx}$ becomes a monotonically decreasing function of $T$ and below which $\sigma_{xx}$ increases as $T$ is increased. Meanwhile, $\sigma_{xy}$ develops a dip at $T_0$. One can see from Fig.~\ref{fig:x3sigmalk} that $T_0$ is not sensitive to the shear deformation. Moreover, $\sigma_{xx}$ at low temperature limit increases by increasing the shear deformation. Therefore, the insulating behavior will be destroyed, yielding a transition from an insulating phase to a metallic phase induced by applied stress.

\begin{figure}[H]
    \centering
    \includegraphics[width=0.49\linewidth]{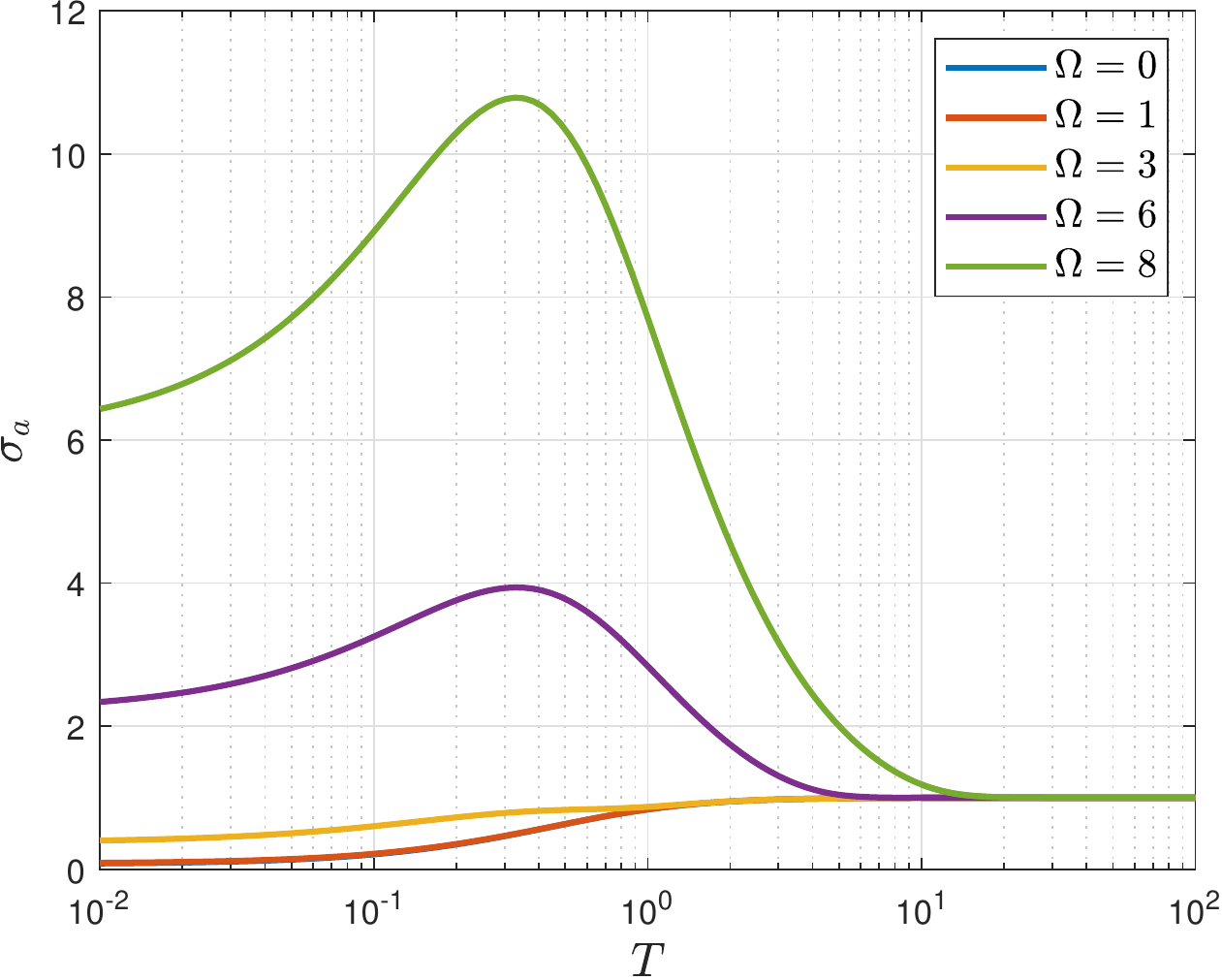}
    \includegraphics[width=0.49\linewidth]{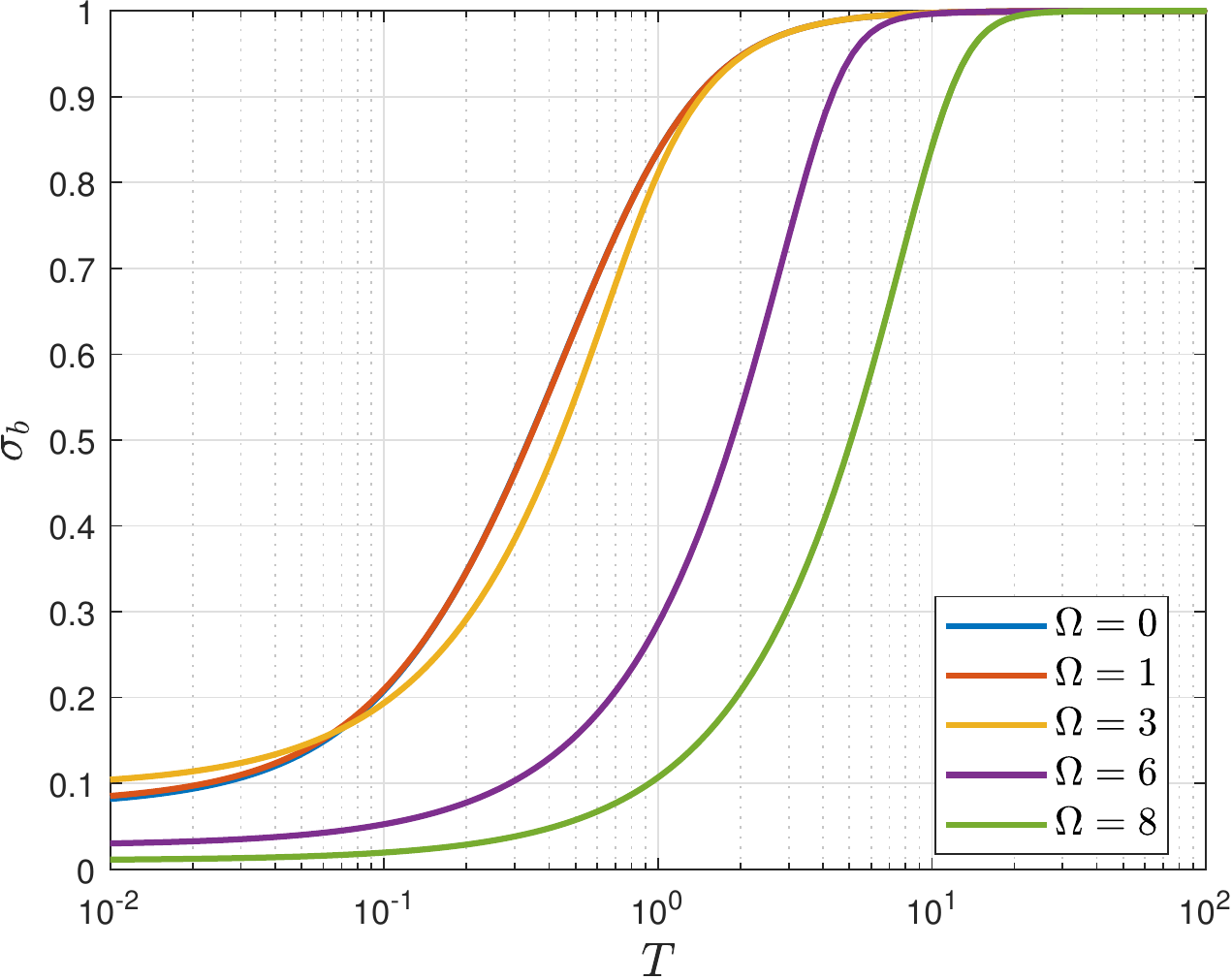}
    \caption{Temperature dependence of $\sigma_{a}$ (left) and $\sigma_{b}$ (right) under different shear strain $\varepsilon=2\sinh (\Omega/2)$ for $W=X^3$ and $k=5$. We choose $\rho=\beta=1$.}\label{fig:x3sigmalkab}
\end{figure}

In order to see the transition clearly, we consider the components of the electric conductivity along the principal axes, \emph{i.e.} $\sigma_a$ and $\sigma_b$ of~\eqref{princpsig}. In our present case, one has $\sigma_a=\sigma_{xx}-\sigma_{xy}$ and $\sigma_b=\sigma_{xx}+\sigma_{xy}$. Note that $\sigma_a\ge\sigma_b$ as $\sigma_{xy}\le 0$ (see Fig.~\ref{fig:x3sigmalk}).
One can see from the right panel of Fig.~\ref{fig:x3sigmalkab} that $\sigma_b$ has a more rapid drop in temperature as the strain is increased, yielding a good insulating phase along the $e_b$ direction.\footnote{Although $\sigma_{xx}$ and $\sigma_{xy}$ in low temperature limit are both monotonic functions of the shear deformation (see Fig.~\ref{fig:x3sigmalk}), the low temperature limit of $\sigma_b=\sigma_{xx}-|\sigma_{xy}|$ is non-monotonic due to the competition between $\sigma_{xx}$ and $\sigma_{xy}$. The low temperature limit of $\sigma_b$ increases with the increase of shear deformation when the applied strain is weak. Nevertheless, when the shear deformation is large enough, it decreases monotonically with the applied strain, yielding a good insulating behavior, see the right panel of Fig.~\ref{fig:x3sigmalkab}. } Meanwhile, as one increases the applied strain, $\sigma_a$ has a slower drop in temperature and then grows quickly above $T_0$ as $T$ is lowered for sufficiently large strain, see the left panel of Fig.~\ref{fig:x3sigmalkab}. Therefore, along the $e_a$ direction, there is a transition from the insulating phase to the metallic phases induced by the applied strain.

\begin{figure}[H]
    \centering
    \includegraphics[width=0.48\linewidth]{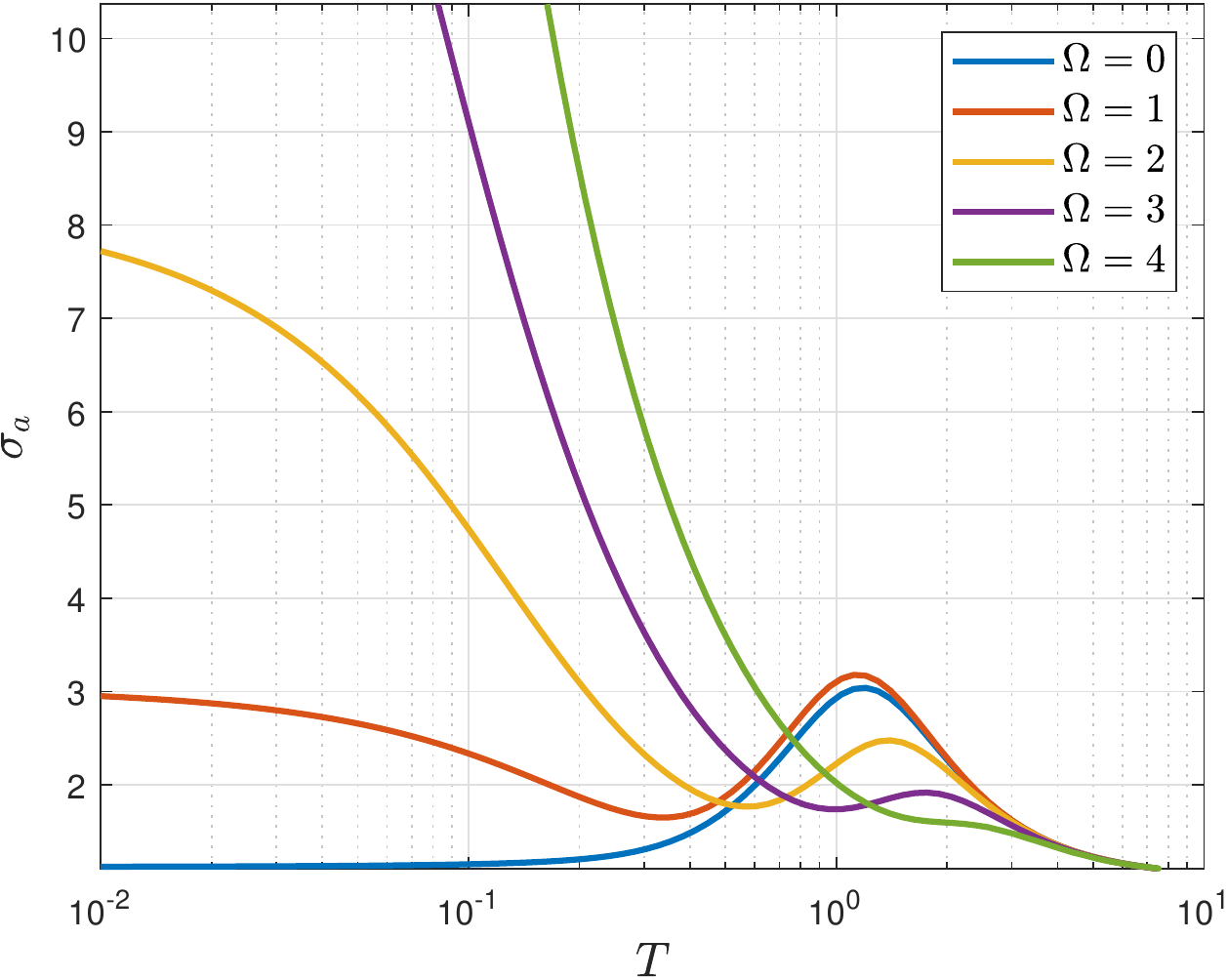}
     \includegraphics[width=0.49\linewidth]{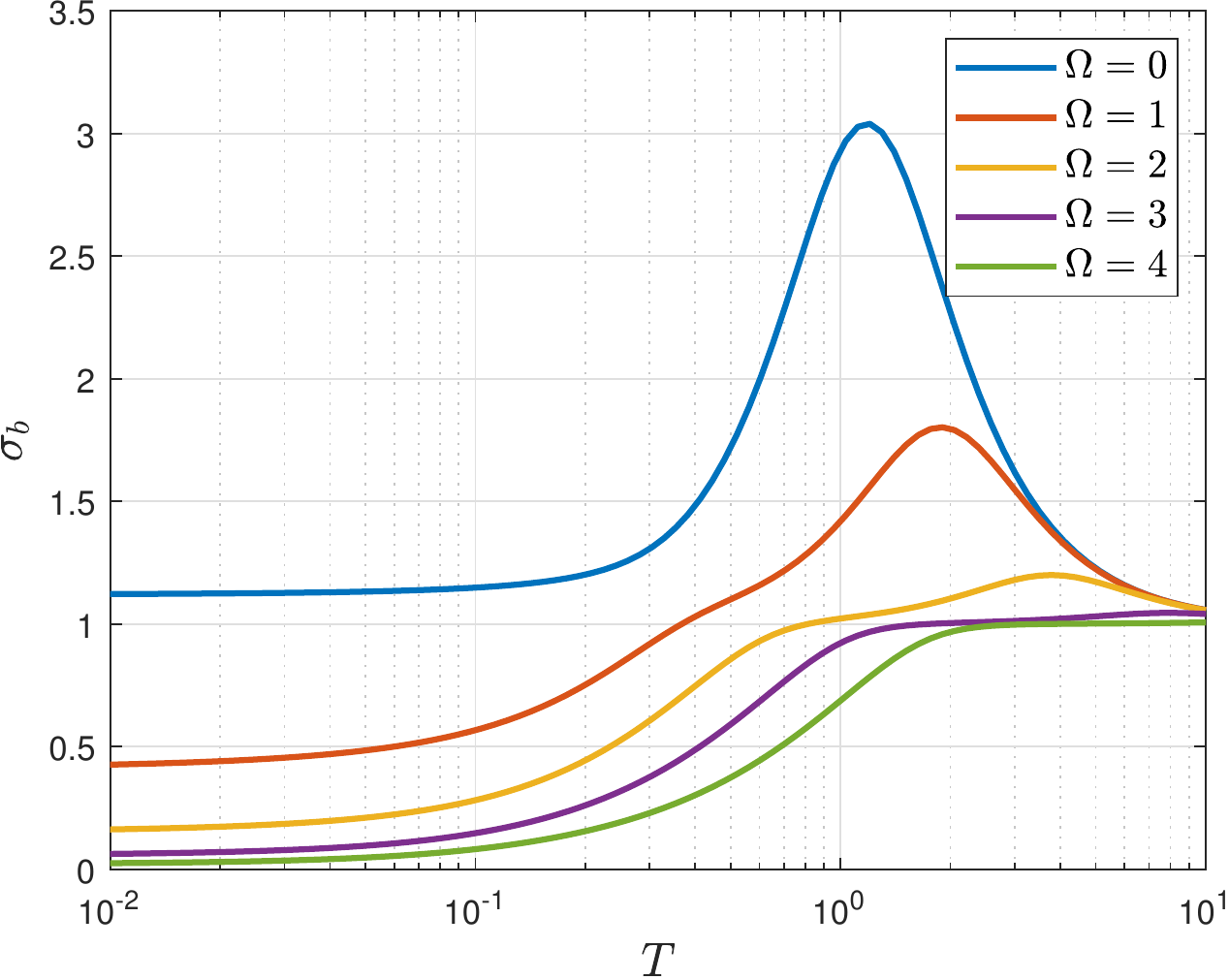}
    \caption{Temperature dependence of $\sigma_{a}$ (left) and $\sigma_{b}$ (right) under different shear strain at weak disorder $k=0.1$. We consider $W=X^3$ and fix other parameters to be $\rho=\beta=1$.
}\label{fig:x3sigmaskab}
\end{figure}

We now consider the electric conductivity at weak disorder. As shown in the left panel of Fig.~\ref{fig:nostrain}, the conductivity at zero strain is non-monotonic in the function of $T$. The temperature dependence of $\sigma_a$ and $\sigma_b$ at $k=0.1$ is presented in Fig.~\ref{fig:x3sigmaskab}. By increasing the shear strain, the peak in $\sigma_b$ is suppressed and then $\sigma_b$ drops rapidly as $T$ is lowered. Thus, one has a good insulating phase along this direction. Meanwhile, the peak along $e_a$ is also suppressed, but $\sigma_a$ grows quickly by decreasing $T$, yielding a good metallic phase. We have therefore arrived at the main result of our paper: strain engineering induces a good insulating phase along one principal axis and a good metallic phase along the other principal axis simultaneously, irrespective of the strength of disorder.

\subsection{Thermoelectric conductivity}
\begin{figure}[H]
    \centering
    \includegraphics[width=0.49\linewidth]{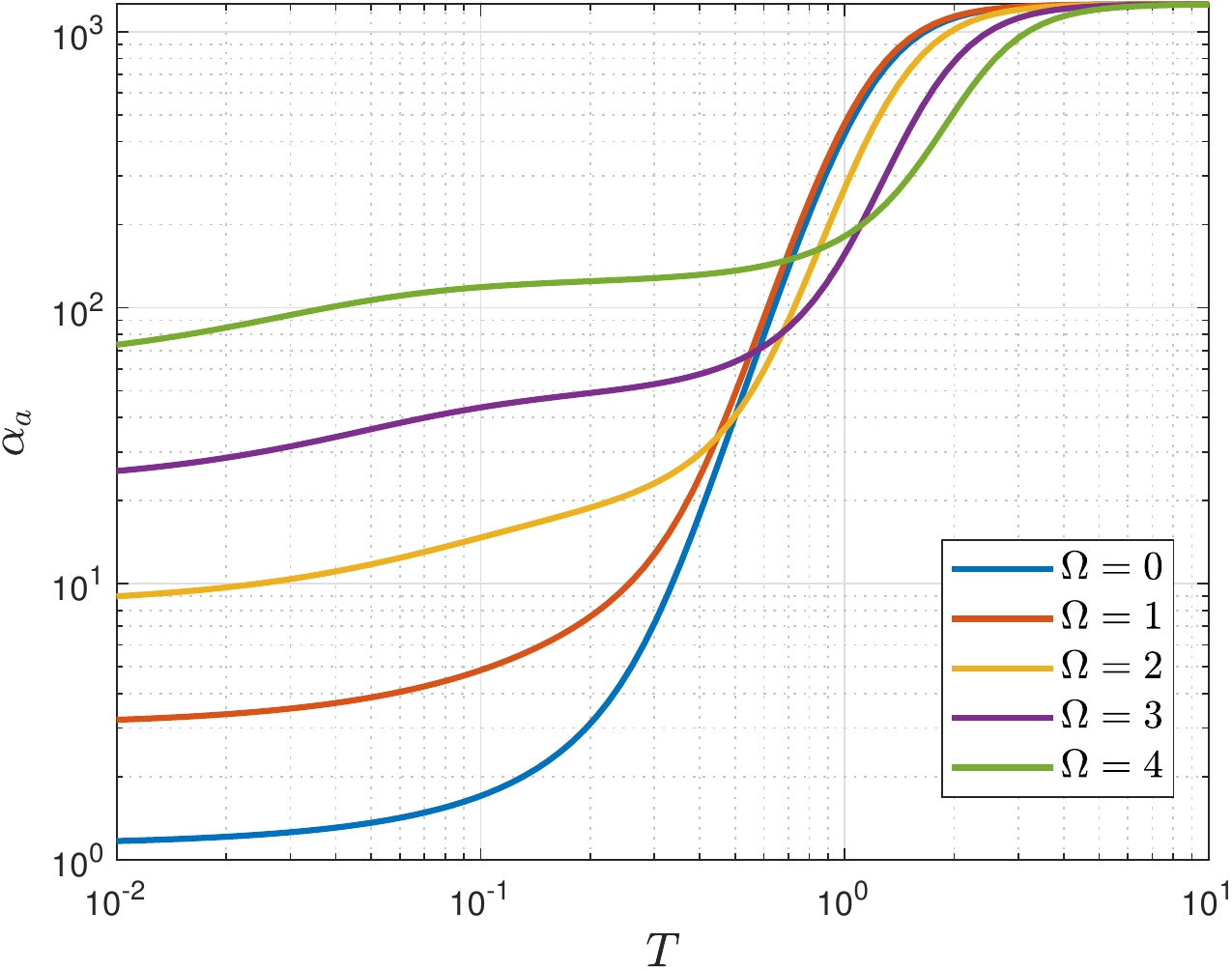}
    \includegraphics[width=0.49\linewidth]{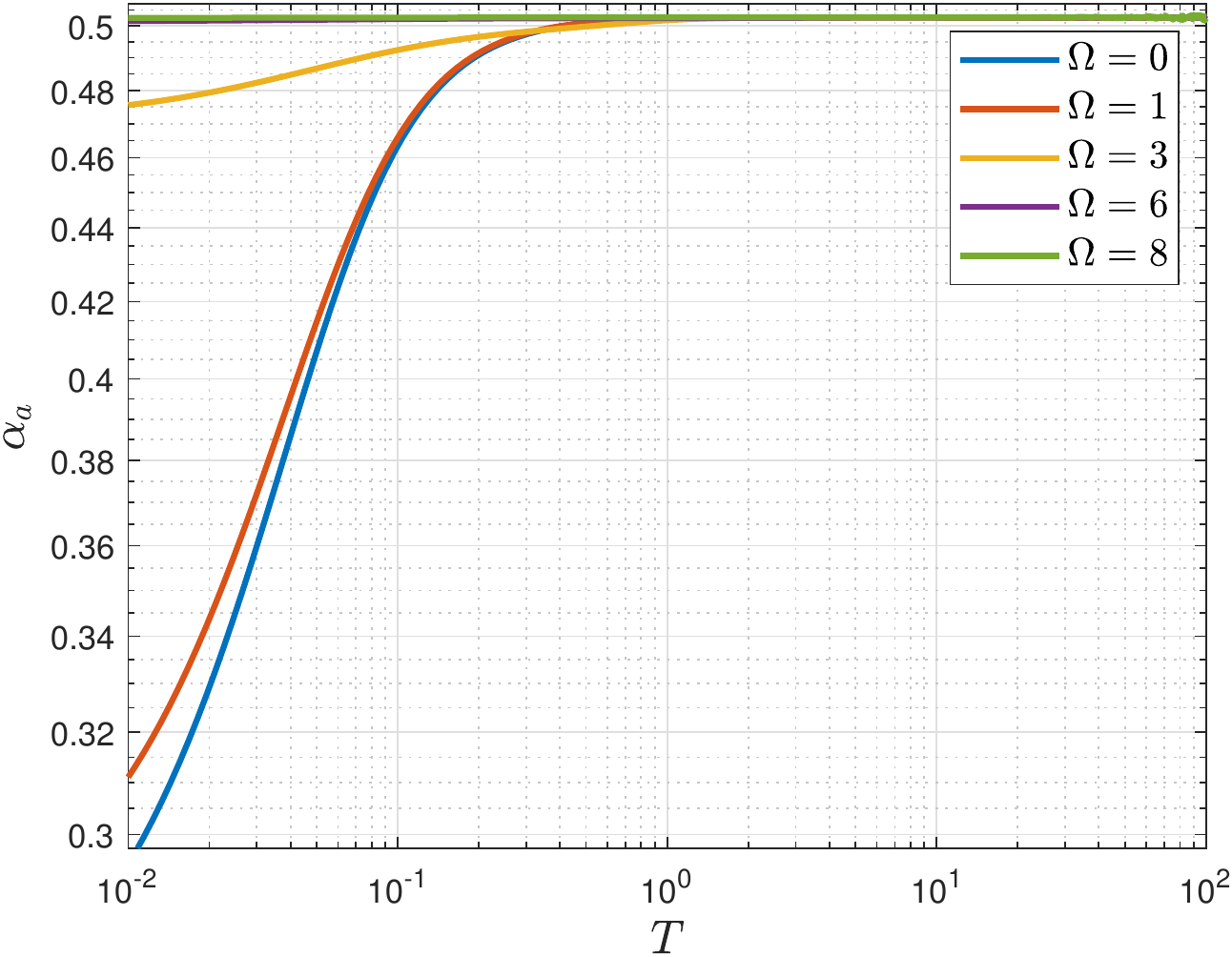}
     \includegraphics[width=0.49\linewidth]{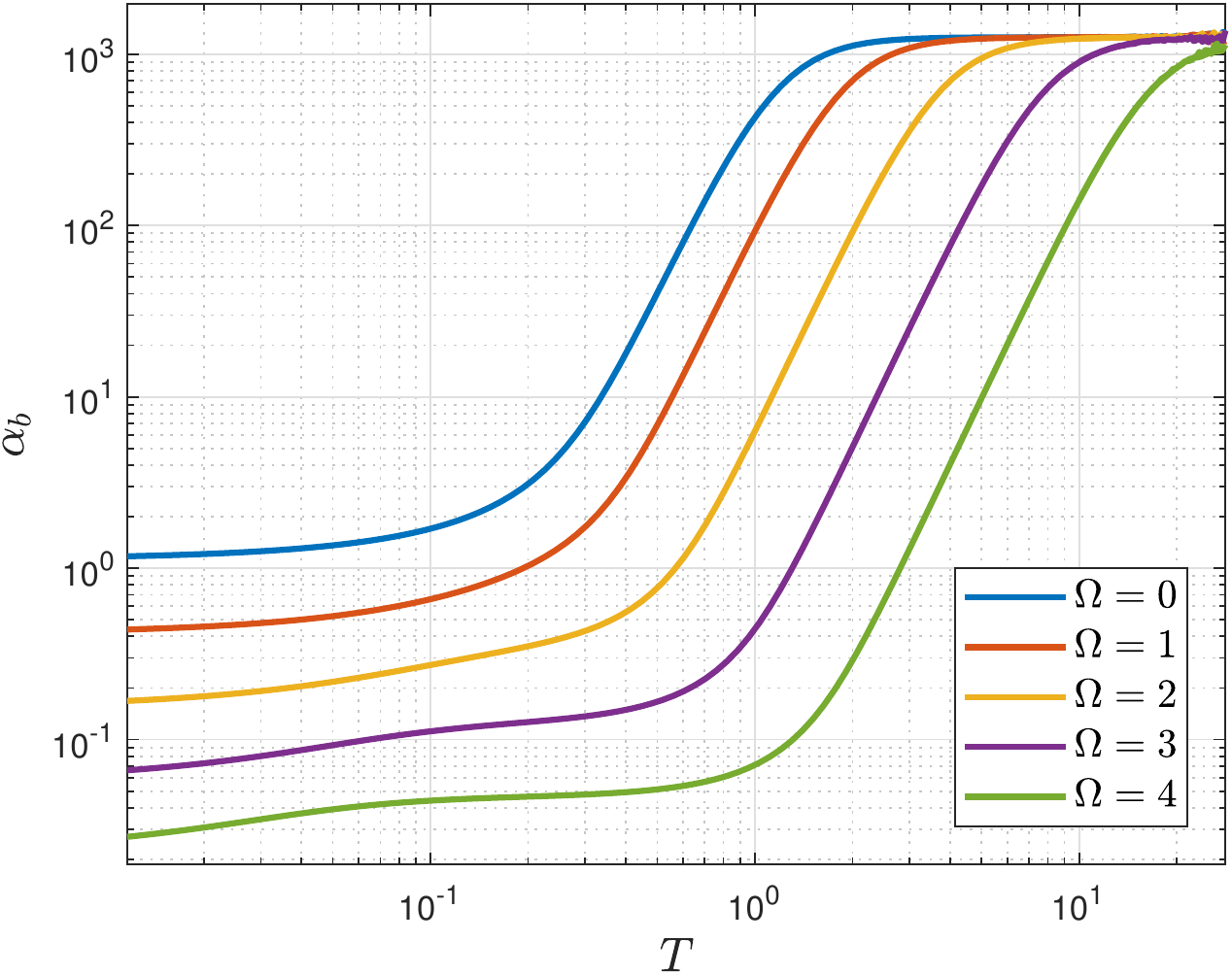}
    \includegraphics[width=0.49\linewidth]{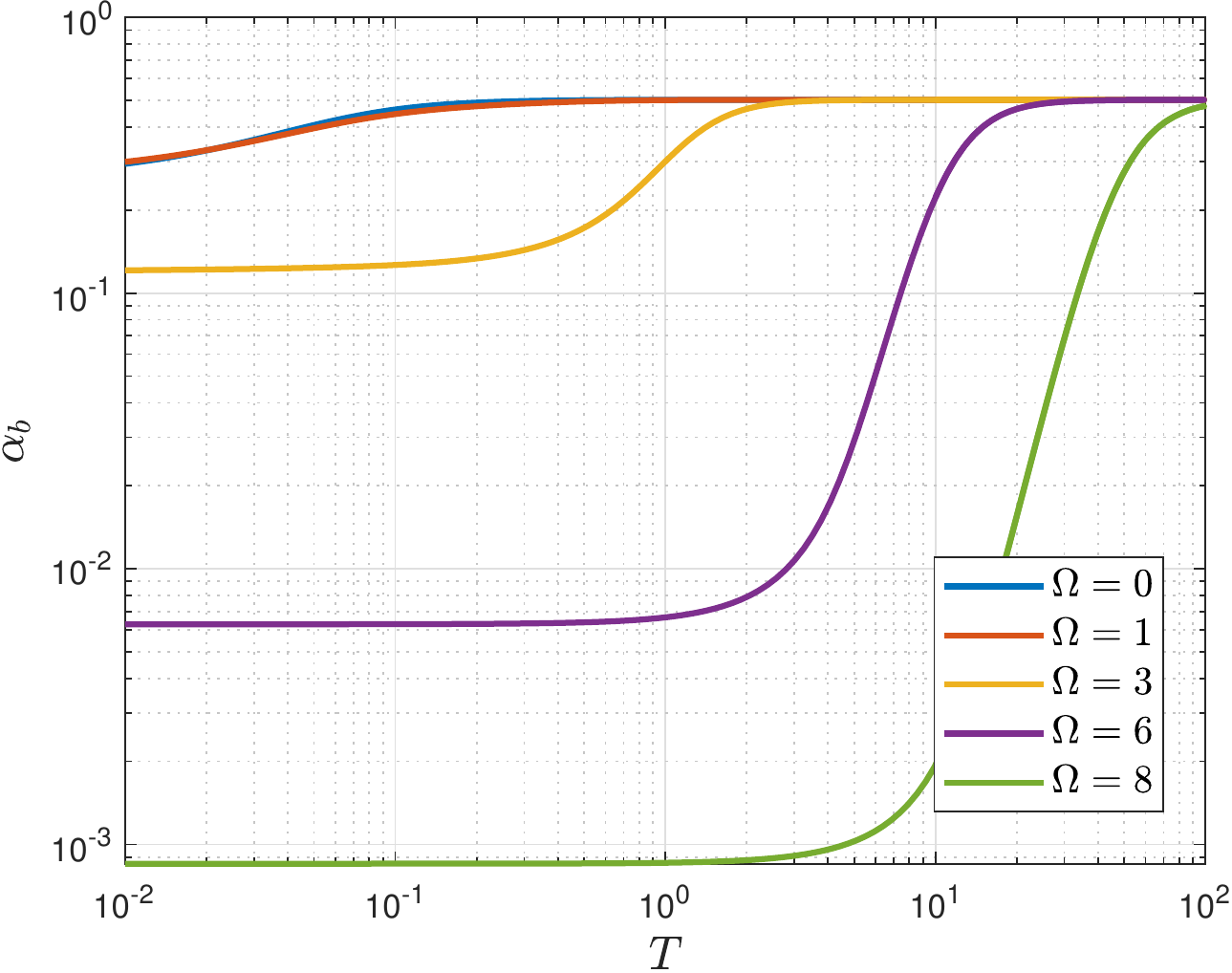}
    \caption{Thermoelectric coefficient with respect to temperature under different shear strain for $W(X,Z)=X^3$. \textbf{Left:} The weak disorder case with $k=0.1$. \textbf{Right:} The strong disorder case with $k=5$. We choose other parameters as $\rho=\beta=1$.
}\label{fig:alphax3}
\end{figure}

We show the temperature dependence of thermoelectric coefficient $\alpha$ under shear strain in Fig.~\ref{fig:alphax3}. For all cases we have checked, both thermoelectric conductivities along two principal axes decrease monotonically as $T$ is lowered.

Irrespective of the disorder parameter $k$, along $e_b$ axis, $\alpha_b=\alpha_{xx}+\alpha_{xy}$ drops more rapidly in temperature as the shear deformation is increased. At low temperature limit, $\alpha_b$ approaches a finite constant which, at large strain, decreases by increasing the shear deformation.
In contrast, the thermoelectric coefficient $\alpha_a=\alpha_{xx}-\alpha_{xy}$ along $e_a$ direction becomes more and more insensitive to the temperature by increasing the shear strain. Moreover, its low temperature value increases with the applied strain. Remarkably, the low temperature value of $\alpha_b$ can be sufficiently small, yielding a good thermoelectric insulator driven by shear deformation.

\begin{figure}[H]
    \centering
    \includegraphics[width=0.49\linewidth]{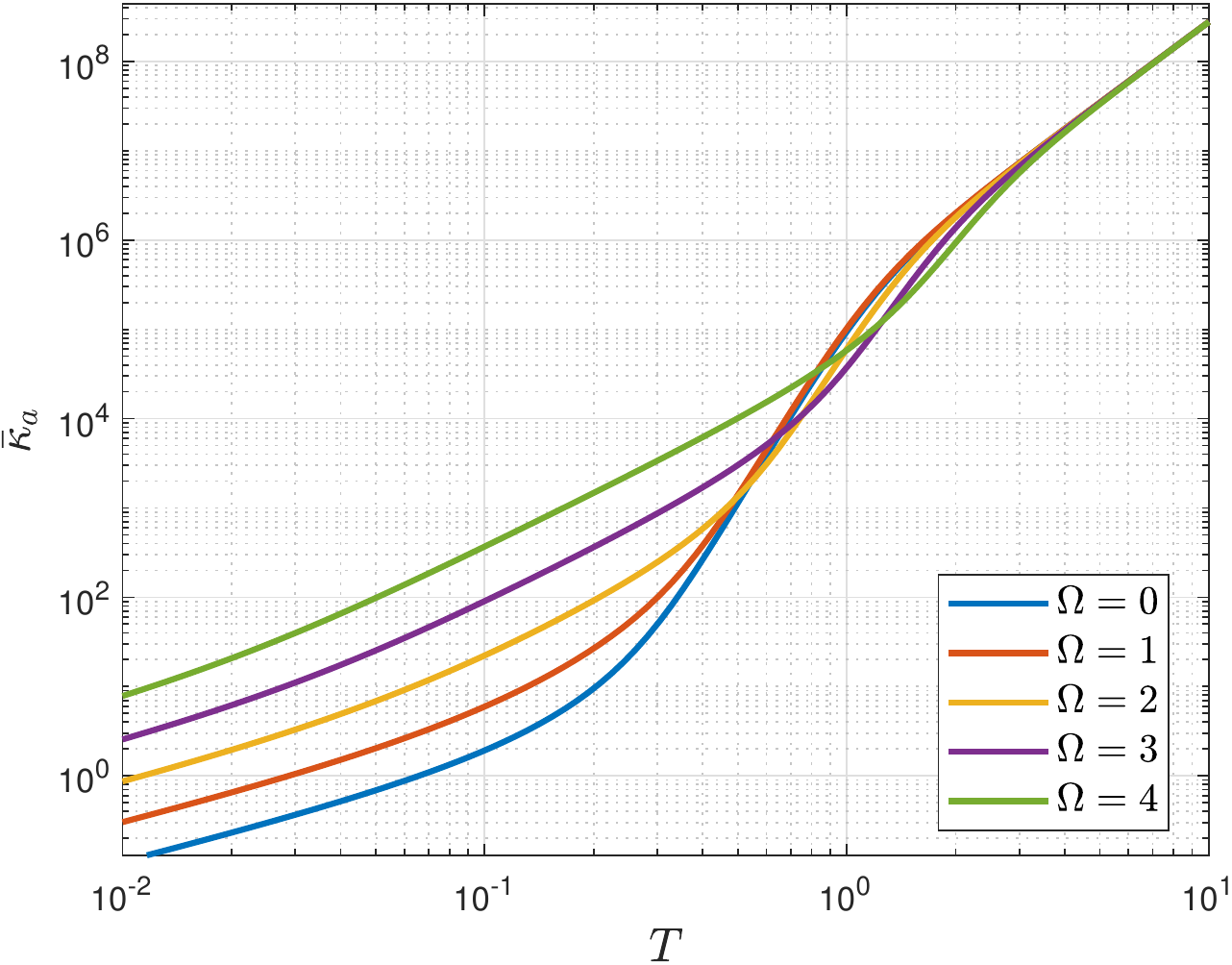}
    \includegraphics[width=0.49\linewidth]{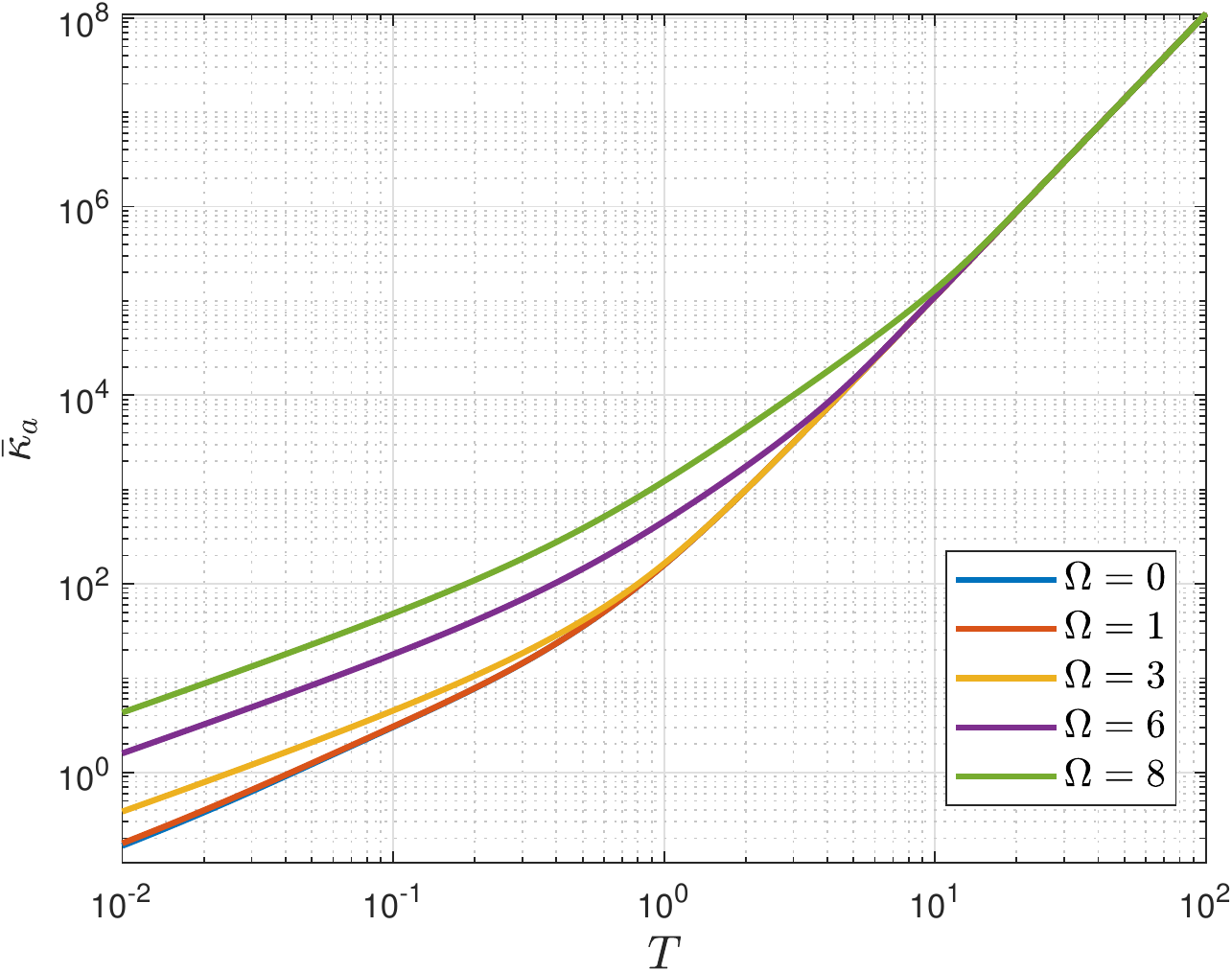}
     \includegraphics[width=0.49\linewidth]{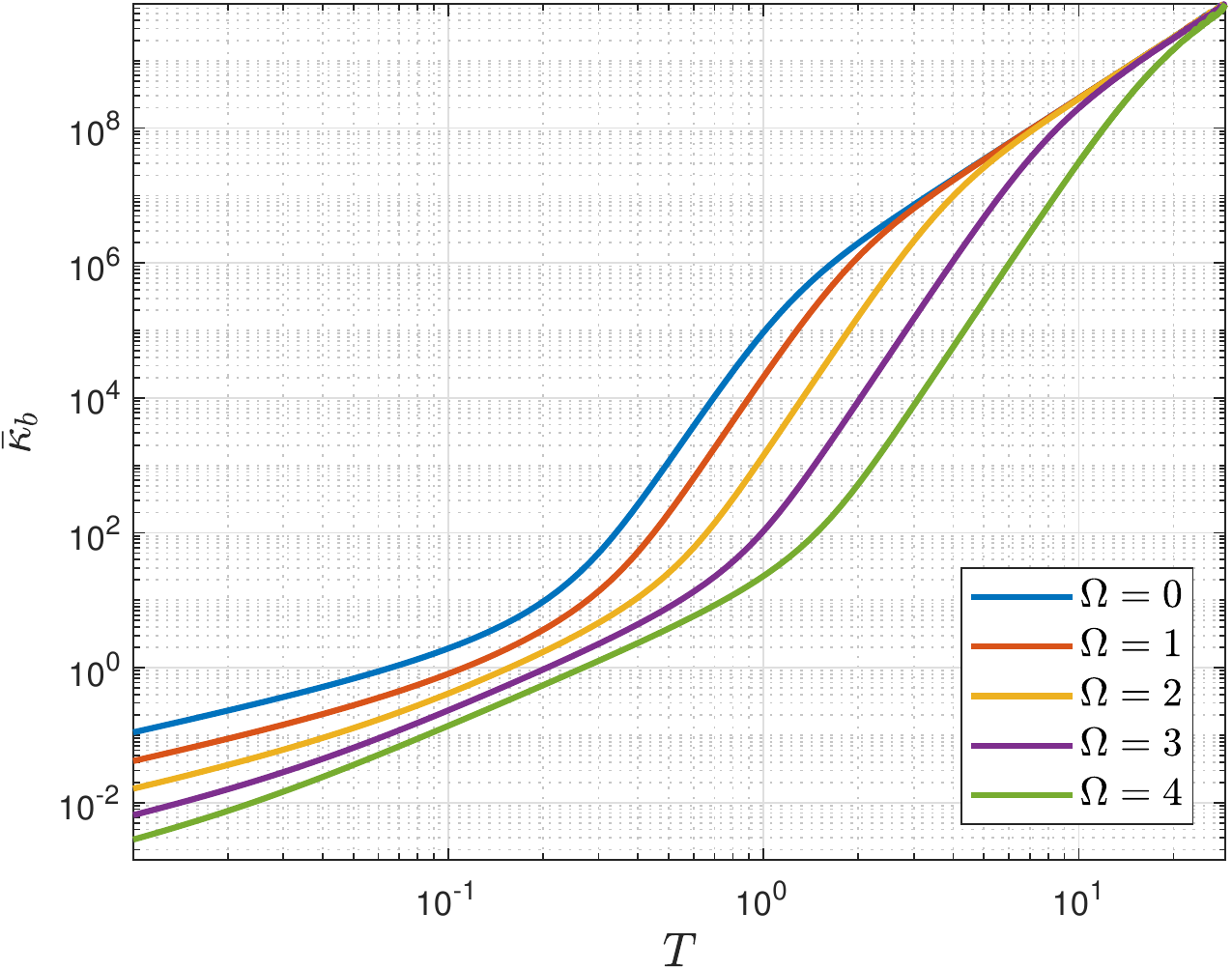}
    \includegraphics[width=0.49\linewidth]{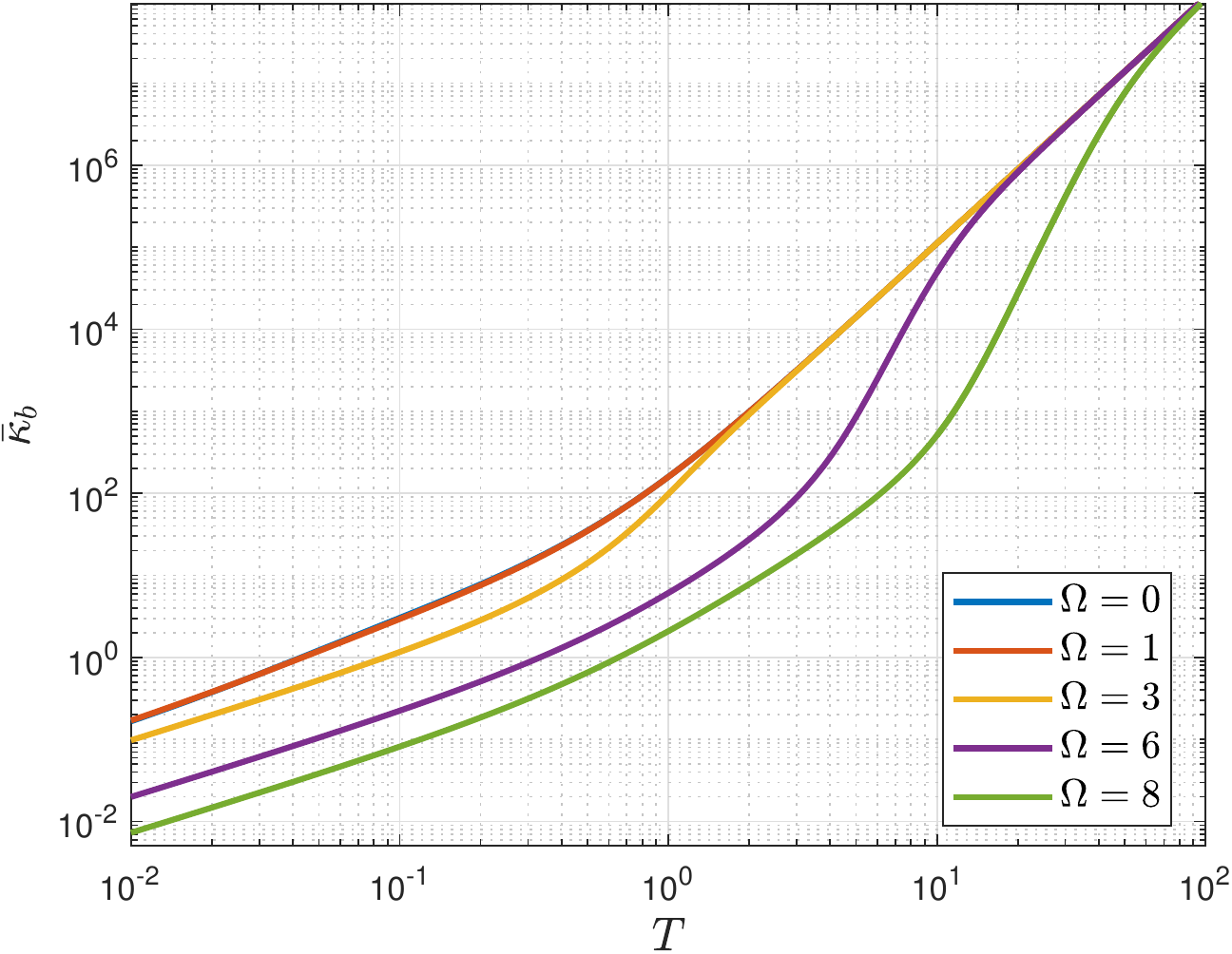}
    \caption{Thermal conductivity with respect to temperature under different shear strain for $W(X,Z)=X^3$. \textbf{Left:} The weak disorder case with $k=0.1$. \textbf{Right:} The strong disorder case with $k=5$. We choose other parameters as $\rho=\beta=1$.
}\label{fig:x3kappak}
\end{figure}

\subsection{Thermal conductivity}
The thermal conductivity along the principal axes with applied strain is presented in Fig.~\ref{fig:x3kappak}. Similar to $\alpha$, for both weak and strong disorder cases, one finds that $\bar{\kappa}$ along $e_b$ direction drops more rapidly in temperature for large strain, while the one along $e_a$ becomes less sensitive to the temperature.

Moreover, we find that both $\bar{\kappa}_a=\bar{\kappa}_{xx}-\bar{\kappa}_{xy}$ and $\bar{\kappa}_b=\bar{\kappa}_{xx}+\bar{\kappa}_{xy}$ become vanishing as $T$ goes to zero, yielding a good thermal insulator. This can be understood from~\eqref{ratio}. Note that $\alpha_a$ and $\alpha_b$ are finite as $T$ goes to zero. Then, one immediately obtains from~\eqref{ratio} that the thermal conductivity $\bar{\kappa}\sim T$ at low temperature limit. Indeed, we have numerically checked that~\eqref{ratio} is satisfied for all cases we have considered.

\begin{figure}[H]
    \centering
    \includegraphics[width=0.49\linewidth]{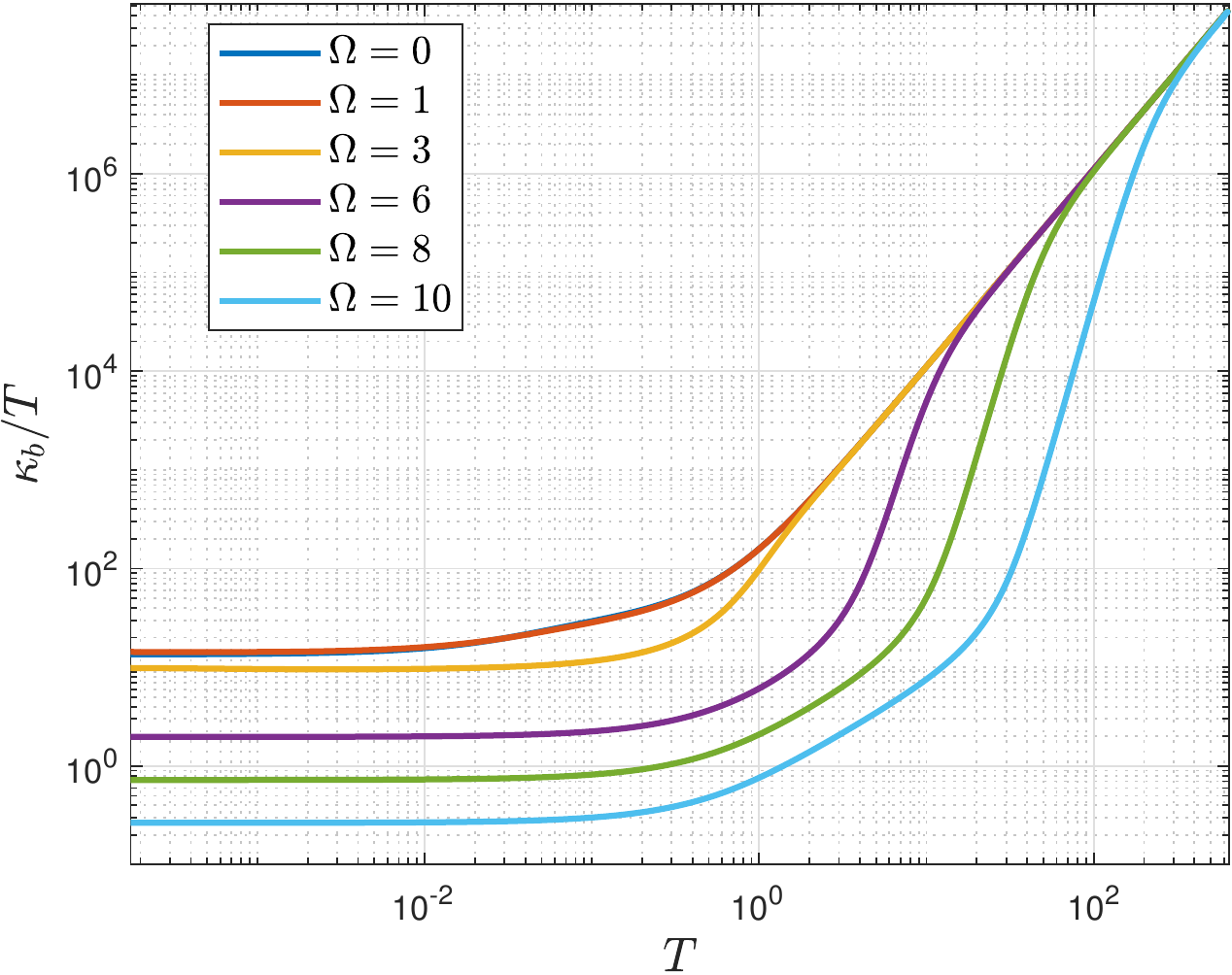}
    \includegraphics[width=0.49\linewidth]{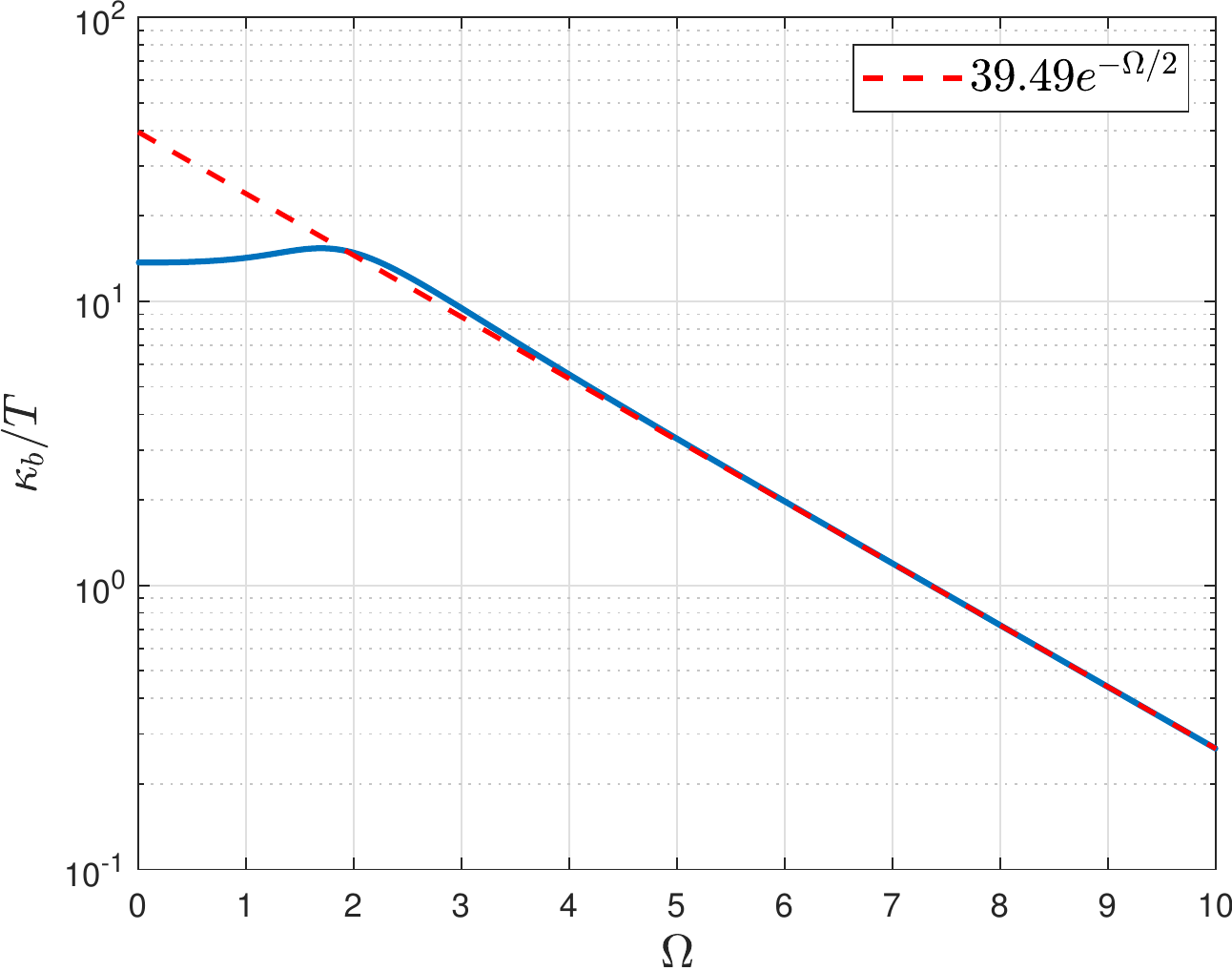}
    \caption{The ratio $\kappa_b/T$ under different shear deformations for $W(X,Z)=X^3$. \textbf{Left:} $\kappa_b/T$ in function of temperature. \textbf{Right:} The low temperature limit of $\kappa_b/T$ versus the shear deformation.  The red dashed line corresponds to the fitted curve on the numerical data with $\Omega>5$.
    We choose $k=5$ and $\rho=\beta=1$.}\label{fig:kappaT}
\end{figure}

Another interesting transport coefficient is the thermal conductivity $\kappa$ at zero electric current flow. One can obtain $\kappa$ from~\eqref{eq:Ohm} that
\begin{equation}
\kappa=\bar{\kappa}-T\bar{\alpha}\,\sigma^{-1}\alpha\,.
\end{equation}
A universal bound was conjectured for the
ratio of the thermal conductivity $\kappa$ over the temperature, which reads~\cite{Grozdanov:2015djs}
\begin{equation}\label{bound}
\frac{\kappa}{T}\ge C_0\,,
\end{equation}
where $C_0$ is a non-zero finite constant. From the left panel of Fig.~\ref{fig:kappaT}, one finds that $\kappa_b/T$ approach a constant at low temperature limit. In the right panel, we show $C_0\equiv \lim_{T\rightarrow 0}(\kappa_b/T)$ as a function of the shear deformation. It is clear that the thermal conductivity bound can be significantly lowered. By fitting the numerical data for $\Omega>5$, one finds $C_0\propto e^{-\Omega/2}\propto 1/\varepsilon$ where we have used~\eqref{deformation}. Thus, the violation of the thermal conductivity bound~\eqref{bound} can be driven by the shear deformation.\,\footnote{The violation of the above bound was found in Horndeski theory where its Lagrangian contains terms that have more than two derivatives~\cite{Liu:2018hzo,Figueroa:2020tya}.}

\begin{figure}[H]
    \centering
    \includegraphics[width=0.49\linewidth]{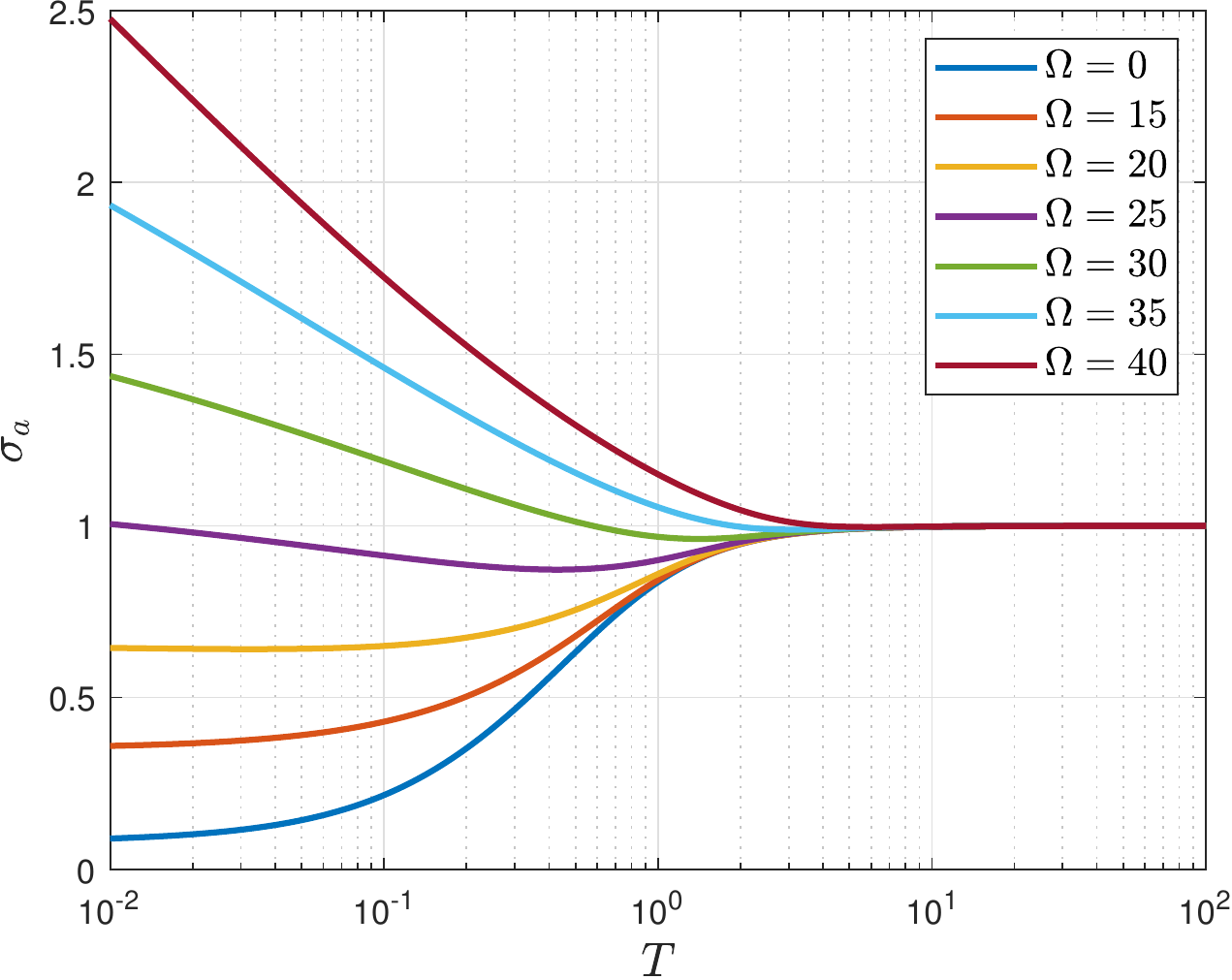}
    \includegraphics[width=0.49\linewidth]{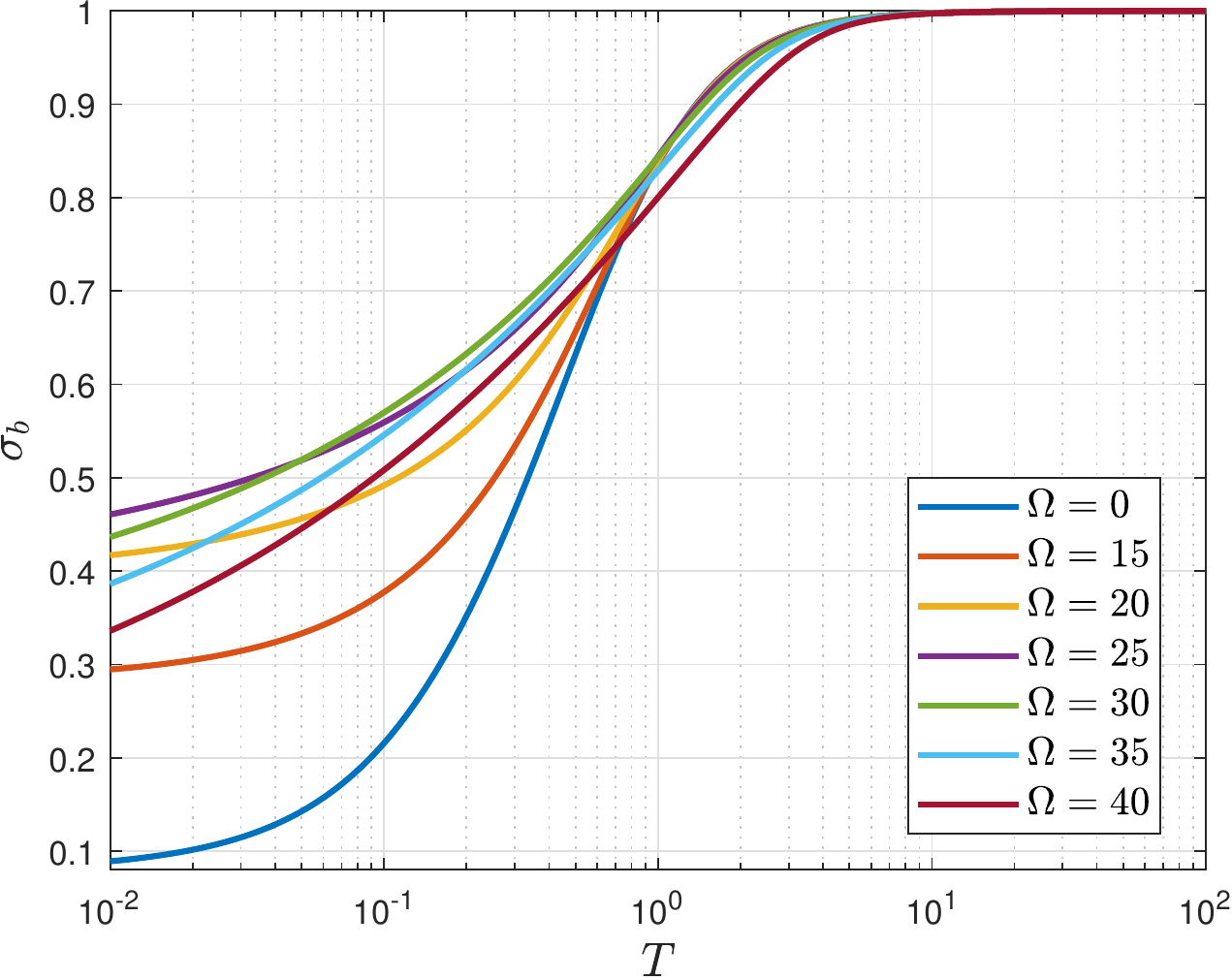}
    \caption{Temperature dependence of $\sigma_a$ (left) and $\sigma_b$ (right) under different shear strain for shear softening $W(X,Z)=X^{\frac{1}{4}}Z^{\frac{7}{8}}$. We have considered $k=5$ with $\rho=\beta=1$.
}\label{fig:xzsigmalkab}
\end{figure}

For the shear softening case, we find similar behaviors as the hardening one we have studied above. For illustration, we show the electric conductivity in function of temperature by increasing the shear stain in Fig.~\ref{fig:xzsigmalkab}. One can see that the unstrained state (the blue curve of Fig.~\ref{fig:xzsigmalkab}) is an insulating phase. Then the strain engineering yields a good metallic phase along $e_a$ direction, while an insulating phase along $e_b$. This feature is also confirmed in Appendix~\ref{app:charge} by considering the case under different charged densities.
Compared to the shear hardening case, a much larger shear deformation $\varepsilon$ is needed to change the conductivity significantly in the softening case.

\begin{figure}[H]
    \centering
    \includegraphics[width=0.49\linewidth]{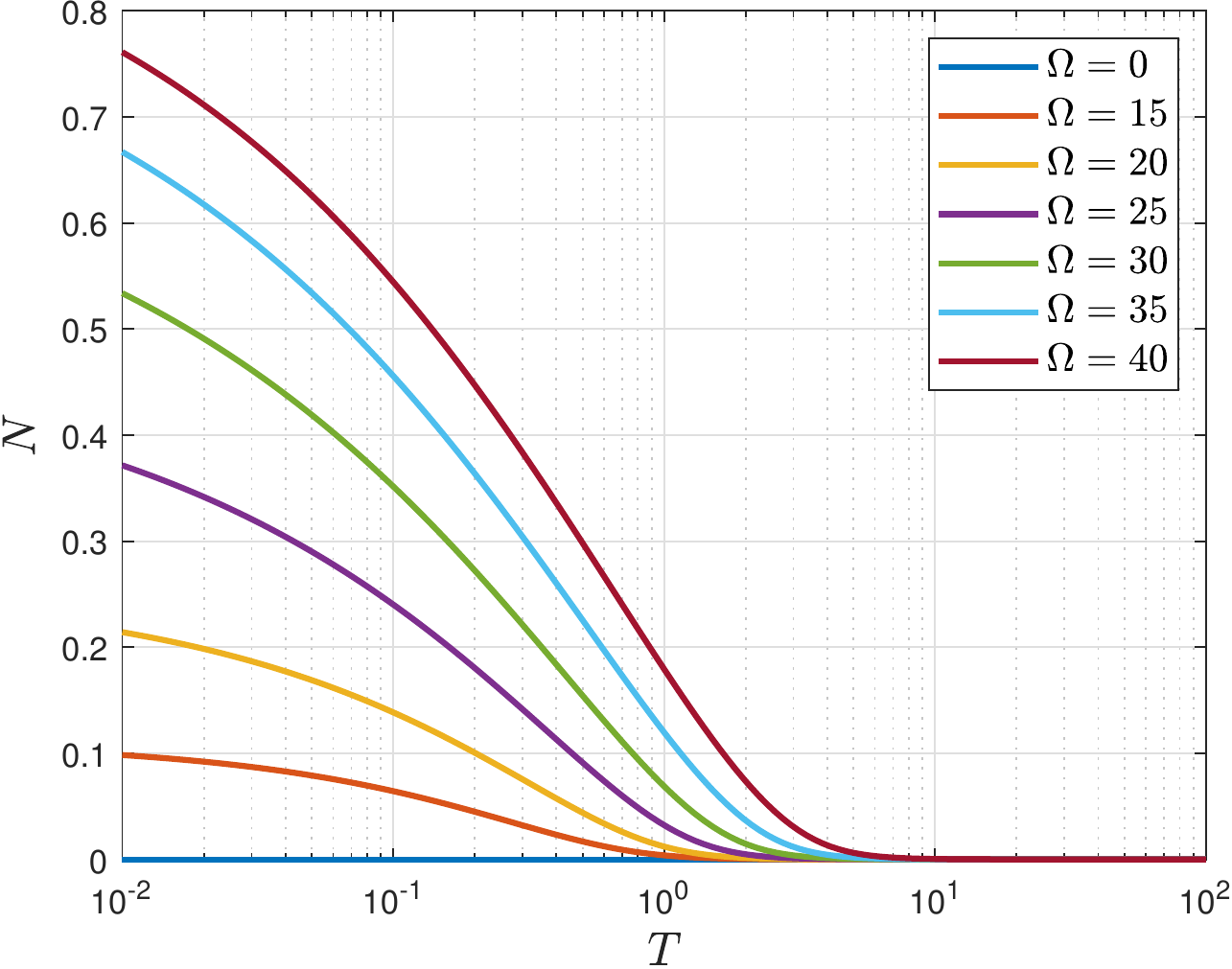}
    \includegraphics[width=0.49\linewidth]{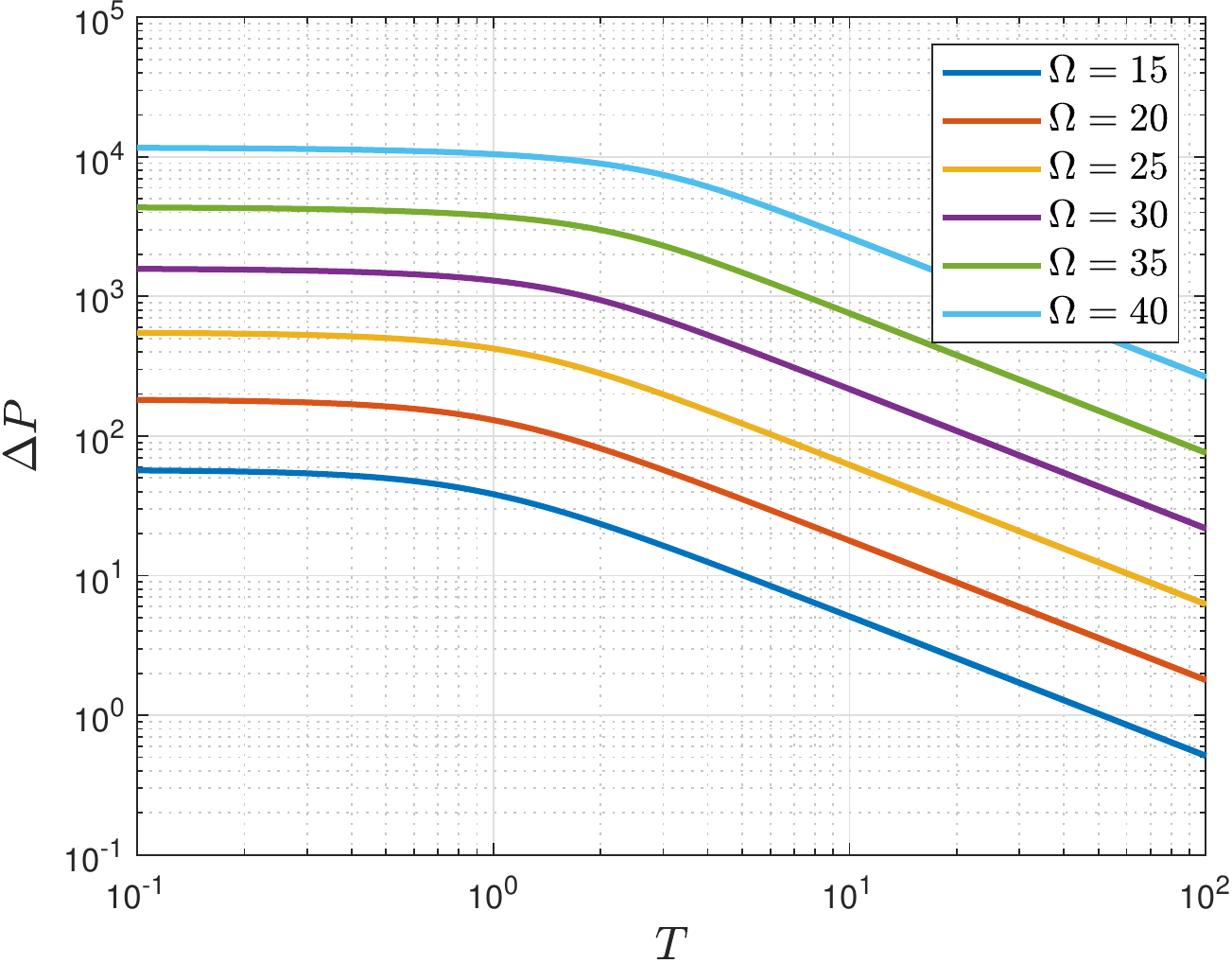}
    \caption{Temperature dependence of the electronic nematicity $N$ (left) and the pressure anisotropy $\Delta P$ (right) under different shear deformation. We choose the same parameters as Fig.~\ref{fig:xzsigmalkab}.
}\label{fig:xzNP}
\end{figure}

In our model, the strained conductivity along $e_a$ direction is always larger than the one along $e_b$ direction. As the strain parameter $\varepsilon$ increases, the pressure difference between the two principal axes, \emph{i.e.} $\Delta P=P_a-P_b=2\Sigma$, becomes larger and larger. Meanwhile, the metallic behavior along $e_a$ and the insulating behavior along $e_b$ become increasingly significant. Thus, the strain-induced transition seems to be closely related to the pressure anisotropy.\,\footnote{The energy density $\mathcal{E}$ as well as the pressure $\mathcal{P}$ defined by the holographic renormalization in Appendix~\ref{appendix} can become negative at
large $k$. This is a common feature of many simple linear axion models. When $\mathcal{E}$ is negative, one has $|P_a|<|P_b|$ although $P_a-P_b=2\Sigma>0$.
There might be some
instability associated with the negative energy density that limits the application of these models, but such instability has never been found. In contrast to the commonly used renormalization in Appendix~\ref{appendix}, one could consider a different holographic renormalization scheme to kill the issue of the negative energy density, see \emph{e.g.}\cite{Li:2020spf}.} Moreover, the anisotropy can be characterized by the electronic nematicity $N$ in~\eqref{eqN}. In Fig.~\ref{fig:xzNP}, we show the temperature dependence of $N$ and $\Delta P$. One finds that both $N$ and $\Delta P$ increase with the increase of shear strain. A better understanding of this shear-induced MIT transition requires further investigation. 

\section{Conclusion and Discussion}\label{conclusion}

In this work we have initiated the study of the transport properties of strongly coupled material under shear deformation using the holographic framework. It corresponds to strained black holes that trigger the spatial anisotropy spontaneously on the gravity side.
As a building block, two sets of scalars have been introduced in our bulk theory, including $\phi^I$ that give momentum dissipation and $\varphi^I$ that incorporate the mechanical deformation. The strain-stress curves have been constructed numerically (see Fig.~\ref{fig:HardeningSoftening}). Depending on the potential $W(X,Z)$, the system can be shear hardening or softening. More complicated behavior of the strain in function of stress can be obtained for an appropriate choice of $W$.

We have obtained the thermoelectric transport subject to nonlinear mechanical deformation in terms of horizon data of the strained black holes, see~\eqref{elecDC}-\eqref{thermalDC}. Interestingly, there are off-diagonal components of the conductivity tensors due to the shear strain. Moreover, they are all symmetric instead of antisymmetric, \emph{i.e.} $\sigma_{xy}=\sigma_{yx}$, $\alpha_{xy}=\alpha_{yx}$ and $\bar{\kappa}_{xy}=\bar{\kappa}_{yx}$, thus should not be anomalous Hall transport. We stress that the time-reversal symmetry is respected in our theory (hence $\alpha=\bar{\alpha}$). We have argued that this feature is due to the anisotropy of our system under shear strain, see Fig.~\ref{fig:sigxyangle}. More precisely, the angle between the coordinate system we used to compute the thermoelectric transport and the one defined by the principal axes is precisely given by $\pi/4$, irrespective of the strain strength. This particular value of the rotation angle explains the off-diagonal components of our conductivity tensors, see~\eqref{offsigma}.

The holographic model admits a rich phenomenology even in the absence of strain. We have studied the regime of large shear deformations and how they do affect the thermoelectric transport. We have found the strain-induced metal-insulator transition. More precisely, by increasing the shear strain, there will be a good insulating phase along one of the principal axes, and meanwhile a good metallic phase along the other principal axis of stress. This feature is found to be generic in our holographic model and is irrespective of the strength of disorder and charge density, see Figs.~\ref{fig:x3sigmalkab},~\ref{fig:x3sigmaskab},~\ref{fig:xzsigmalkab} and~\ref{fig:sigmaomega1}. It resembles a strong electronic nematicity in the system (see Fig.~\ref{fig:xzNP}) associated with the spontaneous breaking of the rotational symmetry due to shear deformation. No similar features have been found in other holographic models with spatial anisotropy, see \emph{e.g.}~\cite{Khimphun:2016ikw,Khimphun:2017mqb,Liu:2021stu}\,\footnote{In contrast to our present case, in these models the spatial anisotropy was introduced by turning on a source explicitly.}. Moreover, it was found that for a large class of black holes, the ratio of thermal conductivity to electric conductivity $\bar{L}=T \sigma/\bar{\kappa}$ has an upper bound $\bar{L}\le s^2/\rho^2$~\cite{Donos:2014cya}. We have explicitly checked many cases in our model and found no violation of the bound. On the other hand, the thermal conductivity bound~\cite{Grozdanov:2015djs} has been found to be violated for large shear deformation (see Fig.~\ref{fig:kappaT}).

We have observed both the metallic and insulating phases along two principal axes for sufficiently large strain. It would be desirable to see if our theoretical results could be checked in lab experiments. To better understand this phenomenon, it will be helpful to find the IR fix point of the model. While our numerical computation suggests an $AdS_2$ geometry at finite deformation, the IR geometry at sufficiently large strain still asks for further investigation. We have limited ourselves to the pure shear deformation, and it will be interesting to consider the opposite scenario of a purely volumetric deformation in which the shear strain $\varepsilon$ is taken to be zero. It would be interesting to extend the discussion to more general cases relevant to transport properties of holographic quantum matter, such as dispersion relations and other transport coefficients. In particular, there might be some instability of the large-strained state, which, in the bulk, is captured by the quasi-normal modes for the solutions~\eqref{metric}~\cite{qnms}. We shall leave these questions to future work.

\section*{Acknowledgement}
We thank Matteo Baggioli, Wei-Jia Li and Zhuo-Yu Xian for their useful comments and suggestions. This work was partially supported by the National Natural Science Foundation of China Grants No.12122513, No.12075298, No.11991052 and No.12047503, and by the Chinese Academy of Sciences Project for Young Scientists in Basic Research YSBR-006.

\appendix

\section{Holographic Renormalization}\label{appendix}
The bulk theory~\eqref{eq:S} in the main text involves both explicit and spontaneous translation symmetry breakings. It is necessary to consider the holographic renormalization procedure which will allow us to confirm the violation of the momentum and provide the precise definition of various observables in the dual field theory. In this appendix, we give more details on the holographic renormalization for our theory~\eqref{eq:S} and refer to~\cite{Bianchi:2001kw,Skenderis:2002wp,Andrade:2013gsa} for more general discussions. To be specific, one needs to substitute the bulk configuration satisfying the equations of motion~\eqref{eom1} into the action~\eqref{eq:S} and obtain the on-shell action $S_{\rm on}$. Since $S_{\rm on}$ is divergent, the renormalization is necessary to obtain the physical results.

It is convenient to work in the  Fefferman-Graham (FG) coordinate
\begin{equation}\label{FG}
    ds^2=\frac{d\rho^2}{\rho^2}+\frac{1}{\rho^2}g_{ij}(x,\rho)dx^i dx^j\,,
\end{equation}
where $i, j$ label all boundary indices and $\rho$ is the radial coordinate that should not be confused with the charge density in the main text. In the FG coordinate, the equations of motion~\eqref{eom1} read  
\begin{align}
&\partial_{i}\left( Y(X_0)\sqrt{g} g^{i j} A_{j}^{\prime}\right)=0\,,\label{eq:Maxwell0} \\
&\partial_{\rho}\left( Y(X_0)\sqrt{g} g^{i j} A_{j}^{\prime}\right)+\partial_{k}\left( Y(X_0)\sqrt{g} g^{k l} g^{i j} F_{l j}\right)=0\,,\label{eq:Maxwelli}\\
&\partial_{\rho}\left(\frac{\partial \mathcal L_{\rm mat}}{\partial X_0}\frac{\sqrt{g}}{\rho^{2}} \partial_{\rho} \phi_{I}\right)+\rho^{-2} \partial_{i}\left(\frac{\partial \mathcal L_{\rm mat}}{\partial X_0}\sqrt{g} g^{i j} \partial_{j} \phi_{I}\right)=0\,,\label{eq:EOMphi}\\
&\partial_\rho\left({\rho^{-2}}{\sqrt{g}}M_{IJ}\varphi_J'\right)+{\rho^{-2}}\partial_i\left(\sqrt{g}g^{ij}M_{IJ}\partial_j\varphi_J\right)=0\,,\label{eq:EOMphi1}\\
&T_{\rho\rho}-\frac{1}{2 \rho^2}T=-\frac{1}{2} \operatorname{Tr}\left(g^{-1} g^{\prime \prime}\right)+\frac{1}{2 \rho} \operatorname{Tr}\left(g^{-1} g^{\prime}\right)+\frac{1}{4} \operatorname{Tr}\left(g^{-1} g^{\prime} g^{-1} g^{\prime}\right)\,,
\label{eq:Eins00} \\
&T_{\rho i }=\frac{1}{2} g^{j k}\left(\nabla_{j} g_{k i}^{\prime}-\nabla_{i} g_{j k}^{\prime}\right)\,, \label{eq:Eins0i}\\
&T_{ij}-\frac{1}{2 \rho^2}g_{ij} T=R_{i j}-\frac{1}{2} g_{i j}^{\prime \prime}+\frac{1}{2 \rho}\left[2 g_{i j}^{\prime}+\operatorname{Tr}\left(g^{-1} g^{\prime}\right) g_{i j}\right]-\frac{1}{4} \operatorname{Tr}\left(g^{-1} g^{\prime}\right) g_{i j}^{\prime}+\frac{1}{2}\left(g^{\prime} g^{-1} g^{\prime}\right)_{i j}\,,\label{eq:Einsij}
\end{align}
where
\begin{equation}
\begin{split}
&\mathcal L_{\rm mat}=-\frac{1}{4}Y(X_0)F^2-V(X_0)-2W(X,Z),\ \frac{\partial \mathcal L_{\rm mat}}{\partial X_0}=-\frac{1}{4}Y'(X_0)F^2-V'(X_0),\\
&M_{IJ}=\left(\begin{array}{cc}
W_X+2W_Z\mathcal{X}_{22}&-2W_Z\mathcal{X}_{12}\\
-2W_Z\mathcal{X}_{21}&W_X+2W_Z\mathcal{X}_{11}
\end{array}\right),\\
&T=-X_0\frac{\partial \mathcal L_{\rm mat}}{\partial X_0}+2XW_X+4ZW_Z-2V(X_0)-4W(X,Z),\\
&T_{\rho\rho}=\frac{\mathcal L_{\rm mat}}{2\rho^2}-\frac{\partial \mathcal L_{\rm mat}}{\partial X_0}\frac{1}{2}\phi_I^{\prime}\phi_I^{\prime}+\frac{1}{2}Y(X_0)\rho^2g^{ij}A_i'A_j'+\varphi'_IM_{IJ}\varphi'_J,\label{eq:EOM00}\\
&T_{\rho i}=-\frac{\partial \mathcal L_{\rm mat}}{\partial X_0}\frac{1}{2}\phi_I^{\prime}\partial_i\phi_{I}+\frac{1}{2}Y(X_0)\rho^2F_{il}g^{lk}A_k'+\partial_i\varphi_IM_{IJ}\varphi'_J\,,\\
&T_{ij}=\frac{\mathcal L_{\rm mat}}{2\rho^2}g_{ij}-\frac{\partial \mathcal L_{\rm mat}}{2\,\partial X_0}\partial_i\phi_{I}\partial_j\phi_{I}+\frac{1}{2}Y(X_0)\rho^2\left(A_i'A_j'+g^{kl}F_{ik}F_{jl}\right)+\partial_i\varphi_IM_{IJ}\partial_j\varphi_J.
\end{split}
\end{equation}
with
\begin{equation}
\begin{split}
X_0&=\frac{1}{2}\rho^2\left(\phi_I^{\prime}\phi_I^{\prime}+g^{ij}\partial_i\phi_{I}\partial_j\phi_{I}\right),\ \mathcal{X}_{IJ}=\rho^2\left(\varphi_I^{\prime}\varphi_J^{\prime}+g^{ij}\partial_i\varphi_{I}\partial_j\varphi_{J}\right)\,,\\
X&=\frac12\rho^2\left(\varphi_I^{\prime}\varphi_I^{\prime}+g^{ij}\partial_i\varphi_{I}\partial_j\varphi_{I}\right)\,,\ 
Z=\mathcal{X}_{11}\mathcal{X}_{22}-\mathcal{X}_{12}\mathcal{X}_{21}\,,\\
F^2&=\rho^4\left(2g^{ij}A_i'A_j'+g^{ij}g^{kl}F_{ik}F_{jl}\right)\,.
\end{split}
\end{equation}
All indices here are raised and lowered with $g_{ij}$ and the trace is taken by using $g_{ij}$. The prime denotes the derivative with respect to $\rho$.

Near the AdS boundary $\rho=0$, the fields can be expanded in the power of $\rho$,
\begin{equation}\label{app:uv}
\begin{split}
&g_{ij}=g^{(0)}_{ij}+\rho g^{(1)}_{ij}+\rho^2 g^{(2)}_{ij}+\rho^3 g^{(3)}_{ij}+\cdots\,,\\
&A_{i}=A^{(0)}_i+\rho A^{(1)}_i+\cdots\,,\\
&\phi_{I}=\phi^{(0)}_I+\rho \phi^{(1)}_I+\rho^2 \phi^{(2)}_I+\rho^3 \phi^{(3)}_I+\cdots\,,\\
&\varphi_{I}=\varphi^{(0)}_I+\rho \varphi^{(1)}_I+\rho^2 \varphi^{(2)}_I+\rho^3 \varphi^{(3)}_I+\cdots\,.
\end{split}
\end{equation}
For simplicity, we have considered $V(X_0)=X_0$ for explicit translation symmetry breaking for which the explicit symmetry breaking source is $\phi^{(0)}_I$. In order to realize the nonlinear mechanical deformation, we require $W\sim \rho^5$ or even faster such that $\varphi_I^{(0)}$ correspond to the vacuum expectation value of the scalar operators dual to the bulk axion fields $\varphi_I$, which matches the field theory description of mechanical deformation.\footnote{We refer to~\cite{Baggioli:2020qdg,Pan:2021cux,Baggioli:2021xuv} for a more detailed description and understanding of this mechanism.} 

According to the holographic dictionary, the leading terms $g_{ij}^{(0)},A^{(0)}_i,\phi^{(0)}_I$ are identified as the sources for the dual field theory operators corresponding to the stress tensor $\hat T_{ij}$, a global $U(1)$ current $\hat J^i$ and a scalar operator $\hat O_{\phi}$, respectively. The sub-leading terms, which turn out to be $g_{ij}^{(3)},A^{(1)}_i,\phi^{(3)}_I$ and $\varphi^{(0)}_I$ will be shown to be the expectation values.  Other coefficients of~\eqref{app:uv} can be expressed in terms of the leading and sub-leading terms. 

In the following discussion, we shall focus on 
\begin{equation}
W(X,Z)=X^mZ^n\,,
\end{equation}
with $m$ and $n$ constants. The consistency of a theory demands~\cite{Baggioli:2019elg}
\begin{equation}\label{app:mn1}
m\ge 0,\quad m+2 n>3/2 \,.
\end{equation}
To ensure the presence of massless phonons in the absence of explicit translation symmetry breaking, one should further impose~\cite{Baggioli:2019elg}
\begin{equation}
m+2 n>5/2 \,.
\end{equation}

We can then obtain some useful relations by substituting the UV expansion~\eqref{app:uv} into~\eqref{eq:Maxwell0}-\eqref{eq:Einsij}.
\begin{itemize}
\item We obtain from the lowest order of~\eqref{eq:EOMphi} and~\eqref{eq:EOMphi1} that
\begin{align}
\phi_I^{(1)}=0,\quad \varphi_I^{(1)}=0\,.
\end{align}
\item Given that $\phi_I^{(1)}=0$ and $\varphi_I^{(1)}=0$, the $\rho^0$ order of~\eqref{eq:Eins00} reads $\operatorname{Tr}\left(g^{-1} g^{(1)} g^{-1} g^{(1)}\right)=0$, which means
\begin{equation}
g^{(1)}_{ij}=0\,.
\end{equation}
\item Since we require $Y(0)=1$, the lowest order of~\eqref{eq:Maxwell0} leads to 
\begin{align}\label{app:J}
\nabla^{i(0)}A_{i}^{(1)}=0\,,
\end{align}
where $\nabla^{(0)}$ is the covariant derivative associated with $g_{ij}^{(0)}$.
\item The $\rho^{-2}$ order of~\eqref{eq:EOMphi} yields
\begin{align}
\phi_I^{(2)}=\frac12\square_0\phi_I^{(0)}\,,
\end{align}
with $\square_0=\nabla^{i(0)}\nabla_i^{(0)}$.
\item 
For the potential $W(X,Z)=X^m Z^n$ with $m+2n>1$, the lowest order of~\eqref{eq:EOMphi1} yields
\begin{align}
\varphi_K^{(2)}=\frac{-1}{2(2m+4n-3)}\big(\tilde M^{-1}\big)_{KI}\nabla_i^{(0)}\left(\tilde M_{IJ}g^{(0)ij}\nabla^{(0)}_j\varphi_J^{(0)}\right)\,,
\end{align}
where $\tilde M_{IJ}=\lim_{\rho\to0}{\rho^{-2m-4n+2}M_{IJ}}$.
\item The $\rho^1$ order of~\eqref{eq:Eins00} yields 
\begin{align}\label{app:trace}
\operatorname{Tr}\left(g^{-1} g^{(3)}\right)=0.
\end{align}
\item The $\rho^2$ order of~\eqref{eq:Eins0i} gives
\begin{align}\label{app:T}
\nabla^{j(0)}g_{ij}^{(3)}=\phi_I^{(3)}\partial_i\phi_I^{(0)}+\frac{1}{3}F_{ij}^{(0)}g^{(0)jk}A_k^{(1)}\,.
\end{align}
\item The $\rho^0$ order of~\eqref{eq:Einsij} yields
\begin{align}\label{app:g2}
g_{i j}^{(2)}=-\left(R_{i j}^{(0)}-\frac{1}{4} g_{i j}^{(0)} R^{(0)}\right)+\frac{1}{2}\left[\partial_{i} \phi_{I}^{(0)} \partial_{j} \phi_{I}^{(0)}-\frac{1}{4} g_{i j}^{(0)}\left(g^{(0) k l} \partial_{k} \phi_{I}^{(0)} \partial_{l} \phi_{I}^{(0)}\right)\right].
\end{align}
\end{itemize}

Substituting the field configurations that satisfy the equations of motion~\eqref{eom1}, we obtain the on-shell action that is divergent near the AdS boundary.
\begin{align}
\mathcal{S}_{\rm 0}=&\int d^{4} x \frac{\sqrt{g(x,\rho)}}{\rho^4}\left[2 \Lambda+2W(X,Z)-2X W_X-4Z W_Z-\frac{Y(X_0)+X_0Y'(X_0)}{4}F^2\right]\notag\\
&-\int_\partial d^{3} x\sqrt{\gamma}2 K\,,\label{eq:Sonshell}
\end{align}
where we have added the Gibbons-Hawking boundary term for a well-defined Dirichlet variational principle. Here $\gamma$ is the induced metric on the boundary and $K$ is the trace of the extrinsic curvature.\footnote{Denoting $n_\mu$ to be the outward pointing unit normal to the boundary, we define the extrinsic curvature $K_{\mu\nu}=-\frac{1}{2}\left(\nabla_{\mu}n_\nu+\nabla_{\nu}n_\mu\right)$.} We have also used the trace of the Einstein's equation $R=4\Lambda-T$ to eliminate the scalar curvature. 
Note that for the potential $W(X,Z)=X^mZ^n$ satisfying~\eqref{app:mn1}, there is no divergence from $\varphi_I$. 

We now proceed to calculate the regularized action by introducing a UV cutoff at $\rho=\epsilon$.
\begin{align}
\mathcal{S}_{\rm reg}&=\int_{\rho\ge\epsilon} d\rho d^{3} x \frac{\sqrt{g(x,\rho)}}{\rho^4}\left[2 \Lambda+2W(X,Z)-2X W_X-4Z W_Z-\frac{Y(X_0)+X_0Y'(X_0)}{4}
F^2\right]\,,\notag\\
&\ \ \ \ -2\int_{\rho=\epsilon} d^{3} x\sqrt{\gamma} K\\
&=\int_{\rho\ge\epsilon} d\rho d^{3} x\frac{\sqrt{g^{(0)}}}{\rho^4}2 \Lambda -2\int_{\rho=\epsilon} d^{3} x\sqrt{\gamma} K+\text{finite terms}\\
&=\int d^{3} x \sqrt{g^{(0)}}\left[\frac{4}{\epsilon^3}-\frac{2}{\epsilon}\operatorname{Tr}g^{(2)}\right]+\text{finite terms}\\
&=\int d^{3} x \sqrt{g^{(0)}}\left[\frac{4}{\epsilon^3}+\frac{1}{4\epsilon}\left(2R^{(0)}[g]-g_{ij}^{(0)}\partial_i\phi_I^{(0)}\partial_j\phi_I^{(0)}\right)\right]+\text{finite terms}\label{eq:Sreg}
\end{align}
where we have used 
\begin{align}
K=-\nabla_\mu n^\mu=\sqrt\frac{{g^{(0)}}}{{\gamma}}\left(-\frac{3}{\rho^3}-\frac{\operatorname{Tr}g^{(2)}}{2\rho}\right)\,,
\end{align}
and
\begin{align}
\operatorname{Tr}g^{(2)}=-\frac{1}{4}R^{(0)}[g]+\frac{1}{8}g_{ij}^{(0)}\partial_i\phi_I^{(0)}\partial_j\phi_I^{(0)}\,,
\end{align}
from~\eqref{app:g2}.

The divergent terms must be cancelled by counter terms in order to obtain the renormalized action. Notice that $\sqrt{g^{(0)}}=\rho^3\sqrt{\gamma}\left(1-\frac{1}{2}\rho^2\operatorname{Tr}g^{(2)}\right)$. We can obtain the counter term $\mathcal{S}_{\rm ct}$ and the renormalized action $\mathcal{S}_{\rm ren}$
\begin{align}
&\mathcal{S}_{\rm ct}=\int_{\partial \mathcal{M}} d^{3} x \sqrt{\gamma}\left[-4-R[\gamma]-\frac{1}{2}\phi_I\square_{\gamma}\phi_I\right]\label{eq:Sct}\,,\\
&\mathcal{S}_{\rm ren}=\mathcal{S}_{0}+\mathcal{S}_{ct}\,.\label{eq:Sren}
\end{align}
The on-shell variation of the renormalized action~\eqref{eq:Sren} directly yields\,\footnote{Note that there is no contribution from $\varphi_I$ since we have already set the source of $\varphi_I$ to be vanishing.}
\begin{align}
\delta\mathcal{S}_{\rm ren}=\int_\partial d^{3} x \sqrt{-g^{(0)}}\left[\frac{3}{2}g^{(3)ij}\delta g_{ij}^{(0)}+3\phi_I^{(3)}\delta\phi_I^{(0)}+A^{(1)i}\delta A_i^{(0)}\right]\,,
\end{align}
from which we have
\begin{equation}
\left\langle \hat T^{i j}\right\rangle=3 g^{(3)ij},\quad \left\langle \hat J^{i}\right\rangle=A^{(1)i},\quad \left\langle \hat O_{(\phi) I}\right\rangle=3\phi_I^{(3)}\,.
\end{equation}
Moreover, using~\eqref{app:J},~\eqref{app:trace} and~\eqref{app:T}, we obtain the following Ward identities:
\begin{align}
\nabla_i^{(0)} \left\langle \hat J^{i}\right\rangle=0,\qquad  \left\langle \hat {T^i}_i \right\rangle=0\,,\label{app:trace0}\\
\nabla_j^{(0)}\left\langle\hat T^{ji}\right\rangle=F^{(0)ij}  \left\langle \hat J^{j}\right\rangle+\left\langle \hat O_{(\phi) I}\right\rangle\nabla_i^{(0)}\phi_I^{(0)}\,.\label{app:Ward}
\end{align}

So far, we have considered the EF coordinate~\eqref{FG}. For our coordinate system~\eqref{metric} in the main text, we can directly obtain the general response of the boundary theory using the renormalized action~\eqref{eq:Sren}.
\begin{align}
\left\langle \hat T_{i j}\right\rangle &=\lim _{u \rightarrow 0}- \frac{2}{u \sqrt{\gamma}} \frac{\delta S_{r e n}}{\delta \gamma^{i j}}\,,\nonumber \\
&=\lim _{u \rightarrow 0} u^{-1}\left[2\left(K_{i j}-K \gamma_{i j}+G_{[\gamma] i j}-2 \gamma_{i j}\right)+\frac{1}{2} \gamma_{i j} \partial \phi_{I} \cdot \partial \phi_{I}-\nabla_{i} \phi_{I} \nabla_{j} \phi_{I}\right]\,,\label{app:Tij}\\
\left\langle \hat J^{i}\right\rangle &=\lim _{u \rightarrow 0} \frac{1}{u^3 \sqrt{\gamma}} \frac{\delta S_{r e n}}{\delta A_{i}}=\lim _{u \rightarrow 0}u^{-3} Y(X_0) n^\mu F_\mu^i\,,\\
\left\langle \hat O_{(\phi) I}\right\rangle&=\lim _{u \rightarrow 0} \frac{1}{u^{3} \sqrt{\gamma}} \frac{\delta S_{r e n}}{\delta \phi_{I}}=-\lim _{u \rightarrow 0} u^{-3}\left[\left(1-\frac{F^2}{4}Y'(X_0)\right)n^{\mu} \partial_{\mu} \phi_{I}+\square_{\gamma} \phi_{I}\right]\,.
\end{align}
Moreover, the transformation between the two coordinate systems are given by
\begin{equation}
 \frac{d\rho^2}{\rho^2} =\frac{d u^2}{u^2f(u)} \,.
\end{equation}
One should obtain the same boundary expectation values using the above two approaches. From the last Ward identity~\eqref{app:Ward}, one can show that, at the linear level in the fluctuations, momentum in the dual field theory is not conserved due to the bulk profile of $\phi_I$~\eqref{metric}. 

We now look at the background~\eqref{metric} and derive the stress-energy tensor $\left\langle T_{i j}\right\rangle$. In this case~\eqref{metric}, we have $G_{[\gamma]i j}=0$ and
\begin{equation}
\begin{split}
&K_{ij}=\frac{\sqrt{f(u)}}{2u^2}\left(
\begin{array}{cccc}
 K_{tt} & 0 & 0  \\
 0 & K_{xx} & K_{xy} \\
 0 & K_{yx} & K_{yy} \\
\end{array}
\right)\,,\\
&K_{tt}=e^{-\chi (u)} \left(-u f'(u)+u f(u) \chi '(u)+2 f(u)\right)\,,\\
&K_{xx}=K_{yy}=u h'(u) \sinh h(u)-2 \cosh h(u)\,,\\
&K_{xy}=K_{yx}=u h'(u) \cosh h(u)-2 \sinh h(u)\,,\\
&K=\frac{1}{2 \sqrt{f(u)}}\left(u f'(u)-f(u) \left(u \chi '(u)+6\right)\right)\,,\\
&\gamma_{i j}=\frac{1}{u^{2}}\left(\begin{array}{ccc}
e^{-\chi(u)} f(u) & 0 & 0 \\
0 & \cosh h(u) & \sinh h(u) \\
0 & \sinh h(u) & \cosh h(u)
\end{array}\right)\,,
\end{split}
\end{equation}
at a given hypersurface with $u$ fixed.
Substituting the UV expansion~\eqref{adsUV} into~\eqref{app:Tij} and taking the limit $u\rightarrow0$, we can obtain the energy density $\mathcal{E}$, the isotropic pressure $\mathcal{P}$ and the shear stress $\Sigma$:
\begin{equation}\label{EP}
\begin{split}
\mathcal{E}\equiv\left\langle \hat T_{tt}\right\rangle=-2f^{(3)}(0)\,,\\
\mathcal{P}\equiv\left\langle \hat T_{xx}\right\rangle=\left\langle \hat T_{yy}\right\rangle=-f^{(3)}(0)\,,\\
\Sigma\equiv\left\langle \hat T_{xy}\right\rangle=\left\langle \hat T_{yx}\right\rangle=3 h^{(3)}(0)\,.\\
\end{split}
\end{equation}
One immediately finds that $\mathcal{E}=2\mathcal{P}$ as required by the Ward identity~\eqref{app:trace0}. The stress tensor $\tau$ is then given by
\begin{equation}\label{straintensor}
\tau=\begin{pmatrix}
\mathcal{P}      &  \Sigma  \\
  \Sigma   &  \mathcal{P}
\end{pmatrix}\,.
\end{equation}
By a simple change of reference frame which diagonalizes the above matrix, one has
\begin{equation}
\tilde{\tau}=\begin{pmatrix}
P_a      &  0  \\
0   &  P_b
\end{pmatrix}\,\quad \text{with}\quad  P_a=\mathcal{P}+\Sigma,\quad P_b=\mathcal{P}-\Sigma\,.
\end{equation}
Then, we can define the pressure anisotropy
\begin{equation}
 \Delta=P_a-P_b=2\Sigma\,,
\end{equation}
which characterizes the anisotropy of our system under shear strain.

\section{Strain Effect on Conductivity by Dialing Charge Density}\label{app:charge}
In order to confirm our findings in the main text, we consider the strain effect on the electric conductivity by dialing the charge density. We focus on the potential $W=X^3$ that corresponds to the shear hardening. 
\begin{figure}[H]
    \centering
    \includegraphics[width=0.48\linewidth]{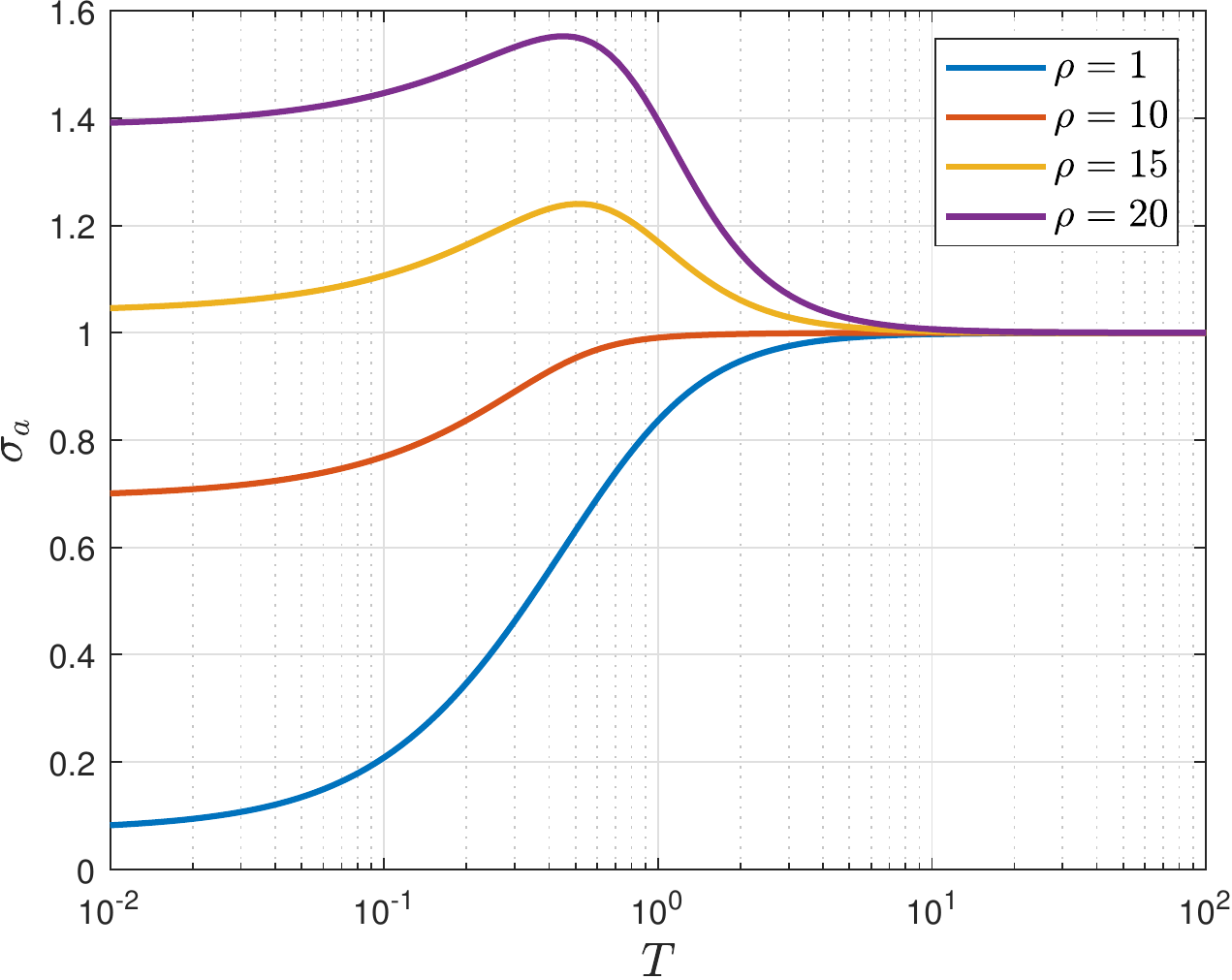}
     \includegraphics[width=0.48\linewidth]{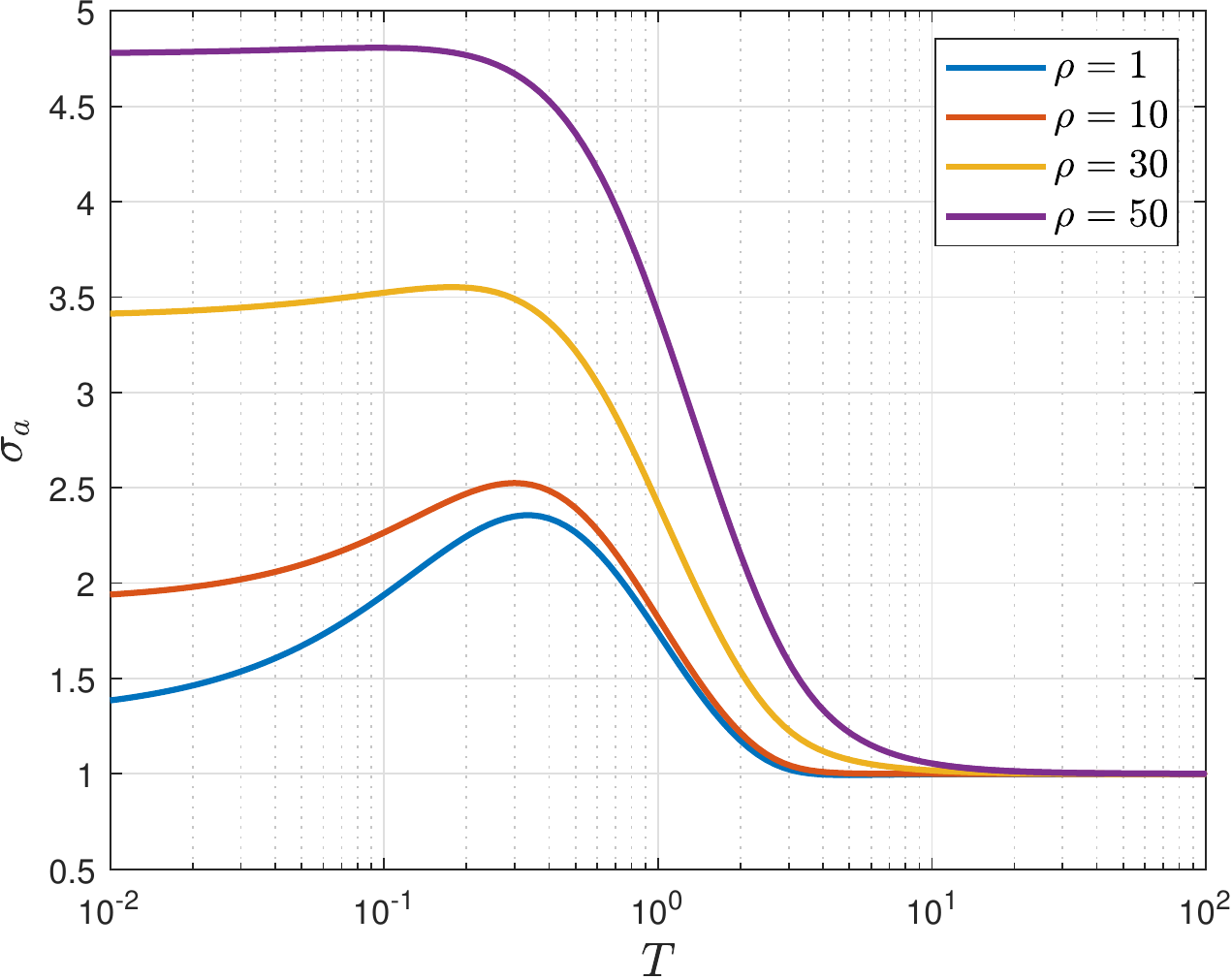}
      \includegraphics[width=0.48\linewidth]{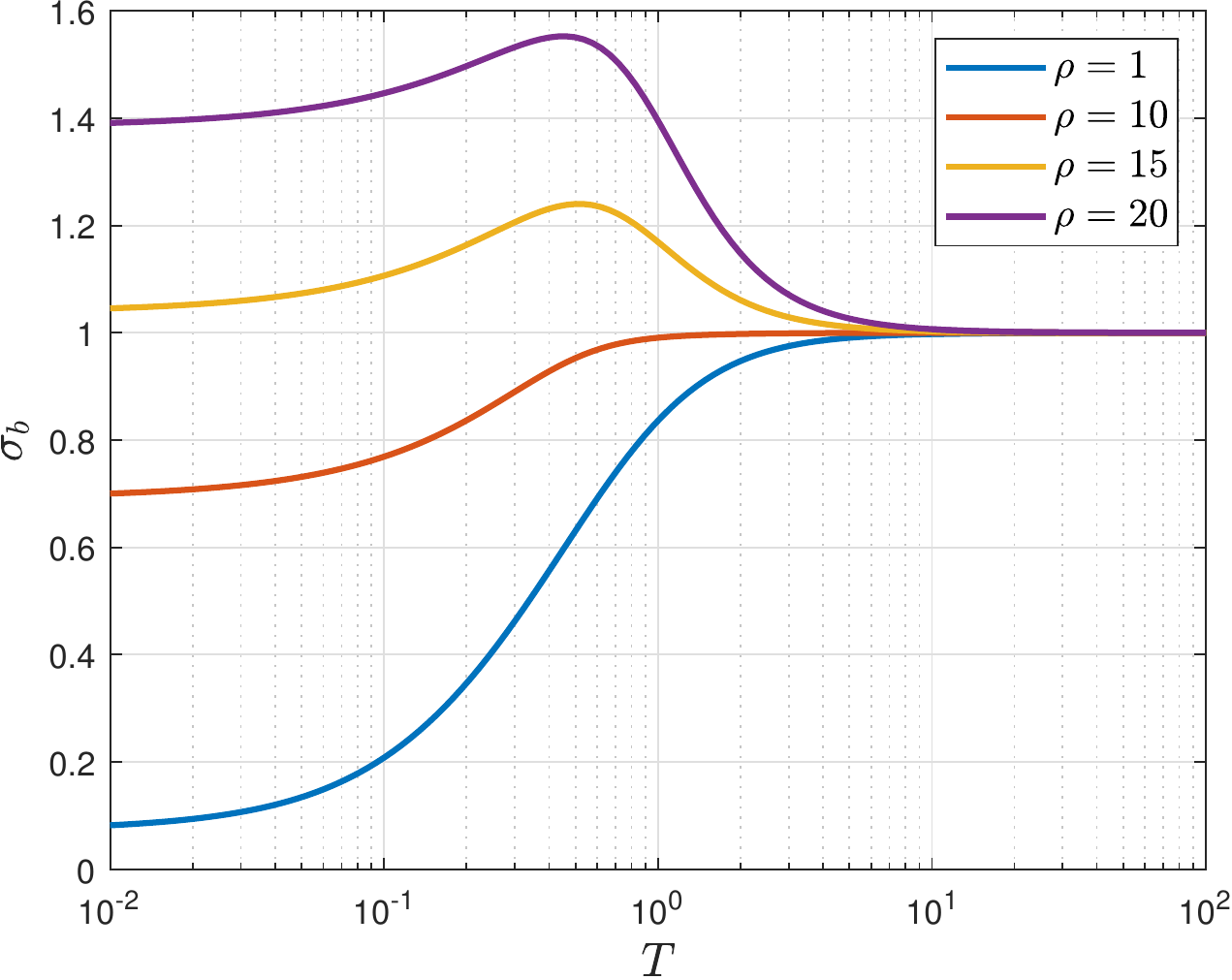}
       \includegraphics[width=0.48\linewidth]{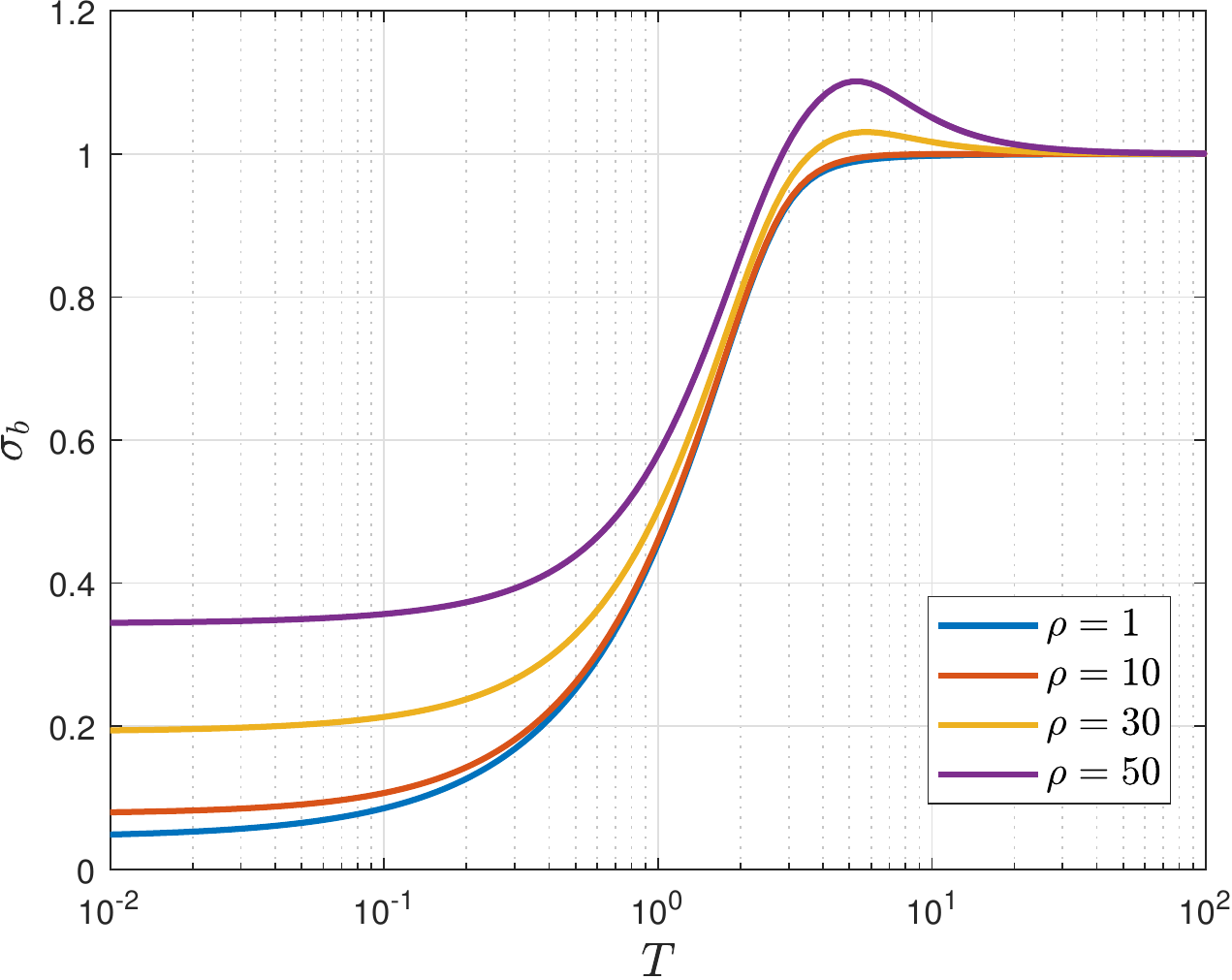}
    \caption{Electric conductivity with respect to temperature under different charge densities for $W(X,Z)=X^3$. \textbf{Left:} The case at small shear deformation with $\Omega=0.1$. \textbf{Right:} The one with large shear deformation $\Omega=5$. We have fixed $k=\beta=1$.}
    \label{fig:sigmarho}
\end{figure}

The electric conductivity in the presence of shear strain with different charge densities is shown in Fig.~\ref{fig:sigmarho}. One can see clearly that increasing the charge density will increase electric conductivity. For sufficiently large charge density, there is a metallic behavior above a particular temperature. When the charge density is small, one has an insulating behavior. Comparing the left panel to the right one, one finds that a metallic behavior along the $\bm{e}_a$ direction will be enhanced or induced as the shear deformation is increased.

\begin{figure}[H]
    \centering
    \includegraphics[width=0.49\linewidth]{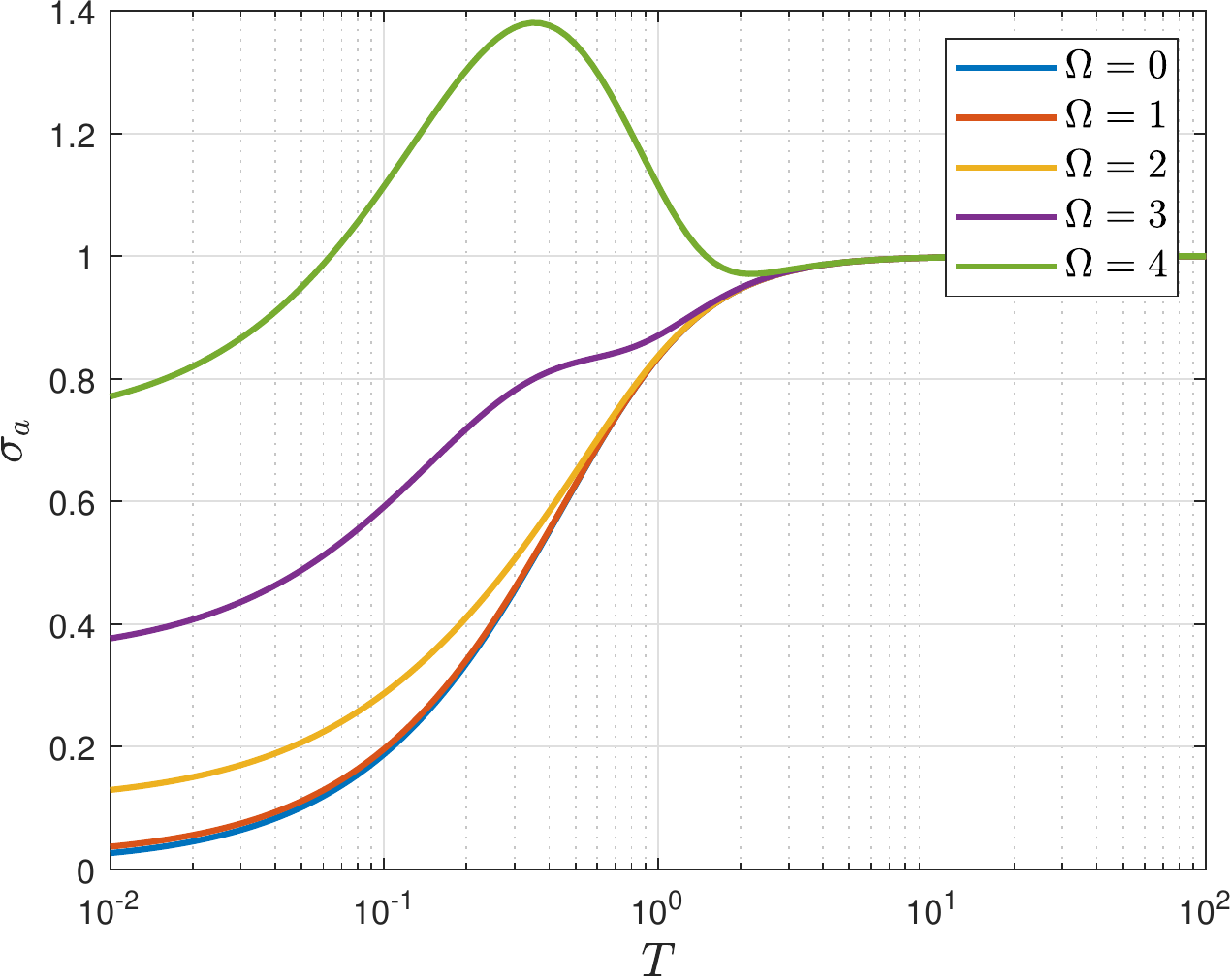}
    \includegraphics[width=0.49\linewidth]{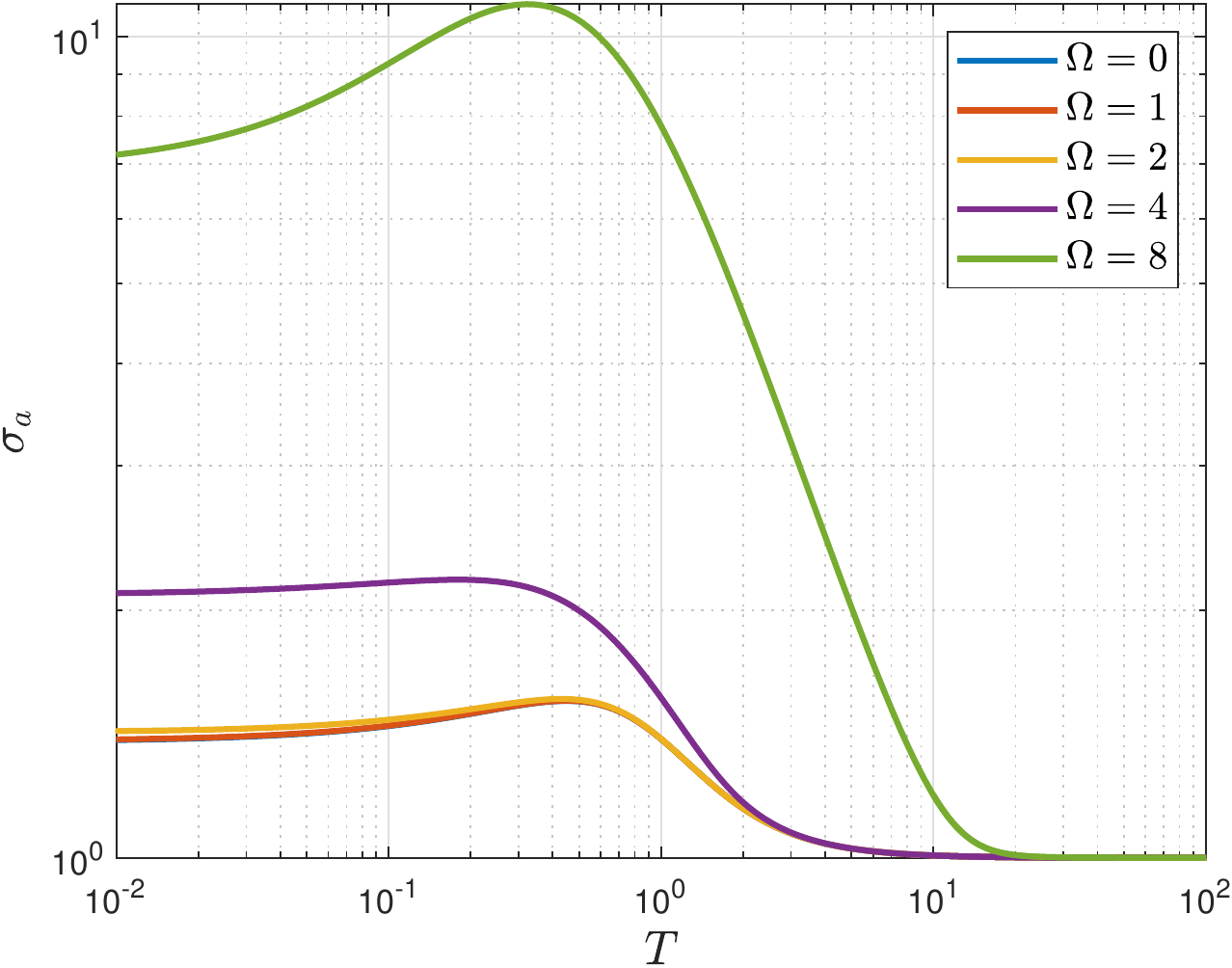}
   \includegraphics[width=0.49\linewidth]{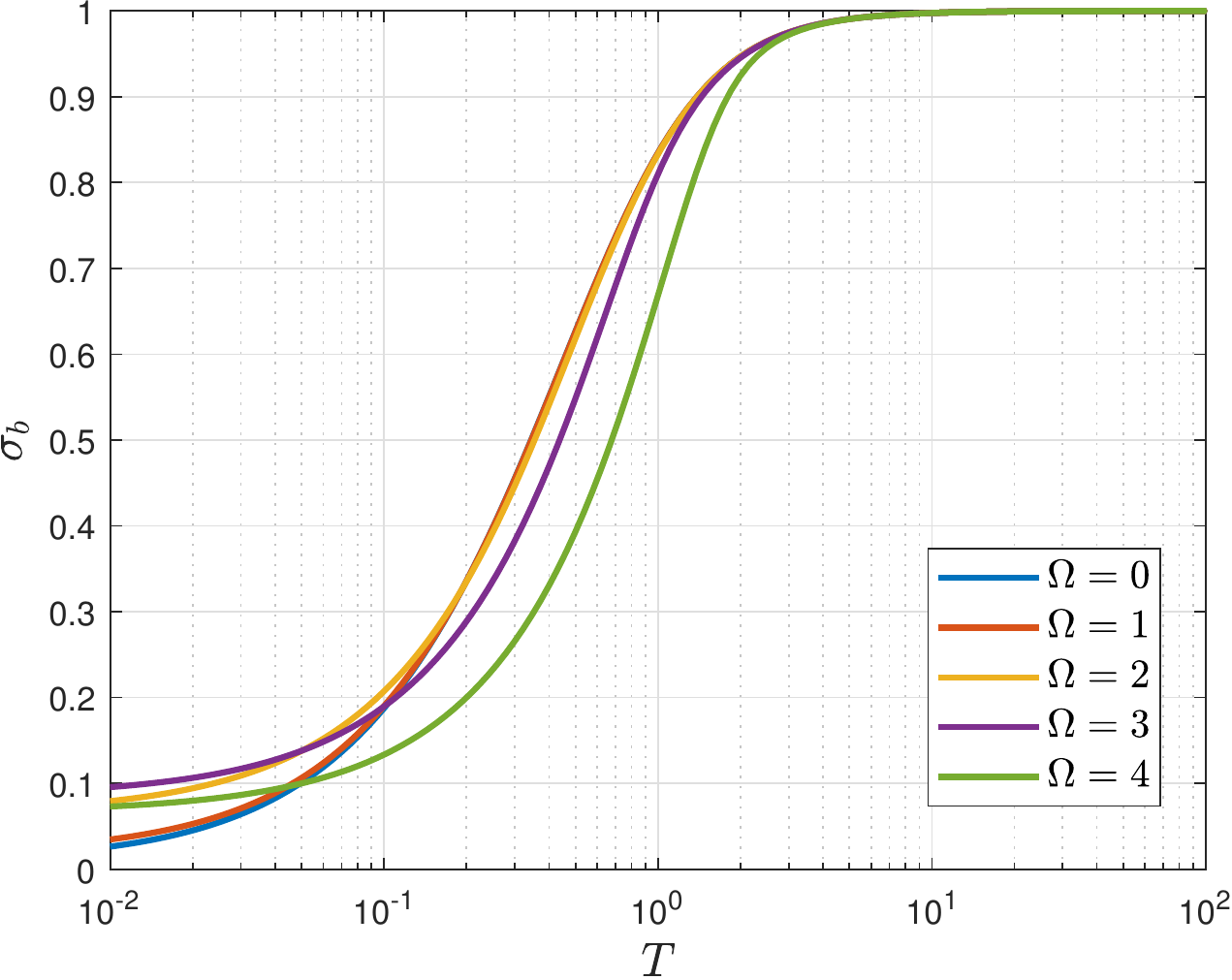}
    \includegraphics[width=0.49\linewidth]{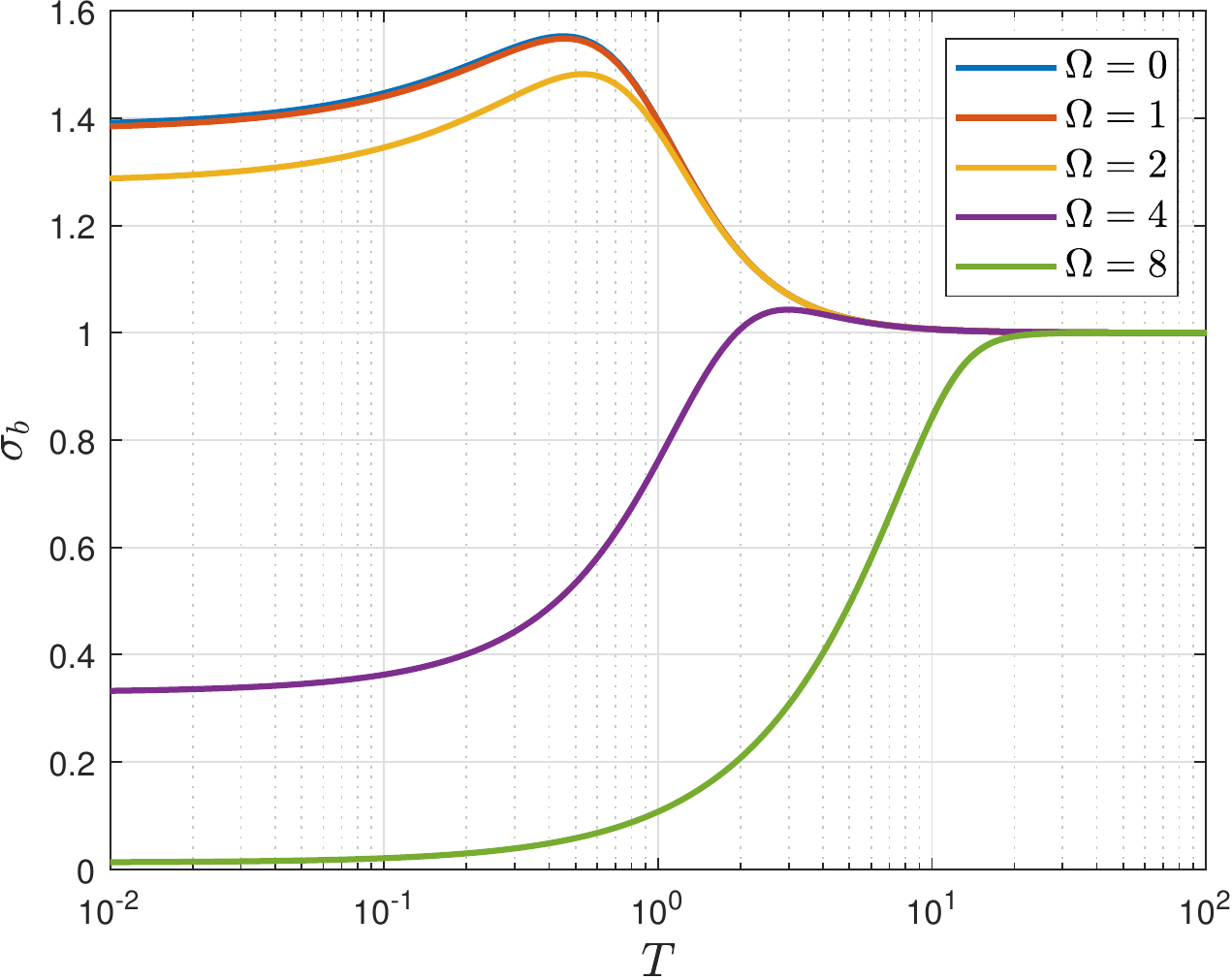}
    \caption{Electric conductivity in function of temperature under different shear strain for $W(X,Z)=X^3$. \textbf{Left:} The case at charge density with $\rho=0.1$. \textbf{Right:} The one with large charge density $\rho=20$. We have fixed $k=\beta=1$.}\label{fig:sigmaomega1}
\end{figure}
In order to see this feature more clearly, we plot the temperature dependence of the conductivity under different shear strain in Fig.~\ref{fig:sigmaomega1}. In the left panel, the unstrained case is an insulating phase. As the shear deformation is increased, while it remains an insulating behavior along $\bm{e}_b$, there induces a metallic behavior along $\bm{e}_a$. In the left panel, the unstrained case has a metallic behavior above a certain temperature. By increasing the shear deformation, the metallic behavior along $\bm{e}_a$ is enhanced, while the one along $\bm{e}_b$ is suppressed and is replaced by an insulating phase for sufficiently large deformation.
It is clear that strain engineering induces a good insulating phase along the principal axis $\bm{e}_b$ and a metallic phase along the other principal axis simultaneously, irrespective of the strength of charge density.

\providecommand{\href}[2]{#2}\begingroup\raggedright\endgroup


\begin{thebibliography}{1}


\bibitem{Kovtun:2004de}
P.~Kovtun, D.~T.~Son and A.~O.~Starinets,
``Viscosity in strongly interacting quantum field theories from black hole physics,''
Phys. Rev. Lett. \textbf{94}, 111601 (2005)
[arXiv:hep-th/0405231 [hep-th]].


\bibitem{Hartnoll:2016apf}
S.~A.~Hartnoll, A.~Lucas and S.~Sachdev,
``Holographic quantum matter,''
[arXiv:1612.07324 [hep-th]].

\bibitem{Baggioli:2022pyb}
M.~Baggioli and B.~Gout\'eraux,
``Effective and holographic theories of strongly-correlated phases of matter with broken translations,''
[arXiv:2203.03298 [hep-th]].

\bibitem{Cai:2015cya}
R.~G.~Cai, L.~Li, L.~F.~Li and R.~Q.~Yang,
``Introduction to Holographic Superconductor Models,''
Sci. China Phys. Mech. Astron. \textbf{58}, no.6, 060401 (2015)
[arXiv:1502.00437 [hep-th]].

\bibitem{Zaanen:2021llz}
J.~Zaanen,
``Lectures on quantum supreme matter,''
[arXiv:2110.00961 [cond-mat.str-el]].

\bibitem{Liu:2020rrn}
H.~Liu and J.~Sonner,
``Quantum many-body physics from a gravitational lens,''
Nature Rev. Phys. \textbf{2}, no.11, 615-633 (2020)
[arXiv:2004.06159 [hep-th]].

\bibitem{Landsteiner:2019kxb}
K.~Landsteiner, Y.~Liu and Y.~W.~Sun,
``Holographic topological semimetals,''
Sci. China Phys. Mech. Astron. \textbf{63}, no.5, 250001 (2020)
[arXiv:1911.07978 [hep-th]].


\bibitem{Chen:2022goa}
Y.~Chen, D.~Li and M.~Huang,
``The dynamical holographic QCD method for hadron physics and QCD matter,''
[arXiv:2206.00917 [hep-ph]].

\bibitem{Kim:2010zq}
B.~S.~Kim, E.~Kiritsis and C.~Panagopoulos,
``Holographic quantum criticality and strange metal transport,''
New J. Phys. \textbf{14}, 043045 (2012)
[arXiv:1012.3464 [cond-mat.str-el]].

\bibitem{Amoretti:2015gna}
A.~Amoretti and D.~Musso,
``Magneto-transport from momentum dissipating holography,''
JHEP \textbf{09}, 094 (2015)
[arXiv:1502.02631 [hep-th]].

\bibitem{Kiritsis:2016cpm}
E.~Kiritsis and L.~Li,
``Quantum Criticality and DBI Magneto-resistance,''
J. Phys. A \textbf{50} (2017) no.11, 115402
[arXiv:1608.02598 [cond-mat.str-el]].


\bibitem{Ge:2016sel}
X.~H.~Ge, Y.~Tian, S.~Y.~Wu and S.~F.~Wu,
``Hyperscaling violating black hole solutions and Magneto-thermoelectric DC conductivities in holography,''
Phys. Rev. D \textbf{96} (2017) no.4, 046015
[arXiv:1606.05959 [hep-th]].


\bibitem{Blauvelt:2017koq}
E.~Blauvelt, S.~Cremonini, A.~Hoover, L.~Li and S.~Waskie,
``Holographic model for the anomalous scalings of the cuprates,''
Phys. Rev. D \textbf{97} (2018) no.6, 061901
[arXiv:1710.01326 [hep-th]].

\bibitem{Cremonini:2018kla}
S.~Cremonini, A.~Hoover, L.~Li and S.~Waskie,
``Anomalous scalings of cuprate strange metals from nonlinear electrodynamics,''
Phys. Rev. D \textbf{99}, no.6, 061901 (2019)
[arXiv:1812.01040 [hep-th]].



\bibitem{Amorim:2015bga}
B.~Amorim, A.~Cortijo, F.~de Juan, A.~G.~Grushin, F.~Guinea, A.~Guti\'errez-Rubio, H.~Ochoa, V.~Parente, R.~Rold\'an and P.~San-Jos\'e, \textit{et al.}
``Novel effects of strains in graphene and other two dimensional materials,''
Phys. Rept. \textbf{617}, 1-54 (2016)
[arXiv:1503.00747 [cond-mat.mes-hall]].



\bibitem{Nicolas:2018}
A.~Nicolas, E.~E.~Ferrero, K.~Martens and J.~L.~Barrat,
``Deformation and flow of amorphous solids: An updated review of mesoscale elastoplastic models,''
Rev. Mod. Phys. \textbf{90}, 45006 (2018)
[arXiv:1708.09194 [cond-mat.dis-nn]].



\bibitem{Abrecht:2003}
M.~Abrecht, D.~Ariosa, D.~Cloetta, S.~Mitrovic, M.~Onellion, X.~X.~Xi, G.~Margaritondo and D.~Pavuna,
``Strain and High Temperature Superconductivity: Unexpected Results from Direct Electronic Structure Measurements in Thin Films,''
Phys. Rev. Lett. \textbf{91}, 057002 (2003).

\bibitem{Malinowski:2020}
P.~Malinowski, Q.~Jiang, J.~J.~Sanchez, J.~Mutch, Z.~Liu, P.~Went, J.~Liu, P.~J.~Ryan, J.-W.~Kim and J.-H.~Chu, 
``Suppression of superconductivity by anisotropic strain near a nematic quantum critical point,” 
Nat. Phys. \textbf{16}, 1189 (2020). 

\bibitem{Kostylev:2019ezg}
I.~Kostylev, S.~Yonezawa, Z.~Wang, Y.~Ando and Y.~Maeno,
``Uniaxial-strain control of nematic superconductivity in Sr$_{x}$Bi$_{2}$Se$_{3}$,''
Nature Commun. \textbf{11}, no.1, 4152 (2020)
[arXiv:1910.03252 [cond-mat.supr-con]].



\bibitem{Hameed:2005}
S.~Hameed, D.~Pelc, Z.~W.~Anderson, A.~Klein, R.~J.~Spieker, L.~Yue, B.~Das, J.~Ramberger, M.~Lukas, Y.~Liu, \textit{et al.}
``Enhanced superconductivity and ferroelectric quantum criticality in plastically deformed strontium titanate,''
Nature Materials \textbf{21}, 54–61 (2022)
[arXiv:2005.00514 [cond-mat.supr-con]].

\bibitem{Baggioli:2020qdg}
M.~Baggioli, V.~C.~Castillo and O.~Pujolas,
``Black Rubber and the Non-linear Elastic Response of Scale Invariant Solids,''
JHEP \textbf{09}, 013 (2020)
[arXiv:2006.10774 [hep-th]].


\bibitem{Pan:2021cux}
D.~Pan, T.~Ji, M.~Baggioli, L.~Li and Y.~Jin,
``Nonlinear elasticity, yielding, and entropy in amorphous solids,''
Sci. Adv. \textbf{8}, no.22, abm8028 (2022)
[arXiv:2108.13124 [cond-mat.soft]].

\bibitem{Baggioli:2022aft}
M.~Baggioli and G.~Frangi,
``Holographic supersolids,''
JHEP \textbf{06}, 152 (2022)
[arXiv:2202.03745 [hep-th]].


\bibitem{Baggioli:2021tzr}
M.~Baggioli, L.~Li and H.~T.~Sun,
``Shear Flows in Far-From-Equilibrium Strongly Coupled Fluids,''
Phys. Rev. Lett. \textbf{129}, no.1, 011602 (2022)
[arXiv:2112.14855 [hep-th]].

\bibitem{Baggioli:2019mck}
M.~Baggioli, S.~Grieninger and H.~Soltanpanahi,
``Nonlinear Oscillatory Shear Tests in Viscoelastic Holography,''
Phys. Rev. Lett. \textbf{124}, no.8, 081601 (2020)
[arXiv:1910.06331 [hep-th]].



\bibitem{Donos:2015gia}
A.~Donos and J.~P.~Gauntlett,
``Navier-Stokes Equations on Black Hole Horizons and DC Thermoelectric Conductivity,''
Phys. Rev. D \textbf{92}, no.12, 121901 (2015)
[arXiv:1506.01360 [hep-th]].

\bibitem{Banks:2015wha}
E.~Banks, A.~Donos and J.~P.~Gauntlett,
``Thermoelectric DC conductivities and Stokes flows on black hole horizons,''
JHEP \textbf{10}, 103 (2015)
[arXiv:1507.00234 [hep-th]].

\bibitem{Andrade:2013gsa}
T.~Andrade and B.~Withers,
``A simple holographic model of momentum relaxation,''
JHEP \textbf{05}, 101 (2014)
[arXiv:1311.5157 [hep-th]].

\bibitem{Baggioli:2021xuv}
M.~Baggioli, K.~Y.~Kim, L.~Li and W.~J.~Li,
``Holographic Axion Model: a simple gravitational tool for quantum matter,''
Sci. China Phys. Mech. Astron. \textbf{64}, no.7, 270001 (2021)
[arXiv:2101.01892 [hep-th]].


\bibitem{Imada:1998zz}
M.~Imada, A.~Fujimori and Y.~Tokura,
``Metal-insulator transitions,''
Rev. Mod. Phys. \textbf{70}, 1039-1263 (1998).

\bibitem{MIT:2011}
V.~Dobrosavljevic,
``Introduction to Metal-Insulator Transitions,''
[arXiv:1112.6166 [cond-mat.str-el]].

\bibitem{Baggioli:2016oqk}
M.~Baggioli and O.~Pujolas,
``On holographic disorder-driven metal-insulator transitions,''
JHEP \textbf{01}, 040 (2017)
[arXiv:1601.07897 [hep-th]].

\bibitem{Gouteraux:2016wxj}
B.~Gout\'eraux, E.~Kiritsis and W.~J.~Li,
``Effective holographic theories of momentum relaxation and violation of conductivity bound,''
JHEP \textbf{04}, 122 (2016)
[arXiv:1602.01067 [hep-th]].


\bibitem{An:2020tkn}
Y.~S.~An, T.~Ji and L.~Li,
``Magnetotransport and Complexity of Holographic Metal-Insulator Transitions,''
JHEP \textbf{10}, 023 (2020)
[arXiv:2007.13918 [hep-th]].

\bibitem{Grozdanov:2015djs}
S.~Grozdanov, A.~Lucas and K.~Schalm,
``Incoherent thermal transport from dirty black holes,''
Phys. Rev. D \textbf{93}, no.6, 061901 (2016)
[arXiv:1511.05970 [hep-th]].

\bibitem{Donos:2019hpp}
A.~Donos, D.~Martin, C.~Pantelidou and V.~Ziogas,
``Incoherent hydrodynamics and density waves,''
Class. Quant. Grav. \textbf{37}, no.4, 045005 (2020)
[arXiv:1906.03132 [hep-th]].

\bibitem{Amoretti:2019kuf}
A.~Amoretti, D.~Are\'an, B.~Gout\'eraux and D.~Musso,
``Gapless and gapped holographic phonons,''
JHEP \textbf{01}, 058 (2020)
[arXiv:1910.11330 [hep-th]].

\bibitem{Alberte:2018doe}
L.~Alberte, M.~Baggioli, V.~C.~Castillo and O.~Pujolas,
``Elasticity bounds from Effective Field Theory,''
Phys. Rev. D \textbf{100}, no.6, 065015 (2019)
[erratum: Phys. Rev. D \textbf{102}, no.6, 069901 (2020)]
[arXiv:1807.07474 [hep-th]].

\bibitem{Baggioli:2016pia}
M.~Baggioli, B.~Gout\'eraux, E.~Kiritsis and W.~J.~Li,
``Higher derivative corrections to incoherent metallic transport in holography,''
JHEP \textbf{03}, 170 (2017)
[arXiv:1612.05500 [hep-th]].

\bibitem{Grandi:2021bsp}
N.~Grandi, V.~Juri\v{c}i\'c, I.~Salazar Landea and R.~Soto-Garrido,
``Towards holographic flat bands,''
JHEP \textbf{05}, 123 (2021)
[arXiv:2103.01690 [hep-th]].

\bibitem{Carmine:2014}
Carmine Ortix, 
``Nonlinear Hall effect with time-reversal symmetry: Theory and material realizations,"
Advanced Quantum Technologies 2100056 (2021)
[arXiv:2104.06690 [cond-mat.meshall]].

\bibitem{Wu:2017}
J.~Wu, A.~T.~Bollinger, X.~He and I.~Božović,
``Spontaneous breaking of rotational symmetry in copper oxide superconductors,''
Nature \textbf{547}, 432–435 (2017).


\bibitem{Liu:2018hzo}
H.~S.~Liu,
``Violation of Thermal Conductivity Bound in Horndeski Theory,''
Phys. Rev. D \textbf{98}, no.6, 061902 (2018)
[arXiv:1804.06502 [hep-th]].

\bibitem{Figueroa:2020tya}
J.~P.~Figueroa and K.~Pallikaris,
``Quartic Horndeski, planar black holes, holographic aspects and universal bounds,''
JHEP \textbf{09}, 090 (2020)
[arXiv:2006.00967 [hep-th]].

\bibitem{Li:2020spf}
L.~Li,
``On Thermodynamics of AdS Black Holes with Scalar Hair,''
Phys. Lett. B \textbf{815}, 136123 (2021)
[arXiv:2008.05597 [gr-qc]].


\bibitem{Khimphun:2016ikw}
S.~Khimphun, B.~H.~Lee and C.~Park,
``Conductivities in an anisotropic medium,''
Phys. Rev. D \textbf{94}, no.8, 086005 (2016)
[arXiv:1604.00156 [hep-th]].

\bibitem{Khimphun:2017mqb}
S.~Khimphun, B.~H.~Lee, C.~Park and Y.~L.~Zhang,
``Anisotropic dyonic black brane and its effects on holographic conductivity,''
JHEP \textbf{10}, 064 (2017)
[arXiv:1705.00862 [hep-th]].

\bibitem{Liu:2021stu}
P.~Liu and J.~P.~Wu,
``Dynamic properties of two-dimensional latticed holographic system,''
JHEP \textbf{02}, 119 (2022)
[arXiv:2104.04189 [hep-th]].

\bibitem{Donos:2014cya}
A.~Donos and J.~P.~Gauntlett,
``Thermoelectric DC conductivities from black hole horizons,''
JHEP \textbf{11}, 081 (2014)
[arXiv:1406.4742 [hep-th]].

\bibitem{qnms}
M.~Baggioli, L.~Li, W.~J.~Li and H.~T.~Sun,
in progress.

\bibitem{Bianchi:2001kw}
M.~Bianchi, D.~Z.~Freedman and K.~Skenderis,
``Holographic renormalization,''
Nucl. Phys. B \textbf{631}, 159-194 (2002)
[arXiv:hep-th/0112119 [hep-th]].

\bibitem{Skenderis:2002wp}
K.~Skenderis,
``Lecture notes on holographic renormalization,''
Class. Quant. Grav. \textbf{19}, 5849-5876 (2002)
[arXiv:hep-th/0209067 [hep-th]].


\bibitem{Baggioli:2019elg}
M.~Baggioli, V.~C.~Castillo and O.~Pujolas,
``Scale invariant solids,''
Phys. Rev. D \textbf{101}, no.8, 086005 (2020)
[arXiv:1910.05281 [hep-th]].

\end{thebibliography}
\end{document}